\documentclass[a4paper,oneside,11pt]{article}
\usepackage{amsmath}

\usepackage{amsmath,amstext,amsfonts,amsbsy,amssymb,amscd,bbm,epsfig,lscape}

\usepackage[titletoc,title]{appendix}
\usepackage{hyperref}

\usepackage{times}
\usepackage{float}

\usepackage{subfig}
\usepackage{graphicx}

\usepackage{cprform}

\usepackage{authblk}

\usepackage{cite}

\usepackage{enumitem}

\newcommand{\Dslash}{\relax{\kern+.25em / \kern-.70em D}}

\newcommand{\Real}{\relax{\mathsf{\Gamma\kern-.35em R}}}
\newcommand{\Int}{\relax{\mathsf{Z\kern-.40em Z}}}

\newcommand{\be}{\begin{equation}}
\newcommand{\ee}{\end{equation}}
\newcommand{\bea}{\begin{eqnarray}}
\newcommand{\eea}{\end{eqnarray}}
\newcommand{\nn}{\nonumber}

\newcommand{\obar}[1]{\kern3pt\overline{\kern-2pt #1\kern-0pt}\kern1pt}

\newcommand{\corrbar}[1]{\kern3pt\overline{\kern-2pt #1\kern-0pt}\kern1pt}

\newcommand{\oVApAVren}[1]{\kern3pt\overline{\kern-2pt #1\kern-0pt}\kern1pt_{\rm\scriptscriptstyle VA+AV;s}}

\newcommand{\zbar}{\kern3pt\overline{\kern-2pt Z\kern-0pt}\kern1pt}

\newcommand{\zbarVApAV}[1]{\kern3pt\overline{\kern-2pt Z\kern-0pt}\kern1pt_{\rm\scriptscriptstyle VA+AV #1}}

\begin{document}

%%%%%%%%%%%%%%%%%%%%%%%%%%%%%%%%%%%%%%%%%%%%%%%%
% TITLE

\begin{titlepage}

%%% Preprint numbers %%%

\vspace*{-20truemm}
%\hspace*{7cm} ROM2F/2015/05, ~~RM3-TH/15-8

\begin{flushright}
\hspace*{-1cm} 
 
%% \vspace{5truemm}\today
\end{flushright}\vspace{5truemm}

%%% Title and authors %%%

\centerline{\Large \bf  $\Delta \mathbf{S}=2$ and $\Delta \mathbf{C}=2$ bag parameters in the Standard Model} 
\vspace*{0.2cm}
\centerline{\Large \bf  and beyond  from $\mathbf{N_f=2+1+1}$  twisted-mass lattice QCD}

\vskip 5 true mm
\centerline{\bigrm N.~Carrasco$^{(a)}$, P.~Dimopoulos$^{(b,\,c)}$, R.~Frezzotti$^{(c,\,d)}$, V.~Lubicz$^{(a,\,e)}$,  }
\vspace*{0.2cm}
\centerline{\bigrm G.C.~Rossi$^{(b,\,c,\,d)}$, S.~Simula$^{(a)}$, C.~Tarantino$^{(e,\,a)}$}

\vspace*{1truemm}
\begin{figure}[!h]
  \begin{center}
    \includegraphics[scale=0.70]{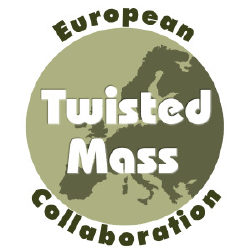}
 \end{center}
\end{figure}
\vspace*{0.5cm}
\centerline{(ETM Collaboration)}
\vspace*{0.5cm}
\vskip 1 true mm
\vskip 0 true mm

\centerline{\it $^{(a)}$ INFN, Sezione di Roma Tre}
\centerline{\it Via della Vasca Navale 84, I-00146 Rome, Italy}
\vskip 2 true mm
\centerline{\it $^{(b)}$ Centro Fermi - Museo Storico della Fisica e Centro Studi e Ricerche Enrico Fermi}
\centerline{\it Compendio del Viminale, Piazza del Viminiale 1, I-00184, Rome, Italy}
\vskip 2 true mm
\centerline{\it $^{(c)}$ Dipartimento di Fisica, Universit\`a di Roma ``Tor Vergata''}
\centerline{\it Via della Ricerca Scientifica 1, I-00133 Rome, Italy}
\vskip 2 true mm
\centerline{\it $^{(d)}$ INFN, Sezione di ``Tor Vergata"}
\centerline{\it Via della Ricerca Scientifica 1, I-00133 Rome, Italy}
\vskip 2 true mm
\centerline{\it $^{(e)}$ Dipartimento di Fisica, Universit\`a  Roma Tre}
\centerline{\it Via della Vasca Navale 84, I-00146 Rome, Italy}
\vskip 2 true mm

\vskip 20 true mm

%%% Abstract %%%

\thicktablerule
\vskip 3 true mm
\noindent{\tenbf Abstract}
\vskip 1 true mm
\noindent
{\tenrm  
We present unquenched lattice QCD results for the matrix elements of four-fermion operators 
relevant to  the description of the neutral $K$ and $D$ mixing
in the Standard Model and its extensions. We have employed simulations with $N_f = 2 + 1 + 1$ 
dynamical sea quarks at three values of the lattice spacings
in the interval 0.06 -- 0.09 fm and pseudoscalar meson masses in the range 210 -- 450 MeV. 
Our results are extrapolated to the continuum limit and to the
physical pion mass. Renormalization constants have been determined 
non-perturbatively in the RI-MOM scheme. In particular, for the Kaon bag-parameter, which is 
relevant for the $\overline{K}^0-K^0$ mixing in the Standard Model, we obtain $B_K^{RGI} = 0.717(24)$.}
\vskip 3 true mm
\thicktablerule
%\vspace{10truemm}
\eject
\end{titlepage}

\section{Introduction}
\label{sec:intro}

Lacking experimental data above production threshold, flavor physics offers the unique possibility for 
an indirect discovery of New Physics (NP)
effects through virtual exchanges of yet-to-be-discovered heavy particles in loop suppressed processes. 
This approach, which is particularly promising
for processes that are highly suppressed within
the Standard Model (SM), proved to be very successful in the past allowing for the indirect determination of the charm and top quark 
mass~\cite{Glashow:1970gm, Gaillard:1974hs, Albrecht:1987dr}.

Moreover flavor physics data play a major r\^{o}le in providing stringent tests of the CKM paradigm and 
allowing the determination of the magnitude of the 
mixing matrix elements. In particular $\Delta S=2$ and $\Delta B=2$ 
flavor changing neutral current processes are crucial to the 
unitarity triangle analysis. 
They are also quite valuable in constraining NP models, see {\it e.g.}~\cite{Bona:2007vi, Isidori:2010kg, 
Mescia:2012fg, Charles:2013aka, Buras:2013ooa, Kersten:2012ed, Charles:2015gya, Bevan:2014cya} with 
data on $\Delta S=2$ oscillations providing the most stringent constraints~\cite{Bona:2007vi, 
Bertone:2012cu, Mescia:2012fg, Isidori:2010kg }.

Of special interest are the $\Delta C=2$ transitions occurring in $\overline{D}^0-D^0$ 
oscillations~\cite{Aubert:2007wf, Staric:2007dt, Aaij:2013wda} and~\cite{Bevan:2014tha, Amhis:2012bh}, 
as this is the only SM process in which mixing involves up-type quarks. 
CP violation through these mixings is expected to be strongly 
suppressed within the SM, because they are dominated by light 
$(d, s)$ quark exchange entailing also important long range interactions. 
Thus any experimental signal of CP violation in the neutral $D$ meson sector would be a strong indication for 
the existence of NP~\cite{Blaylock:1995ay, Petrov:2006nc, Golowich:2007ka, Gedalia:2009kh, Ciuchini:2007cw}. 
Even in the absence of CP-violation, our determination 
of $\Delta C=2$ operator matrix elements allows to put constraints on 
Beyond the Standard Model (BSM) models. 

In this paper we present a determination of the bag-parameters relevant for the description 
of the $\Delta S=2$ and $\Delta C=2$ transitions. We compute meson--anti-meson matrix elements 
of the whole basis of dimension-six four-fermion operators contributing the most general form 
of the effective $\Delta F=2$ Hamiltonian~\cite{Beall:1981ze, Gabbiani:1996hi,  Gabbiani:1988rb, 
Gabrielli:1995bd}. Beyond the ``left-left" operator, relevant for the SM, flavor-changing extra 
terms appear. The full effective $\Delta F=2$ Hamiltonian reads  
\begin{equation}
{\cal H}_{\textrm{eff}}^{\Delta F=2}={\displaystyle \sum_{i=1}^{5}C_{i}(\mu)\widehat {\cal O}_{i}(\mu)}+{\displaystyle \sum_{i=1}^{3}\widetilde C_{i}(\mu)\widehat {\widetilde{\cal O}}_{i}(\mu)}\, ,
\label{eq:Heff}
\end{equation}
where $C_i$ and $\widetilde C_i$ are the Wilson coefficients 
and the bare operators, 
${\cal O}_{i}$ and $\widetilde{\cal O}_{i}$, corresponding to renormalized operators appearing in Eq.~(\ref{eq:Heff}), are 
{\renewcommand{\arraystretch}{1.3}
\begin{equation} \label{eq:operators}
\begin{array}{ll}
%\multicolumn{2}{l} 
{{\cal O}_{1}=\left[\bar{h}^{\alpha}\gamma_{\mu}(1-\gamma_{5}){\ell}^{\alpha}\right]\left[\bar{h}^{\beta}\gamma_{\mu}(1-\gamma_{5}){\ell}^{\beta}\right],} &
\tilde{\cal O}_{1}=\left[\bar{h}^{\alpha}\gamma_{\mu}(1+\gamma_{5}){\ell}^{\alpha}\right]\left[\bar{h}^{\beta}\gamma_{\mu}(1+\gamma_{5}){\ell}^{\beta}\right] ,  \\
{\cal O}_{2}=\left[\bar{h}^{\alpha}(1-\gamma_{5}){\ell}^{\alpha}\right]\left[\bar{h}^{\beta}(1-\gamma_{5}){\ell}^{\beta}\right]\, , & 
\tilde{\cal O}_{2}=\left[\bar{h}^{\alpha}(1+\gamma_{5}){\ell}^{\alpha}\right]\left[\bar{h}^{\beta}(1+\gamma_{5}){\ell}^{\beta}\right]\, , \\
{\cal O}_{3}=\left[\bar{h}^{\alpha}(1-\gamma_{5}){\ell}^{\beta}\right]\left[\bar{h}^{\beta}(1-\gamma_{5}){\ell}^{\alpha}\right] \, , &
\tilde{\cal O}_{3}=\left[\bar{h}^{\alpha}(1+\gamma_{5}){\ell}^{\beta}\right]\left[\bar{h}^{\beta}(1+\gamma_{5}){\ell}^{\alpha}\right]\, , \\
\multicolumn{2}{l}{{\cal O}_{4}=\left[\bar{h}^{\alpha}(1-\gamma_{5}){\ell}^{\alpha}\right]\left[\bar{h}^{\beta}(1+\gamma_{5}){\ell}^{\beta}\right]\, ,} \\
\multicolumn{2}{l}{{\cal O}_{5}=\left[\bar{h}^{\alpha}(1-\gamma_{5}){\ell}^{\beta}\right]\left[\bar{h}^{\beta}(1+\gamma_{5}){\ell}^{\alpha}\right] , \, }  \\
\end{array}
\end{equation}
where $\alpha, \beta$ are color indices. The operators $\widetilde{{\cal O}}_{1-3}$ are obtained from ${\cal O}_{1-3}$ with the replacement 
of $(1-\gamma_5) \to (1+\gamma_5)$. Since $\widetilde{{\cal O}}_{1-3}$ and ${\cal O}_{1-3}$ have 
identical parity conserving parts, parity invariance of QCD allows to restrict our 
attention to only the set of operators ${\cal O}_{i}, i=1,\ldots,5$. 
In the present work we focus on the cases $(h, \ell) \equiv (s, d)$ and $(h, \ell) \equiv (c, u)$. 

The Wilson coefficients describe short distance effects. Accordingly, they will also depend on the 
heavy degrees of freedom possibly circulating in loops. 
The low energy dynamics is incorporated in the matrix element of the operators $\widehat{\cal O}_{i}$. 
The renormalization scale $\mu$ gets 
compensated between the Wilson coefficients and the matrix elements of the renormalized operators.  

Lattice QCD provides a first principles determination of the bag-parameters $B_i$. 
These are dimensionless quantities defined as the ratio of the non-perturbatively computed 
four-fermion matrix element over the value this matrix element takes in the vacuum saturation approximation. 
The reason for working with ratios is that they offer the advantage of a substantial cancellation of systematic 
and statistical uncertainties between the numerator and the denominator. 
For their definition (see for example Ref.~\cite{Allton:1998sm}) 
\begin{equation}
\begin{array}{l}
\langle\overline{P}^{0}|{\cal O}_{1}(\mu)|P^{0}\rangle= \xi_1 B_{1}(\mu)\, m_{P^0}^{2}f_{P^0}^{2} \,\\
\langle\overline{P}^{0}|{\cal O}_{i}(\mu)|P^{0}\rangle= \xi_{i} B_{i}(\mu)\, 
\dfrac{m_{P^0}^{4} f_{P^0}^{2}}{\left(m_{\ell}(\mu)+m_{h}(\mu)\right)^{2}} 
~~~ {\mbox{for}}~~ i =2, \ldots, 5 \, ,
\label{eq:Bi-def}
\end{array}
\end{equation}
where $\xi_i=\{8/3, -5/3,1/3, 2, 2/3\}$ and $P^0$ stands for either a $K^0$ or a $D^0$ pseudoscalar meson. 
The corresponding mass and decay constant are denoted by\footnote{In our formulae we use the notation for a neutral pseudoscalar meson although 
we work in the isospin symmetric limit; so in practice we make no distinction 
between the masses and decay constants of the neutral and charged pseudoscalar mesons.} $m_{P^0}$ and $f_{P^0}$, respectively. 
The quantities $m_{\ell}(\mu)$ and $m_{h}(\mu)$ are the light and heavy quark masses of the neutral $K$ and 
$D$ pseudoscalar mesons are made of, renormalized at the scale $\mu$. 

We see from Eq.~(\ref{eq:Bi-def}) that, while the matrix element 
$\langle\overline{P}^{0}|{\cal O}_{1}(\mu)|P^{0}\rangle$ vanishes 
as the pseudoscalar mass goes to zero, the four other matrix elements do not.
We recall that $B_1$ parametrises the SM operator while the $B_{i}, {i \ge 2},$ parametrise the BSM ones. 
  
The computations presented in this paper have been performed making use of the $N_f=2+1+1$ dynamical quark gauge 
configurations generated by the European Twisted Mass Collaboration (ETMC)~\cite{Baron:2010bv, Baron:2010th} at 
three values of the lattice spacing, $a \simeq 0.06 - 0.09$ fm, 
with the lightest pseudoscalar mass values in the range $M_{ps} \sim 210 - 450$ MeV. Spatial lattice sizes are 
$L \simeq 2.1 - 3.0$ fm with $M_{ps}L \simeq 3.1 - 4.5$. Operator renormalization has been performed non-perturbatively 
in the RI-MOM scheme~\cite{renorm_mom:paper1}. 
 
{\bf Results} -  For the reader's convenience we immediately summarize our main results for $K^0$ and $D^0$ meson bag-parameters. 

\noindent We collect in Table~\ref{tab:Bi-K-all} the values of the five bag-parameters that are required 
to describe the neutral Kaon mixing in the SM and beyond. We give the numbers in the $\overline{\rm{MS}}$ 
renormalization scheme of Ref.~\cite{mu:4ferm-nlo} and in the RI$'$ scheme 
at the scale of $\mu = $ 3~GeV. 
For results given in the $\overline{\rm{MS}}$ scheme the second quoted error 
provides our estimates for the systematic uncertainty coming from the perturbative matching between RI$'$ 
and $\overline{\rm{MS}}$ schemes, which range from 0.5\% to about 4\%. 
The uncertainties on the central values stemming exclusively from our lattice computations are given by the first error and range from about
3.3\% to 7.5\%. 

Neglecting the tiny over-unquenching error due to the presence of
the charm in the sea (see discussion below) we adopt the continuum
limit results in the  $\overline{\rm{MS}}$  scheme at $\mu = $ 3~GeV as our best estimate
of the desired Kaon mixing bag parameters in QCD with $u$, $d$ and $s$ 
active flavors for the same scheme and scale. Then for $B_K \equiv B_1$ we find in the RGI scheme the 
value~\footnote{If for converting our continuum limit $B_1$ in the $\overline{\rm{MS}}$  scheme
at $\mu = $ 3~GeV to its RGI counterpart we had taken $N_f=4$ we would
have obtained $B_K^{RGI (Nf =4)} = 0.728(24)$.} 
\begin{equation}
B_K^{RGI (N_f=3)} = 0.717(24)\, ,
\label{eq:BKRGI}
\end{equation}  
with a total uncertainty of about 3.4\%.

{\renewcommand{\arraystretch}{1.3}
\begin{table}[!ht]
\begin{centering}
\begin{tabular}{|c c c c c c|}
\hline 
\multicolumn{6}{|c|}{$\overline{K}^0-K^0$}\tabularnewline
\hline 
\hline
$\overline{\rm{MS}}$ (3 GeV) & 0.506(17)(3)& 0.46(3)(1)& 0.79(5)(1) & 0.78(4)(3) & 0.49(4)(1) \tabularnewline
\hline
RI$'$ (3 GeV)                & 0.498(16) &0.62(3) & 1.10(7) & 0.98(5) & 0.66(5) \tabularnewline
\hline
\end{tabular}
\par\end{centering}
\caption{\label{tab:Bi-K-all} Continuum limit results for the bag-parameters $B_{i}$ ($i=1, \ldots, 5$) relevant to the 
$\overline{K}^0-K^{0}$ mixing 
renormalized in the $\overline{\rm{MS}}$ scheme of Ref.~\cite{mu:4ferm-nlo} 
and in the RI$'$ scheme at the scale of $\mu =$ 3~GeV. For results given in the $\overline{\rm{MS}}$ scheme the second error 
indicates an estimate for the systematic uncertainty owing to the perturbative matching of RI$'$ and $\overline{\rm{MS}}$ schemes.}
\end{table}

In Table~\ref{tab:Bi-D-all} we summarize the results for the bag-parameters 
relevant for the case of the $\overline{D}^0-D^{0}$ oscillations. 
For results given in the $\overline{\rm{MS}}$ scheme the second error we quote represents
our estimate of the systematic uncertainty coming from the perturbative matching between RI$'$ and $\overline{\rm{MS}}$ schemes. 
The uncertainties stemming only from our lattice computations are given by the first quoted error and
range from about 4\% to 8\%.
{\renewcommand{\arraystretch}{1.3}
\begin{table}[!ht]
\begin{centering}
\begin{tabular}{|c c c c c c|}
\hline 
\multicolumn{6}{|c|}{$\overline{D}^0-D^0$}\tabularnewline
\hline 
\hline
 $\overline{\rm{MS}}$ (3 GeV) & 0.757(27)(4)& 0.65(3)(2)& 0.96(8)(2) & 0.91(5)(4) & 0.97(7)(1) \tabularnewline
\hline
RI$'$ (3 GeV)                & 0.744(27) &0.87(5) & 1.34(11) & 1.14(6) & 1.39(9) \tabularnewline
\hline
\end{tabular}
\par\end{centering}
\caption{\label{tab:Bi-D-all}  Same as in Table~\ref{tab:Bi-K-all} for the $\overline{D}^0-D^{0}$ mixing. }
\end{table}

The results of this paper are compared with the existing unquenched 
determinations~\footnote{For recent reviews see Ref~\cite{Aoki:2013ldr, Carrasco:2014nda}.} in Figs.~\ref{fig:BK-compar}, 
\ref{fig:Bi-compar} and~\ref{fig:BD-compar}.

In Fig.~\ref{fig:BK-compar} we present a compilation of recent (unquenched)  RGI values of $B_K$. 
Lattice computations are quite accurate with a total uncertainty of only a few percent. 
It is worth noting the rather weak dependence of $B_K^{RGI}$ on the number of dynamical flavors. 
Our current $B_K^{RGI}$ value compares well with $N_f=2+1$ and $N_f=2$ lattice computations. \\
Our result is also found to be in agreement with the estimate obtained using a model based on the dual representation of QCD 
as a theory of weakly interacting mesons for large $N$, which predicts negative sign corrections 
to the large $N$ limit estimate given by the value $B_K=0.75$~\cite{Buras:2014maa, Bardeen:1987vg}.

\begin{figure}[!h]
\begin{center}
\includegraphics[scale=1.0,angle=-0]{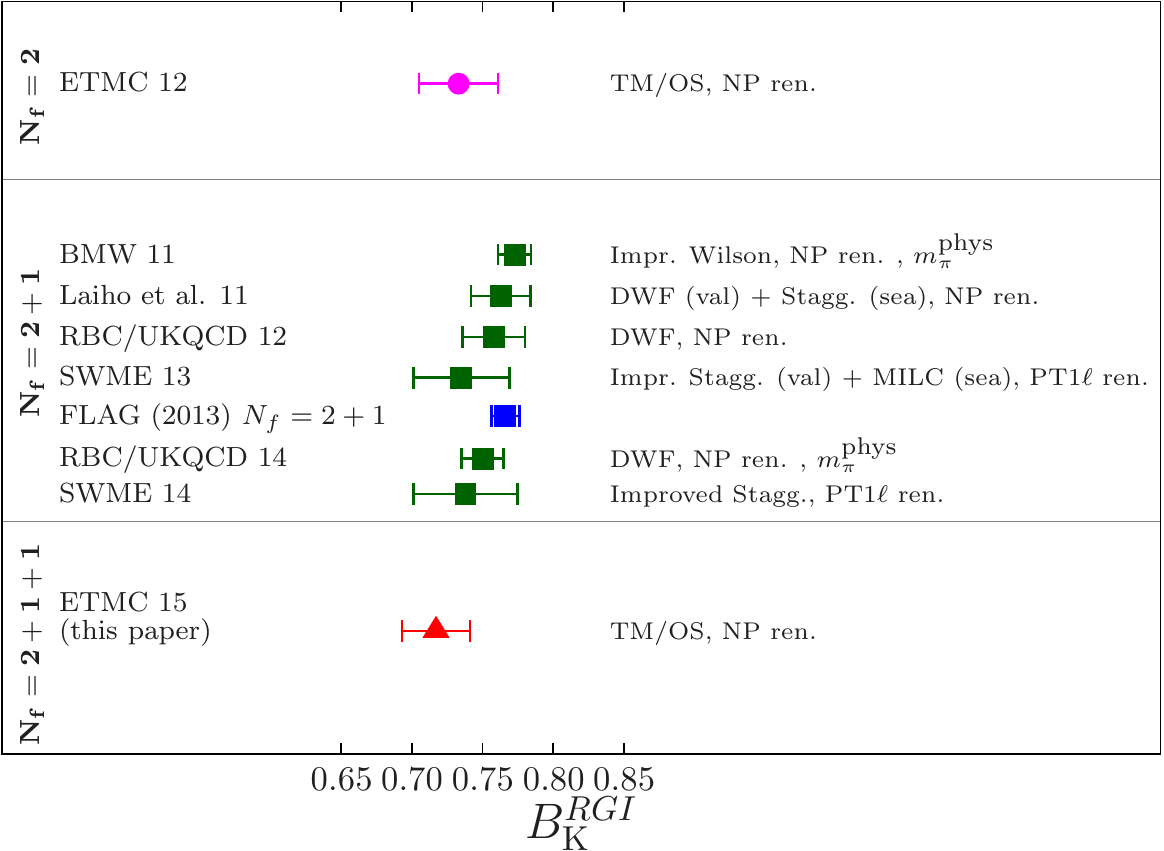}
\end{center}
 \caption{ \label{fig:BK-compar}  A compilation of unquenched lattice results for the RGI 
 value of the $B_K$ parameter. From top to bottom 
 results are taken from 
 Refs.~\cite{Bertone:2012cu},~\cite{Durr:2011ap},~\cite{Laiho:2011np}, ~\cite{Arthur:2012opa},~\cite{Bae:2013tca}
 \cite{Aoki:2013ldr},~\cite{Blum:2014tka},~\cite{Bae:2014sja}. Circle, 
 squares and triangle correspond to $N_f=2$, $N_f=2+1$ and $N_f=2+1+1$ dynamical quark 
 computations, respectively. The full blue square indicates the  FLAG average~\cite{Aoki:2013ldr} over $N_f=2+1$ data. 
 For reader's convenience some information on the basic features of each computation is also given.}
\end{figure}

A comparison of recent determinations
of the $\overline{K}^0-K^0$ bag-parameters $B_i, {i=2, \ldots,  5}$ 
is presented in Fig.~\ref{fig:Bi-compar}~\footnote{For older quenched computations 
of the BSM $B_i$ see Refs.~\cite{Donini:1999nn, Babich:2006bh}.} .
The ETM, RBC/UKQCD and SWME collaborations give for $B_2$ and $B_3$   
results that are well compatible within the errors. 
A tension of up to 3 standard deviations is visible, instead, 
in the case of $B_4$ and $B_5$ after the updated (preliminary) work 
of SWME~\cite{Jang:2014aea}. 

\begin{figure}[!h]
\begin{center}
\includegraphics[scale=0.6,angle=-0]{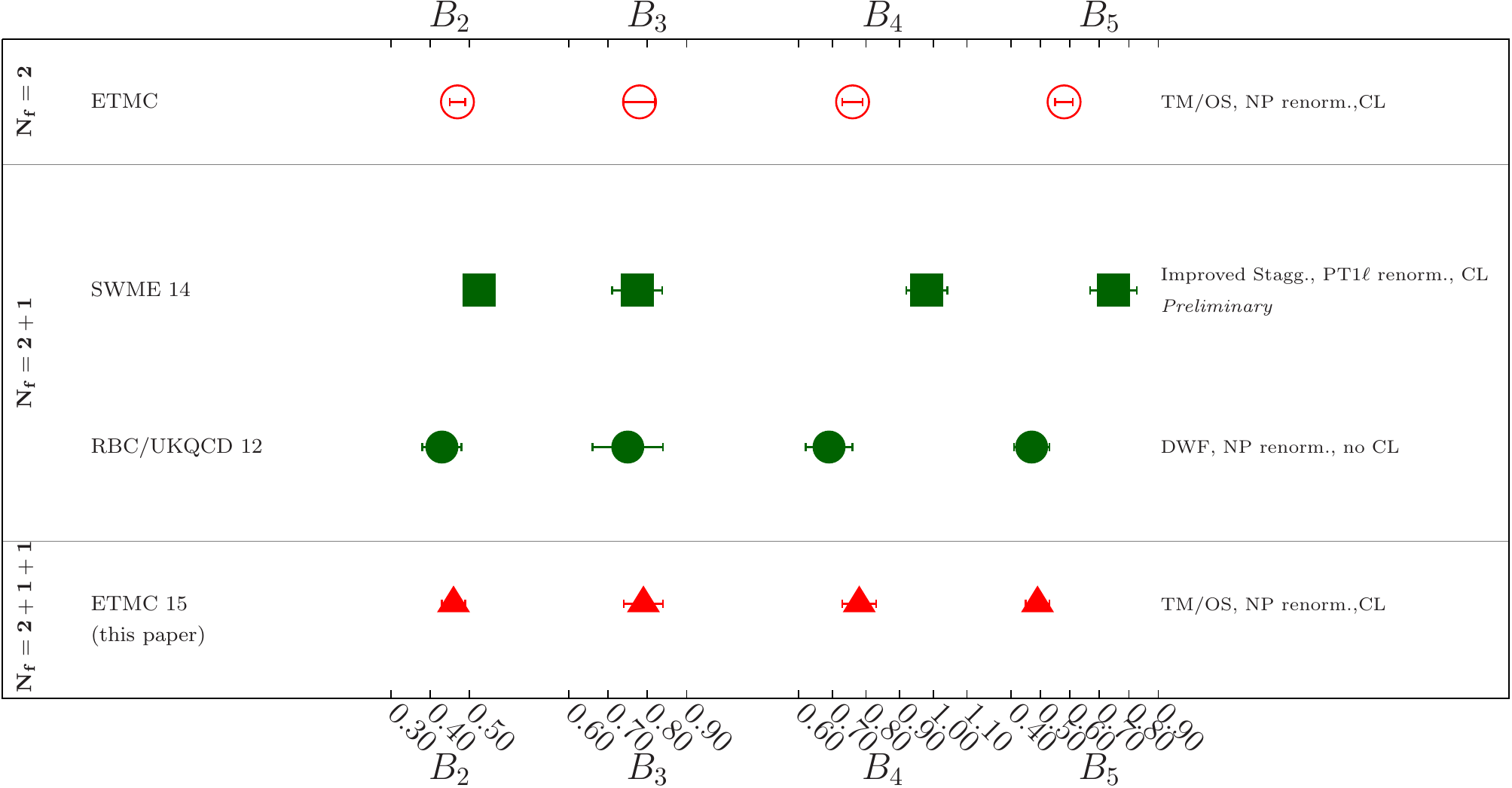}
\end{center}
\caption{A compilation of $K^0$ meson bag-parameters $B_i$, $i=2, \ldots, 5$. 
From top to bottom data have been taken from Refs~\cite{Bertone:2012cu, Jang:2014aea, Boyle:2012qb}.
The work reported in Ref.~\cite{Jang:2014aea} is an  updated computation 
of Ref.~\cite{Bae:2013tca}.   
The label ``CL" stands for continuum limit computation. Work in progress by the RBC/UKQCD has been reported in~\cite{Lytle:2014tsa}.}
\label{fig:Bi-compar}
\end{figure}

Finally, in Fig.~\ref{fig:BD-compar} we show the comparison of the available 
results for the $D^0$ bag-parameters coming from ETMC computations with $N_f=2$ and $N_f=2+1+1$ 
gauge configurations~\footnote{Work in progress of an unquenched $N_f=2+1$ computation for the $D$-mixing 
is reported in Ref.~\cite{Chang:2014cea}. For older works using quenched simulations see Refs~\cite{Lin:2006vc, Becirevic:2001xt}.}.

\begin{figure}[!h]
\begin{center}
\includegraphics[scale=0.6,angle=-0]{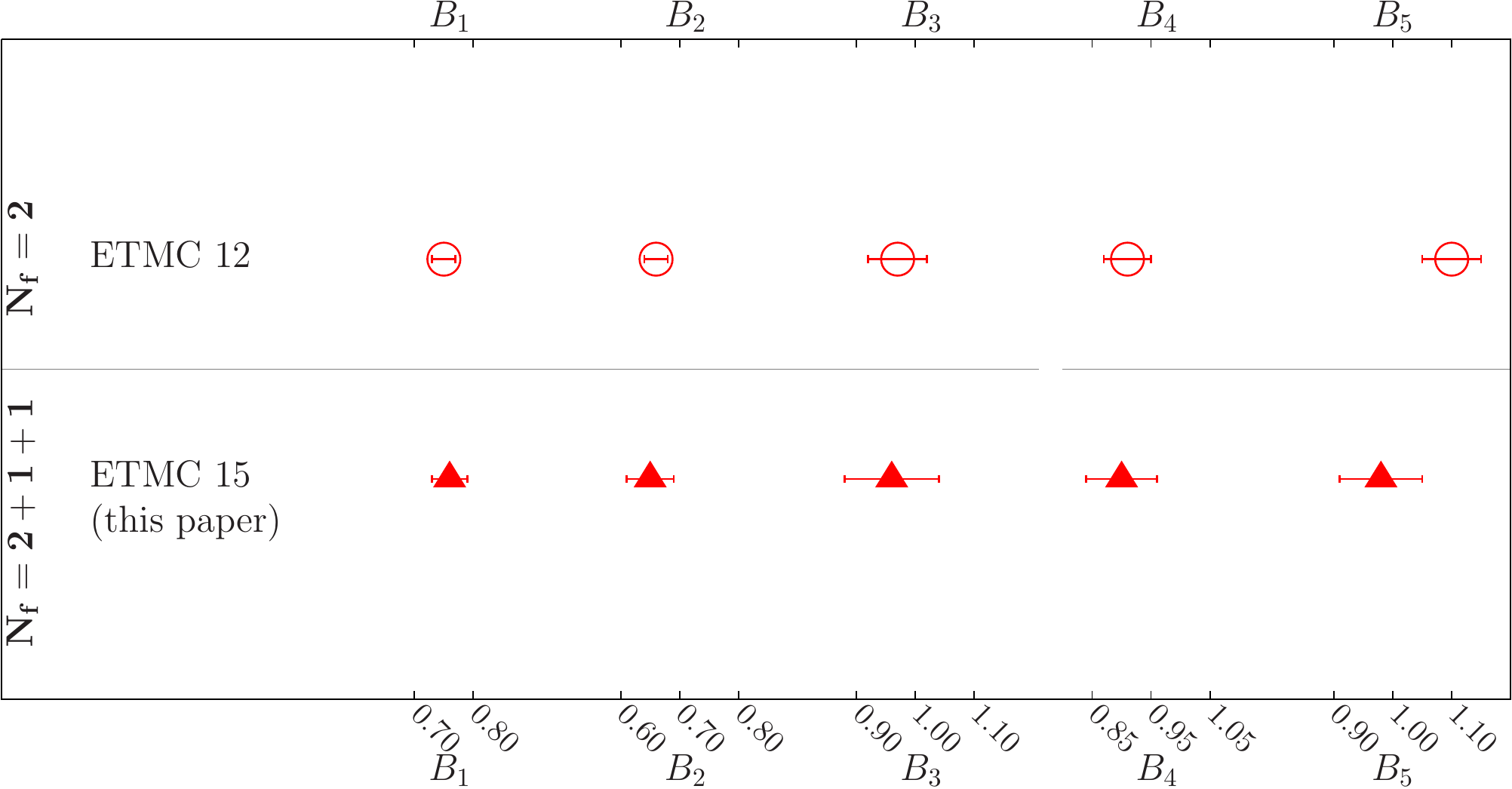}
\end{center}
\caption{Results for the $D^0$ meson bag-parameters $B_i$, $i=1, \ldots, 5$  obtained by the ETMC with $N_f=2+1+1$ (this paper) and $N_f=2$~\cite{Carrasco:2014uya} dynamical flavor lattice simulations.
}
\label{fig:BD-compar}
\end{figure}

Comparing the results for the bag-parameters collected in the 
Tables~\ref{tab:Bi-K-all} and~\ref{tab:Bi-D-all} with  the 
ETMC results published in Refs~\cite{Bertone:2012cu} and~\cite{Carrasco:2014uya}, the latter obtained with $N_f=2$ 
dynamical quark simulations, we notice that they are all compatible among themselves and have similar total uncertainties. 
Therefore, {\it ceteris paribus}, 
the main conclusions presented in these works concerning model-independent constraints 
on the NP scale from $\Delta S=2$ and $\Delta C=2$ operators 
within the unitarity triangle analysis, remain unchanged.  

Before concluding this discussion, we find it useful to comment further about the dependence of $B_K$ and of the other $B$-parameters for $\overline{K}^0-K^0$ 
mixing on the number of dynamical quarks. As known, the standard theoretical formula that provides the indirect CP violation parameter $\epsilon_K$ is obtained, 
through the low-energy effective weak Hamiltonian, after integrating out the heavy degrees of freedom including the charm quark. The reason is that it is only 
when the charm quark is integrated out, i.e. at scales $\mu \sim m_c$, that the imaginary part of the effective Hamiltonian for $\overline{K}^0-K^0$ mixing 
becomes local (at the leading order in the $1/m_c$ expansion). The advantage of this approach is that the long distance contributions to the amplitude, being related
to the matrix elements of local operators, are more easily accessible to lattice computations. The price to pay, however, is that perturbation theory is 
uncertain at scales around the charm mass and, in addition, subleading corrections proportional to powers of $p_K^2/m_c^2$ (where $p_K={\cal O}(m_K,\Lambda_{QCD})$) 
may not be negligible, particularly when aiming at a theoretical prediction for $\epsilon_K$ with percent precision.

In the standard approach, both short-distance Wilson coefficient and long-distance matrix  elements of the effective Hamiltonian have to be computed in 
the presence of three active quarks. In this respect, therefore, the lattice computation of $B$-parameters for $\overline{K}^0-K^0$ mixing presented in this 
paper, being based on simulations performed with $N_f=2+1+1$ dynamical quarks, introduces a systematic error. Previous experience with $N_f=2+1+1$ lattice 
calculations suggests that the effect of the dynamical charm quark is presumably tiny, so that its impact in the determination of physical observables, 
which is undesired in this particular case, is likely too small to be detected at the level of the current precision. 

It should be also noted that a similar source of systematic error is introduced in the lattice calculations of $B_K$ performed with only $N_f=2+1$ dynamical 
quarks. In the latter case, indeed, the effect of the dynamical charm, which properly is not introduced in the determination of the matrix elements, 
is missing however in the lattice computation of the hadronic observables which are needed to fix the action parameters. Therefore, the determination of 
the lattice scale as 
well as of the strange and light quark masses within any $N_f=2+1$ lattice calculation is affected by the systematic error due to the quenching of the charm 
quark. This error then propagates into the calculation of $B_K$. These effects, namely the latter one and the error introduced in the $N_f=2+1+1$ calculations 
of $B_K$, have the same physical origin and are presumably comparable in size. As already noted, both effects are likely to be currently negligible, as also 
indicated, a posteriori, by the good consistency observed in Fig. 1 among the lattice determinations of $B_K$ based on different numbers of active quarks.

In order to bypass this error, the theoretical determination of $\epsilon_K$ should be performed by keeping an active charm in the calculation. With the 
advances of the lattice technique, the computation of the matrix elements of non-local operators has becoming feasible and a first, exploratory lattice 
calculation of 
the real part of the effective Hamiltonian for $\overline{K}^0-K^0$ mixing has been presented in Refs~\cite{Christ:2012se,Bai:2014cva}. The same technique can be 
also applied to the calculation of the imaginary part of the Hamiltonian, which is relevant for $\epsilon_K$. While these lattice studies are not yet as 
accurate as the standard computations of local operator matrix elements, they are opening a new perspective and are likely to allow, in a near future, 
a significant improvement in the accuracy of the theoretical predictions of $\epsilon_K$, for both the long distance and short distance contributions.

{\bf Plan of the paper} --  In Section~\ref{sec:analysis} we review the simulation details and 
discuss our computational and analysis setup. 
Final results and a full account of the error budget is given in Section~\ref{sec:results}. Finally, in the 
Appendices~\ref{sec:RCs-comput-setup} and~\ref{app:RCs4f} we discuss the procedure we employed to determine 
in RI-MOM the full $5 \times 5$ renormalization constants (RCs) matrix  for the four-fermion operators.

\section{Computational details}
\label{sec:analysis}

In this work we have employed the mixed action twisted mass/Osterwalder-Seiler setup proposed in Ref.~\cite{Frezzotti:2004wz} 
which provides automatic ${\cal O}(a)$-improvement  and continuum-like renormalization pattern for the four-fermion 
operators, with only O($a^2$) unitarity violations.
  
\subsection{Lattice setup}

For the action of the light mass-degenerate sea quark doublet we have used the expression of Ref.~\cite{FrezzoRoss1} which reads
\begin{equation}\label{eq:tmlight}
S_{\ell}=a^4\sum_{x}\bar{\psi}_{\ell}(x)\left\{ \dfrac{1}{2}\gamma_{\mu}\left(\nabla_{\mu}+
\nabla_{\mu}^{*}\right)-i\gamma_{5}\tau^{3}\left[M_{\textrm{cr}}-\dfrac{a}{2}\sum_{\mu}\nabla_{\mu}^{*}
\nabla_{\mu}\right]+\mu_{\textrm{sea}}\right\} \psi_{\ell}(x), 
\end{equation}
where it is intended that the untwisted mass has been tuned to its critical value, $M_{cr}$. As usual, 
the symbols $\nabla_{\mu}$ and $\nabla_{\mu}^{*}$ represent the nearest neighbour forward and backward covariant 
derivatives, we define the quark doublet $\psi_{\ell} = (\psi_u~ \psi_d)^{T}$ and $\mu_{sea}$ 
is the (light) sea twisted quark mass.    

With similar notations we take the action for the strange and charm quark doublet in the sea~\cite{Frezzotti:2003xj} to be
\begin{equation}\label{eq:tmheavy}
S_{h}=a^4\sum_{x} \bar{\psi}_{h}(x)\left\{ \dfrac{1}{2}\gamma_{\mu}\left(\nabla_{\mu}+\nabla_{\mu}^{*}\right)- 
i\gamma_{5}\tau^{1}\left[M_{\textrm{cr}}-
\dfrac{a}{2}\sum_{\mu}\nabla_{\mu}^{*}\nabla_{\mu}\right]+{\mu}_{\sigma}+{\mu}_{\delta}{\tau}^{3} \right\}  {\psi}_{h}(x), 
\end{equation}
where $\mu_{\sigma}$ and $\mu_{\delta}$ are the bare twisted mass parameters from which the renormalized strange and charm  
masses can be derived. Pauli matrices in Eqs~(\ref{eq:tmlight}) and~(\ref{eq:tmheavy}) act in flavor space. For more details 
on the twisted mass setup see Refs.~\cite{FrezzoRoss1, Frezzotti:2003xj, Baron:2010bv, Baron:2010th, Boucaud:2007uk, 
Boucaud:2008xu, Baron:2009wt}.
 
Valence quarks are introduced via Osterwalder-Seiler (OS) fermions~\cite{Osterwalder:1977pc}. 
The valence action is written as the 
sum of the {\it individual} quark flavor contributions in the form 
\begin{equation}\label{eq:OS}
S^{\textrm{OS}}=a^4\sum_{x}\sum_{f={\ell},{\ell}',h,h'}\bar{q}_{f}\left\{ \dfrac{1}{2}\gamma_{\mu}\left(\nabla_{\mu}+\nabla_{\mu}^{*}\right)-i\gamma_{5}r_{f}
\left[M_{\textrm{cr}}-\dfrac{a}{2}\sum_{\mu}\nabla_{\mu}^{*}\nabla_{\mu}\right]+\mu_{f}\right\} q_{f}(x) \, . 
\end{equation}
where the label $f$ is let to run over the different valence flavors  $f=\ell, \ell', h,h'$. In the neutral $K$ case light ($\ell$) 
and heavy ($h$) flavors denote down and strange quarks, respectively, while in the neutral $D$-case they  stand for up and charm quarks. 
With the choice $r_h=r_{\ell}=r_{h'}=-r_{\ell'}$ one can prove~\cite{Frezzotti:2004wz} that at maximal twist automatic O($a$)
improvement and absence of wrong chiral mixings~\cite{Bochicchio:1985xa}
is guaranteed. Flavor by flavor bare valence and sea quark masses are  set equal to each other which is enough 
to keep unitarity violations to O($a^2$). 
The multiplicative mass renormalization constant is $Z_P$ for all fermions.

The lattice setup described above has been already successfully applied to  determine the 
full set of four-fermion operator matrix elements relevant for the  
$\overline{K}^0-K^0$, $\overline{D}^0-D^0$ and $\overline{B}_{(s)}^0-B_{(s)}^0$ oscillations in 
Refs.~\cite{Dimopoulos:2009es, Constantinou:2010qv, Bertone:2012cu, Carrasco:2014uya, Carrasco:2013zta}.

\subsection{Simulation Details}

We have used $N_f=2+1+1$ gauge configuration ensembles, produced with the Iwasaki gluon action~\cite{Iwasaki:1985we} 
and maximally twisted Wilson fermions, generated by the ETM Collaboration~\cite{Baron:2010bv,Baron:2010th}.  

In Table~\ref{tab:runs} we summarise the main simulation details relevant for the sea and valence sector. 
Simulation data have been taken at three values of the lattice spacing, 
namely $a=0.0885(36)$, $0.0815(30)$ and $0.0619(18)$~fm, 
corresponding to $\beta=1.90$, $1.95$ and $2.10$, respectively (see Ref.~\cite{Carrasco:2014cwa}). 

As we said, light valence and sea quark masses are set equal, leading to pion masses in the range between 210 and 
450~MeV. Strange and charm sea quark masses are chosen close to their physical value. 
To allow for a smooth interpolation to the physical values of the strange and charm quark mass, we have inverted 
the heavy valence Dirac matrix for three  values, $\mu_{``s"}$, of the strange quark mass and three values, $\mu_{``c"}$, 
of the charm mass, around the corresponding physical mass values.

The lattice scale has been fixed using $f_{\pi}$. The $u/d$, strange and charm quark masses have been determined 
comparing with the experimental values of the pion, $K$ and $D_{(s)}$ meson mass, respectively. Further details of 
our simulation setup can be found in Ref.~\cite{Carrasco:2014cwa}.  

Valence light and strange quark propagators have been computed by employing spatial stochastic sources at a 
randomly chosen time-slice, adopting the ``one-end" trick stochastic method of Ref.~\cite{Foster:1998vw, McNeile:2006bz}. 
In correlators where the charm quark is involved, Gaussian smeared interpolating quark fields~\cite{Gusken:1989qx} are used 
in order to suppress the contribution of excited states. This allows ground state identification at precocious Euclidean time
separations.

For the values of the smearing parameters we take $k_{G}=4$ and $N_{G}=30$ Gaussian. In addition, we apply APE-smearing 
to the gauge links~\cite{Albanese:1987ds} in the interpolating fields with parameters $\alpha_{APE}=0.5$ and $N_{APE}=20$.

\begin{center}
\begin{table}[!h]
\begin{centering}
\scalebox{0.9}{
\begin{tabular}{cclll}
\hline 
{\small $\beta$} & {\small $L^{3}\times T$} & {\small $a\mu_{\ell}=a\mu_{\textrm{sea}}$} & {\small $a\mu_{``s"}$ } & {\small $a\mu_{``c"}$ }\tabularnewline
\hline 
{\small 1.90 ($a^{-1}\sim2.19$ GeV )} & {\small $24^{3}\times48$} &  {\small 0.0040 } & {\small 0.0145\, 0.0185\, 0.0225} 
&{\small 0.21256\, 0.25\, 0.29404}\tabularnewline
{\small $\mu_{\sigma}=0.15$ $\mu_{\delta}=0.19$} &    & {\small 0.0060} & &  \tabularnewline
 &  &   {\small 0.0080 } & &\tabularnewline
 &  &   {\small 0.0100} &  &\tabularnewline
\cline{2-5}
 & {\small $32^{3}\times64$} & {\small 0.0030} & {\small 0.0145\, 0.0185\, 0.0225}   
 & {\small 0.21256\, 0.25\, 0.29404}\tabularnewline
 &    & {\small 0.0040 }  & & \tabularnewline
 &    & {\small 0.0050}  & &\tabularnewline
\hline 
{\small 1.95 ($a^{-1}\sim2.50$ GeV)} & {\small $24^{3}\times48$} & {\small 0.0085} & {\small 0.0141\, 0.0180\, 0.0219} 
&{\small 0.18705\, 0.22\, 0.25875 } \tabularnewline
 \cline{2-5}
{\small $\mu_{\sigma}=0.135$ $\mu_{\delta}=0.17$} & {\small $32^{3}\times64$}  & {\small 0.0025 } &  
{\small 0.0141\, 0.0180\, 0.0219}&{\small 0.18705\, 0.22\, 0.25875 }\tabularnewline
 &    & {\small 0.0035 }& &\tabularnewline
 &    & {\small 0.0055} & &\tabularnewline
 &    & {\small 0.0075} & &\tabularnewline
\hline 
{\small 2.10 ($a^{-1}\sim3.23$ GeV)} & {\small $48^{3}\times96$} & {\small 0.0015 } &{\small 0.0118\, 0.0151\, 0.0184}&
{\small 0.14454\, 0.17\, 0.19995}\tabularnewline
{\small $\mu_{\sigma}=0.12$ $\mu_{\delta}=0.1385$}   &  & {\small 0.0020 } & & \tabularnewline
 &    & {\small 0.0030}  & & \tabularnewline
\hline 
\end{tabular}
}
\par\end{centering}{\small \par}
\caption{\label{tab:runs} Details of the simulation setup. Sea and valence fermion actions are displayed in Eqs.~(\ref{eq:tmlight}), 
(\ref{eq:tmheavy}) and~(\ref{eq:OS}).}
\end{table}
\end{center}

\subsection {Lattice operators and bag-parameters}
\label{subsec:bag-parameters}

A detailed account of the lattice operators entering two- and three-point correlation functions was 
presented in Appendix~\ref{sec:RCs-comput-setup} of Ref.~\cite{Constantinou:2010qv}. 
For the reader's convenience and to fix the notation we recall here some basic information. 
In our mixed action setup one needs to consider the following set of four-fermion operators
\begin{eqnarray}
&&O^{MA}_{1[\pm]}= 2\big{\{}\big{(}[\bar q_h^\alpha\gamma_\mu q_{\ell}^\alpha][\bar q_{h'}^\beta\gamma_\mu q_{\ell'}^\beta]+
[\bar q_{h}^\alpha\gamma_\mu \gamma_5 q_{\ell}^\alpha]
[\bar q_{h'}^\beta\gamma_\mu \gamma_5 q_{\ell'}^\beta]\big{)} \pm \big{(}\ell\leftrightarrow \ell'\big{)}\big{\}}\nonumber \\
&&O^{MA}_{2[\pm]}=2\big{\{}\big{(}[\bar q_h^\alpha q_{\ell}^\alpha][\bar q_{h'}^\beta q_{\ell'}^\beta]+
[\bar q_{h}^\alpha\gamma_5 q_{\ell}^\alpha][\bar q_{h'}^\beta\gamma_5 q_{\ell'}^\beta]\big{)} \pm 
\big{(}\ell\leftrightarrow \ell'\big{)}\big{\}} \nonumber \\
&&O^{MA}_{3[\pm]}= 2\big{\{}\big{(}[\bar q_h^\alpha q_{\ell}^\beta]
[\bar q_{h'}^\beta q_{\ell'}^\alpha]+[\bar q_{h}^\alpha\gamma_5 q_{\ell}^\beta]
[\bar q_{h'}^\beta\gamma_5 q_{\ell'}^\alpha]\big{)}\pm \big{(}\ell\leftrightarrow \ell'\big{)}\big{\}}\nonumber \\
&&O^{MA}_{4[\pm]}= 2\big{\{}\big{(}[\bar q_h^\alpha q_{\ell}^\alpha]
[\bar q_{h'}^\beta q_{\ell'}^\beta]-[\bar q_{h}^\alpha\gamma_5 q_{\ell}^\alpha][\bar q_{h'}^\beta\gamma_5 q_{\ell'}^\beta]\big{)}
\pm \big{(}\ell\leftrightarrow \ell'\big{)}\big{\}}\nonumber \\
&&O^{MA}_{5[\pm]}=2\big{\{}\big{(}[\bar q_h^\alpha q_{\ell}^\beta][\bar q_{h'}^\beta q_{\ell'}^\alpha]-
[\bar q_{h}^\alpha\gamma_5 q_{\ell}^\beta][\bar q_{h'}^\beta\gamma_5 q_{\ell'}^\alpha]\big{)} \pm \big{(}\ell\leftrightarrow \ell'\big{)}\big{\}}\, ,
\label{OMAPM_v2}
\end{eqnarray}
where $\alpha$ and $\beta$ are color indices, the square parentheses denote spin 
covariant operator factors and the label ``MA" stands for ``Mixed Action".  

We have  set periodic boundary conditions for all fields, except for the quark fields which obey anti-periodic boundary conditions in the time direction. 
Two ``wall" operators with $P^0$-meson quantum numbers (recall $P^0$ can be either $K^0$ or $D^0$) are introduced at time 
slices $y_0$ and $y_0+T_{sep}/2$. The first operator is constructed in terms of $q_{\ell}$ and $q_h$ quark fields 
and the second in terms of $q_{\ell'}$ and $q_h'$ quark fields. Explicitly they are given by 
\begin{eqnarray}{\cal P}^{\ell h}_{y_0} &=&   
\Big{(}\dfrac{a}{L}\Big{)}^3\, \sum_{\vec y} \bar q_{\ell}(\vec y, y_0) \gamma_5 q_{h}(\vec y , y_0) \nonumber \\ 
{\cal P}^{\ell' h'}_{y_0+T_{sep}} &=&   \Big{(}\dfrac{a}{L}\Big{)}^3\, \sum_{\vec y} 
\bar q_{\ell'}(\vec y, y_0+T_{sep}) \gamma_5 q_{h'}(\vec y , y_0+T_{sep}) \label{K-WALL}\end{eqnarray}
In terms of them, the correlation functions we need to calculate are 
\begin{eqnarray}
\hspace{-1.5cm}&&C_i(x_0) = \Big{(}\dfrac{a}{L}\Big{)}^3\sum_{\vec x} \langle 
{\cal P}^{\ell' h'}_{y_0 + T_{sep}} \,O_{i[+]}^{MA} (\vec x,x_0) \, {\cal P}^{\ell h}_{y_0} \rangle\, , \quad i=1, \ldots, 
5\, ,
\label{PQPi-correl}\\
\hspace{-1.5cm}&&C_{XP}(x_0) = \Big{(}\dfrac{a}{L}\Big{)}^3\sum_{\vec x} 
\langle X^{h \ell}(\vec x,x_0) \,{\cal P}^{\ell h}_{y_0} \rangle\, ,
\label{PP-12}\\
\hspace{-1.5cm}&&C_{XP}^\prime(x_0) = \Big{(}\dfrac{a}{L}\Big{)}^3 \sum_{\vec x} \langle {\cal P}^{\ell' h'}_{y_0 + T_{sep}}  
\,X^{h'\ell'}(\vec x,x_0) \rangle \, 
\label{PP-34}
\end{eqnarray}
where $X$ can be either the axial current, $A_{0}$, or the pseudoscalar density, $P$. 
 
For three-point correlation functions with heavy (charm  or heavier) 
quarks we can achieve reduced statistical uncertainties by  
decreasing the time separation between the two sources and using smearing 
techniques (see Refs.~\cite{Carrasco:2013zta, Carrasco:2014uya}). 
Therefore, while in the $\overline{K}^0-K^0$ case we have set $T_{sep}=T/2$, 
in the $\overline{D}^0-D^0$ case we have instead produced correlation 
functions setting $T_{\textrm{sep}}/a=18$ at $\beta=1.9$, $T_{\textrm{sep}}/a=20$ 
at $\beta=1.95$ and $T_{\textrm{sep}}/a=26$ at $\beta=2.10$.

Estimators for the bare bag-parameters are extracted from the asymptotic time 
behaviour of the ratios of the three- to two-point correlators 
\begin{equation}
\begin{array}{ccc}
{\cal R}^{(b)}_{1}(x_{0})=\dfrac{C_{1}(x_{0})}{C_{AP}(x_{0})C'_{AP}(x_{0})}, &  \,\,\,\,\,\,\,\,\,\,\,\,\,\,\,\,\,\,\,\,\,\, & 
{\cal R}^{(b)}_{i=2, \ldots, 5}(x_{0})=\dfrac{C_{i=2, \ldots, 5}(x_{0})}{C_{PP}(x_{0})C'_{PP}(x_{0})}\end{array} ,
\end{equation}
which for large time separations, $y_0\ll x_0 \ll y_0+T_{\textrm{sep}}$, tend to the desired (bare) bag-parameters 
\begin{equation}
\hspace{-3.cm}{\cal R}^{(b)}_{1}(x_0) \, \mathop {\xrightarrow{\hspace*{1.2cm}}} \limits^{y_0 \ll x_0 \ll y_0+T_{sep}} 
\, \left.  \dfrac{\langle \overline{P}^0 | O_{1[+]}^{MA} | P^0 \rangle}{\langle \overline{P}^0 | A_0^{h \ell} | 0 \rangle \, 
\langle 0 | A_0^{h' \ell'} | P^0 \rangle} \right|^{(b)} \equiv  B_1^{(b)} \,
\label{eq:bareB1}
\end{equation}

\begin{equation}
{\cal R}^{(b)}_{i}(x_0) \, \mathop {\xrightarrow{\hspace*{1.2cm}}} \limits^{y_0 \ll x_0 \ll y_0+T_{sep}} 
\, \left.  \dfrac{\langle \overline{P}^0 | O_{i[+]}^{MA} | P^0 \rangle}{\langle \overline{P}^0 | P^{h \ell} | 0 \rangle \, 
\langle 0 | P^{h' \ell'} | P^0 \rangle} \right|^{(b)} \equiv  B_i^{(b)} \, ,\quad i=2, \ldots, 5
\label{bareBi}
\end{equation}

Figs~\ref{fig:Data-and-time-plateaux-K} and~\ref{fig:Data-and-time-plateaux-D} refer to 
the neutral $K$ and $D$ meson case, respectively. They  illustrate the 
quality of the plateaux from which the estimates for the bare $B_i$ ($i=1, \ldots, 5$) 
bag-parameters are extracted. The three panels correspond to three values of the 
lattice spacing at which simulations were performed.

We note that our plateau choices are rather conservative and by reasonably varying the plateau interval (e.g. considering 
a plateau's length as long as the double of the size of our principal choice) we find that the maximal 
systematic uncertainty is at the sub-percent level (with maximal estimates being 0.2\% for $B_2$ and 0.5\% for $B_3$ 
for the neutral kaon and D cases, respectively).
We therefore conclude that this systematic uncertainty is so smaller than the statistical one, indicated by the label ``stat+fit+RCs" 
in Tables 4 and 5 (see below for details) that it can be safely neglected.

\begin{figure}[!h]
\begin{center}
\vspace*{0.2cm}
\begin{tabular}{ccc}
\hspace*{0.5cm}\vspace*{0.3cm}\includegraphics[bb=171bp 497bp 496bp 705bp,width=0.33\linewidth , 
keepaspectratio=true]{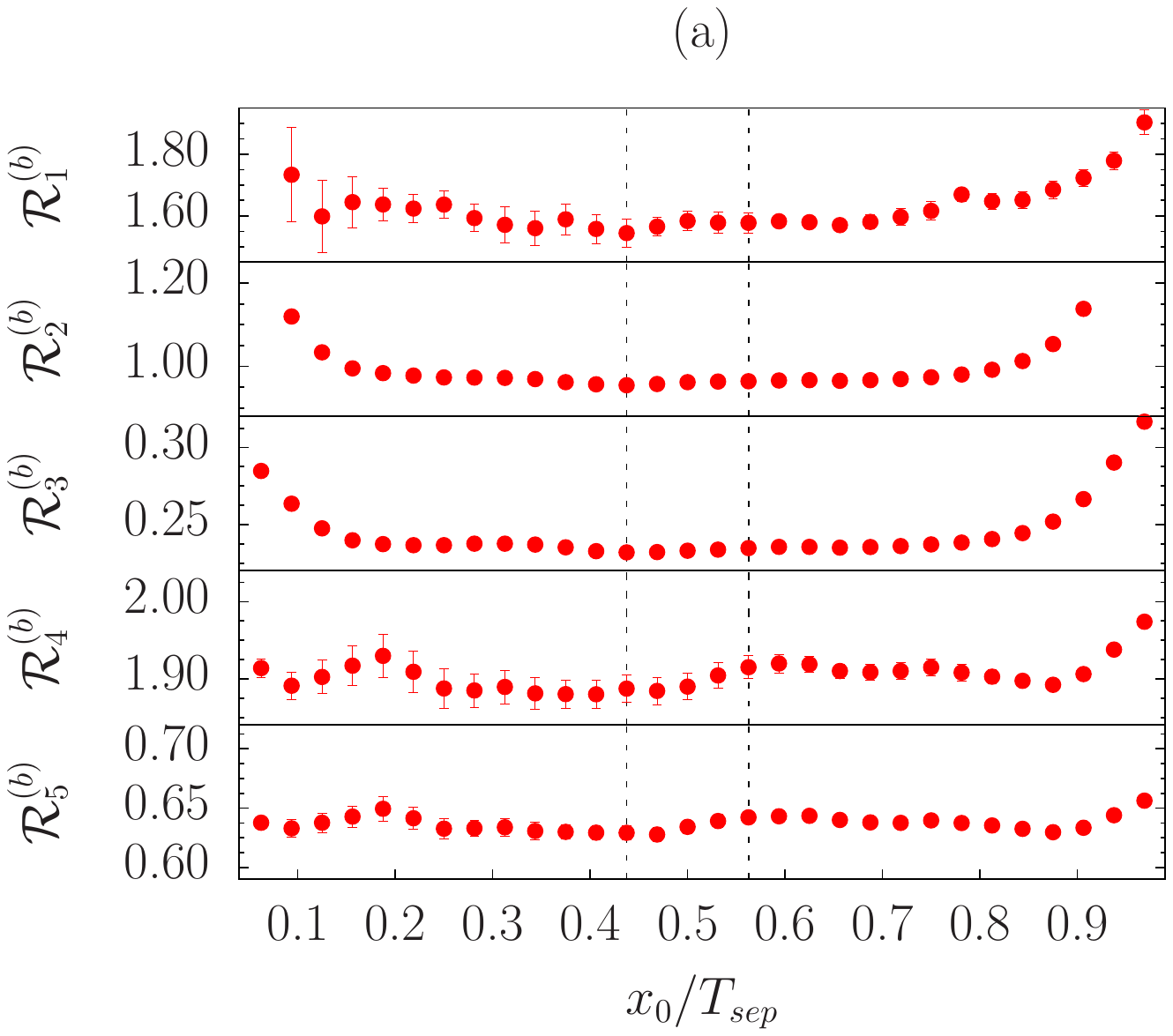} & 
\hspace*{-0.6cm}\vspace*{0.3cm}\includegraphics[bb=171bp 497bp 496bp 705bp,width=0.33\linewidth ,
keepaspectratio=true]{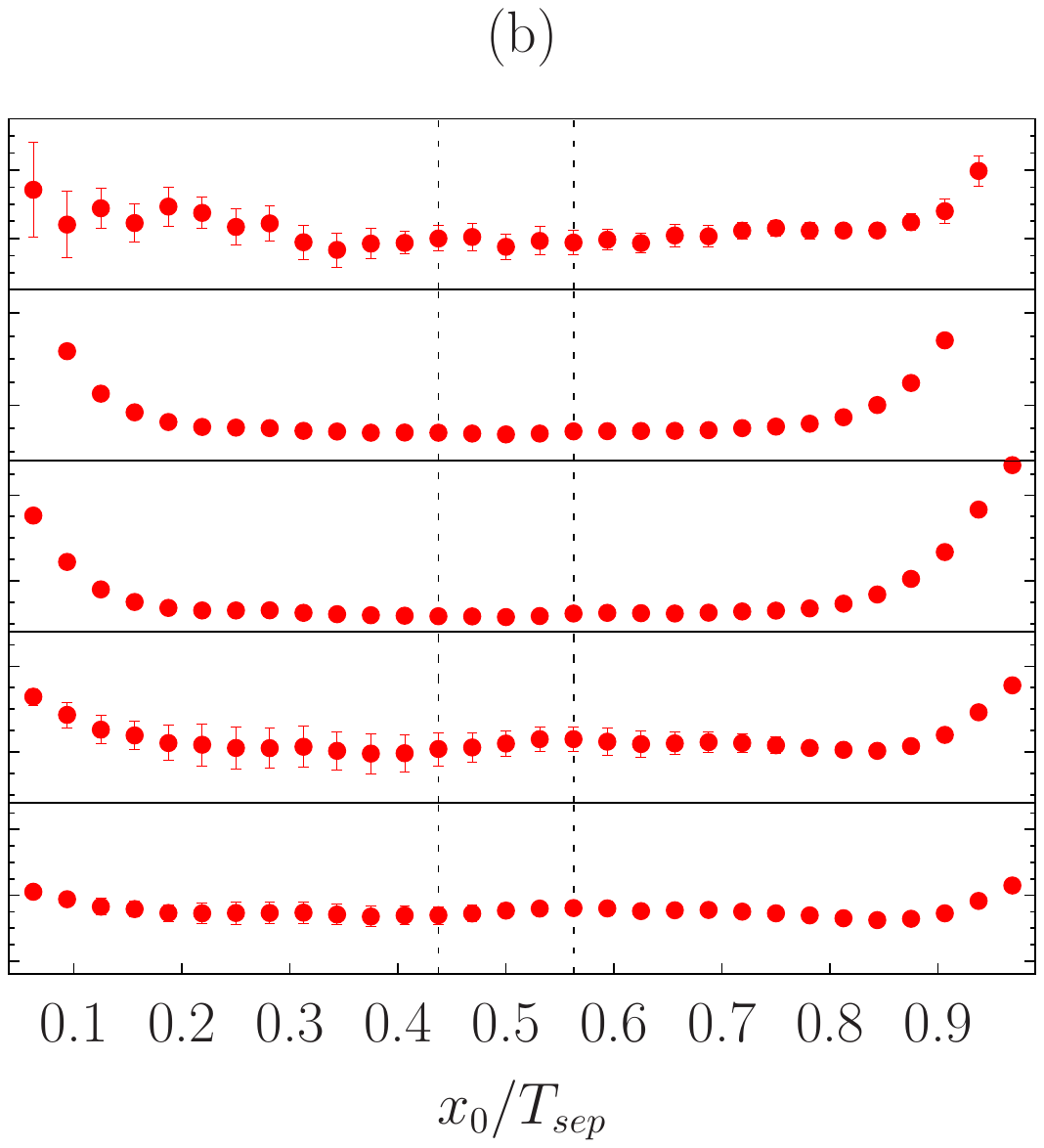} & 
\hspace*{-0.6cm}\vspace*{0.3cm}\includegraphics[bb=171bp 497bp 496bp 705bp,width=0.33\linewidth , 
keepaspectratio=true]{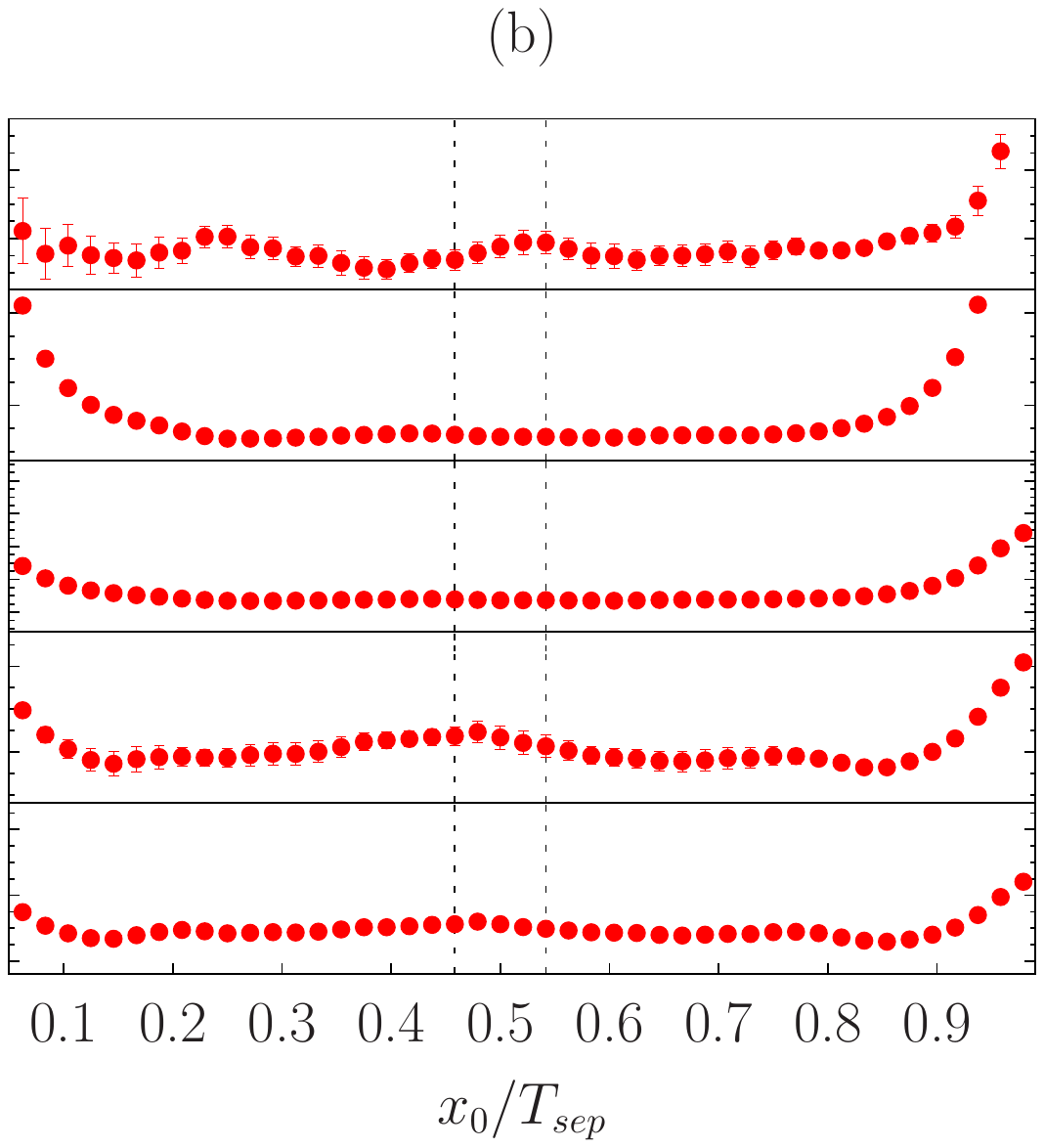}\tabularnewline
\end{tabular}
\end{center}
\caption{\label{fig:Data-and-time-plateaux-K}${\cal R}^{(b)}_{i}(x_0)$ $(i=1,...,5)$ 
simulation points plotted against $x_0/T_{sep}$ for the $\overline{K}^0-K^0$ case. 
Panel (a) $\beta=1.90$, $(a\mu_{\ell},a\mu_{s})=(0.0030,0.0185)$, volume $=32^{3}\times64$. 
Panel (b) $\beta=1.95$, $(a\mu_{\ell}, a\mu_{s})=(0.0025,0.0180)$, volume $= 32^{3}\times64$ . 
Panel (c)  $\beta=2.10$, $(a\mu_{\ell},a\mu_{s})=(0.0015,0.0.0151)$, volume $= 48^{3}\times96$. 
The dotted lines delimit the plateau region. 
Points for ${\cal R}^{(b)}_{1}, \ldots, {\cal R}^{(b)}_4$ at $\beta=2.10$ have been slightly 
shifted upward by +0.05 for accommodating data 
from all three $\beta$'s in the same 
plotting scale.}
\end{figure}

\begin{figure}[!h]
\begin{center}
\vspace*{-0.2cm}
\begin{tabular}{ccc}
\hspace*{0.5cm}\vspace*{0.2cm}\includegraphics[bb=171bp 497bp 496bp 805bp,width=0.33\linewidth , 
keepaspectratio=true]{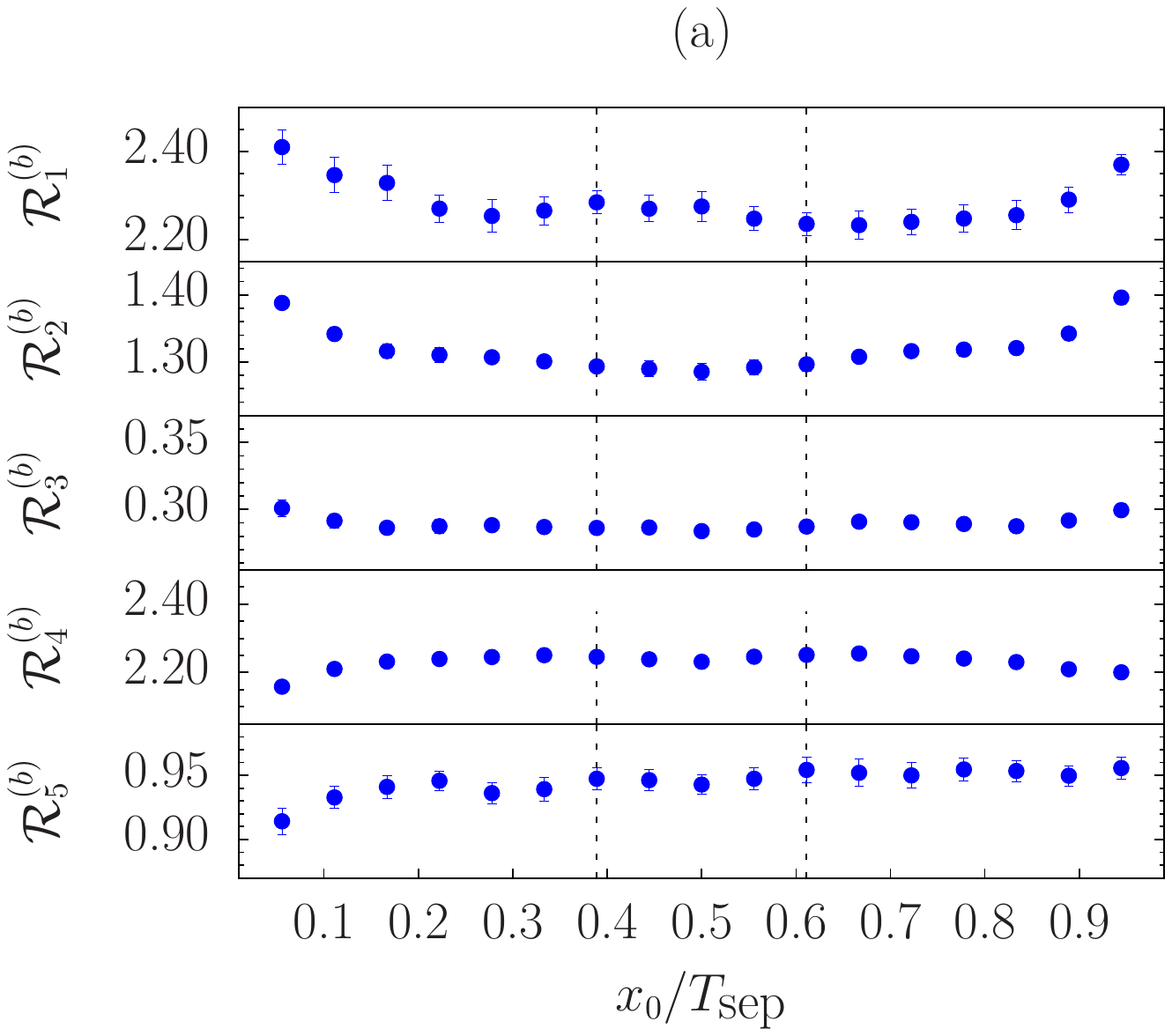} & 
\hspace*{-0.5cm}\vspace*{0.2cm}\includegraphics[bb=171bp 497bp 496bp 705bp,width=0.33\linewidth , 
keepaspectratio=true]{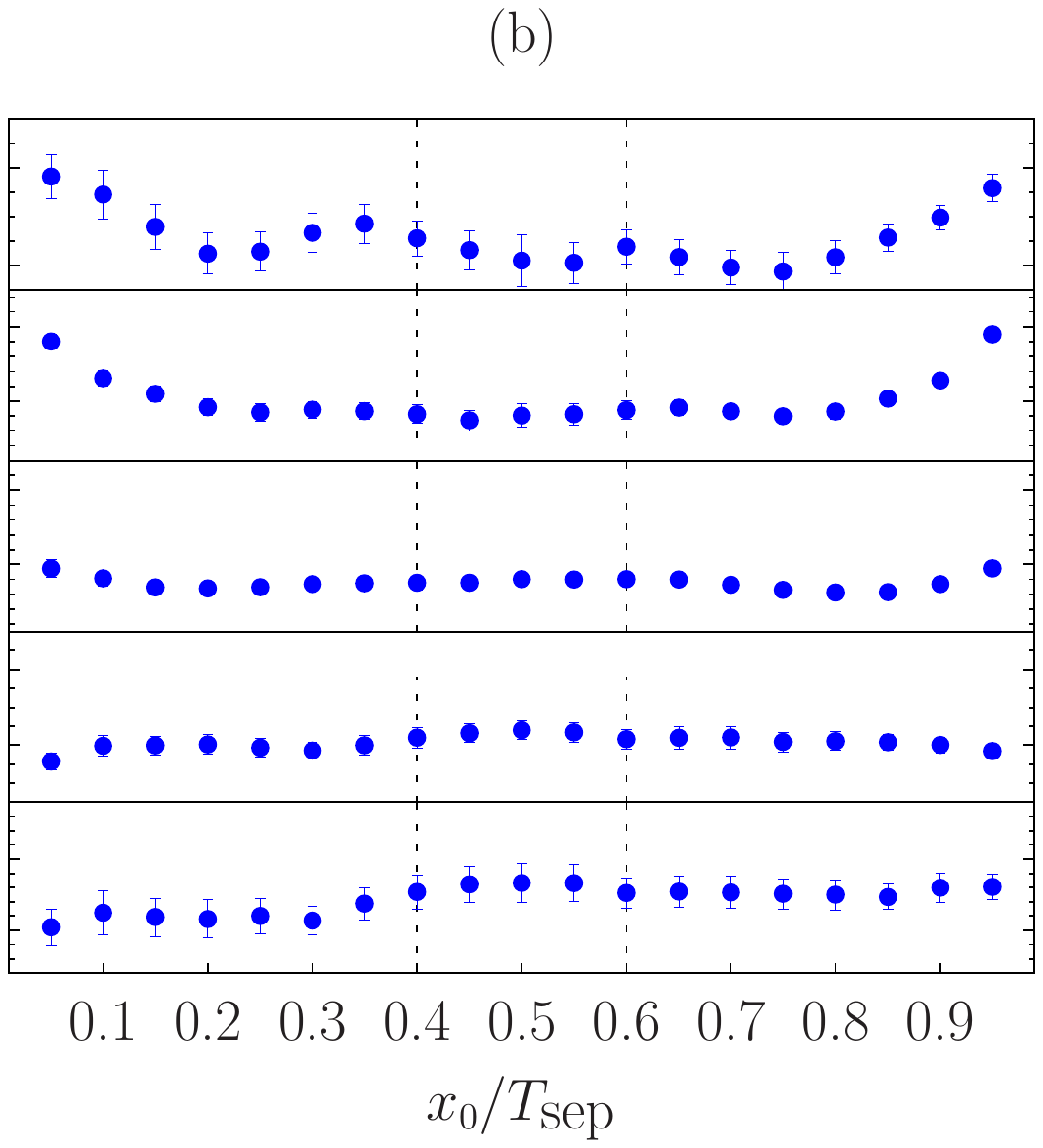}&
\hspace*{-0.5cm}\vspace*{0.2cm}\includegraphics[bb=171bp 497bp 496bp 705bp,width=0.33\linewidth , 
keepaspectratio=true]{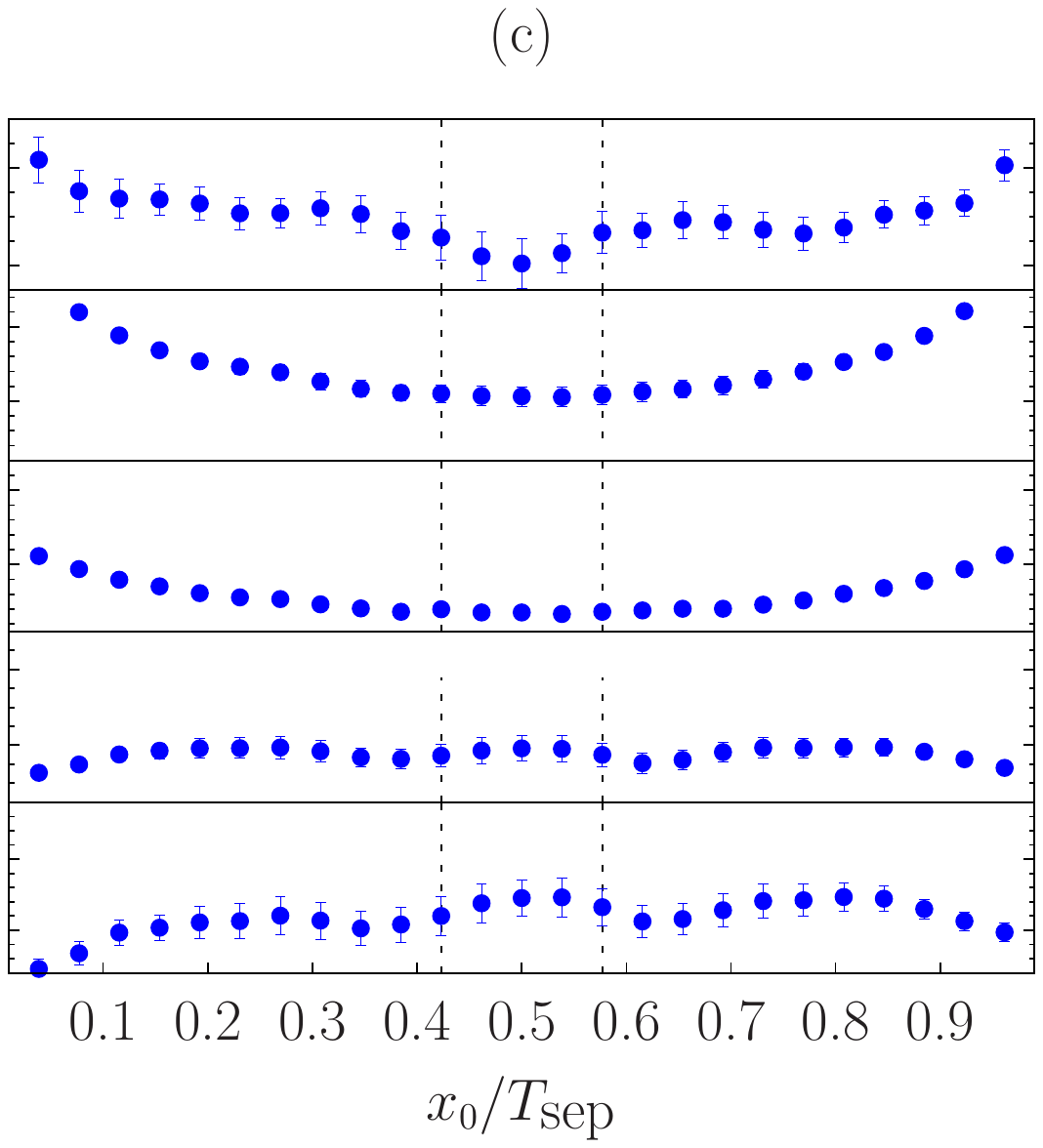} \tabularnewline
\end{tabular}
\end{center}
\caption{\label{fig:Data-and-time-plateaux-D}${\cal R}^{(b)}_{i}(x_0)$ $(i=1,...,5)$  simulation points plotted against 
$x_0/T_{sep}$ for the $\overline{D}^0-D^0$ case. Panel (a) $\beta=1.90$, $(a\mu_{\ell},a\mu_{c})=(0.0030,0.25)$, 
volume $=32^{3}\times64$. Panel (b) $\beta=1.95$, $(a\mu_{\ell},a\mu_{c})=(0.0025,0.22)$, volume $=32^{3}\times64$. 
Panel (c) $\beta=2.10$, $(a\mu_{\ell}, a\mu_{c})=(0.0015,0.17)$, volume $=48^{3}\times96$. 
The dotted lines delimit the plateau region. Points for ${\cal R}^{(b)}_{1}$ and ${\cal R}^{(b)}_2$ 
at $\beta=2.10$ have been slightly shifted upward by +0.1 for accommodating 
data from the three $\beta$'s in the same plotting scale.}
\end{figure}

\subsection{Computation of the renormalized bag-parameters}
\label{sec:ren-chir}

The renormalization pattern of the bag-parameters in our mixed action setup has been discussed in 
detail in Refs.~\cite{Frezzotti:2004wz,Bertone:2012cu,Constantinou:2010qv}. The renormalized bag-parameters are given by
\vspace{-0.1cm}
\begin{equation}
\begin{array}{ccc}
B_{1}=\dfrac{Z_{11}}{\xi_1 Z_{A}Z_{V}}B_{1}^{(b)}, & \,\,\,\,\,\,\,\,\,\,\,\,\,\,\,\,\,\,\,\,\,\, & 
B_{i}=\dfrac{Z_{ij}}{\xi_i Z_{P}Z_{S}}B_{j}^{(b)}\,\,\,\,\,\,\,\,\,\,\,\,\, i,j=2,..,5 ,\end{array} 
\end{equation}
where $(\xi_1, \xi_2,\, \xi_3,\, \xi_4,\, \xi_5)=(8/3, -5/3,\, 1/3,\, 2,\, 2/3)$. 
The RCs of bilinear operators, namely $Z_{V}$, $Z_{A}$, $Z_{P}$ and $Z_{S}$, have been determined non-perturbatively 
in the RI$'$-MOM scheme in Ref.~\cite{Carrasco:2014cwa}. The four-fermion RCs, $Z_{ij}$, have also been computed non-perturbatively 
in the same scheme. The calculation is presented in the Appendices~\ref{sec:RCs-comput-setup} and~\ref{app:RCs4f}.
   
At each value of the light quark mass $\mu_{\ell}=\mu_{sea}$ our estimates of the 
bag-parameters are linearly interpolated to the physical strange (for neutral $K$-mixing) or charm 
(for neutral $D$-mixing) quark mass. In both cases the interpolation turns out to be very smooth. 
A simultaneous chiral and lattice spacing extrapolation to the physical value of the pion mass and 
the continuum limit is finally performed. The $u/d$, strange and charm quark masses 
have been evaluated in the continuum limit in Ref.~\cite{Carrasco:2014cwa}.  

Both for neutral $K$ and $D$ meson mixing studies and for all the $B_i$'s we have employed a linear 
fit ansatz of the general form 
\begin{equation}
B_{i}=B_i^{\chi}+ b_i \hat{\mu}_{\ell}+ D_i a^{2}, \label{eq:linfit}
\end{equation}
which in all cases nicely fits the data. In our notation the hat ( $\widehat{} $ ) symbol denotes 
renormalization in the $\overline{\rm{MS}}$ scheme at the 3~GeV scale. We have also considered  
fit ans\"atze based on NLO ChPT~\cite{Becirevic:2004qd} for the $K$ bag-parameters, of the kind 
\begin{equation}
B_{i}=\overline{B}_{i}^{\chi}\left[1+\overline{b}_i 
\hat{\mu}_{\ell}\mp\dfrac{2\hat{B}_{0}\hat{\mu}_{\ell}}{16\pi^{2}f_{0}^{2}}\log\dfrac{2\hat{B}_{0}\hat\mu_{\ell}}
{16\pi^{2}f_{0}^{2}}\right]+ \overline{D}_i a^{2}
\label{eq:ChPT}
\end{equation}
and for the $D$ bag-parameters the NLO HMChPT fit ans\"atze~\cite{Becirevic:2006me} 
\begin{equation}
\begin{array}{l}
B_{1}=\tilde{B}_{1}^{\chi}\left[1+\tilde{b}_1 \hat{\mu}_{\ell}-
\dfrac{(1-3\hat{g}^{2})}{2}\dfrac{2\hat{B}_{0}\hat{\mu}_{\ell}}{16\pi^{2}f_{0}^{2}}
\log\dfrac{2\hat{B}_{0}\hat\mu_{\ell}}{16
\pi^{2}f_{0}^{2}}\right]+\tilde{D}_1 a^{2}\\
\\
B_{i}=\tilde{B}_{i}^{\chi}\left[1+\tilde{b}_i \hat{\mu}_{\ell}\mp
\dfrac{(1\mp 3\hat{g}^{2}Y)}{2}\dfrac{2\hat{B}_{0}\hat{\mu}_{\ell}}{16\pi^{2}f_{0}^{2}}
\log\dfrac{2\hat{B}_{0}\hat\mu_{\ell}}
{16\pi^{2}f_{0}^{2}}\right]+\tilde{D}_i a^{2} \, \quad i=2, 4,  5\, .
\end{array}
\label{eq:HMChPT}
\end{equation}
In Eq.~(\ref{eq:ChPT}) the sign in front of the 
logarithm is minus for $i=1,2,3$ and plus for $i=4,5$, whereas in the second of the 
Eqs.~(\ref{eq:HMChPT}) the sign is minus for $i=2$ and plus for $i=4,5$. 
In the fit procedure we use the determinations of $\hat{B}_0$ and $f_0$ reported in 
Ref.~\cite{Carrasco:2014cwa}.
We also make use of the value $Y=1$ derived in Ref.~\cite{Becirevic:2006me} 
and of the estimate $\hat{g}=0.53(4)$ obtained from the lattice measurement of 
the $g_{D^{*}D\pi}$ coupling~\cite{Becirevic:2012pf}. 
In HQET the bag parameter $B_3$ is related to $B_1$ and $B_2$. In particular,  
by setting $Y=1$, $B_3$ and $B_2$ acquire identical logarithmic terms.

\begin{figure}[!h]
\vspace*{1cm}
\begin{tabular}{cc}
\hspace*{0.1cm}\vspace*{0.1cm}\includegraphics[bb=171bp 497bp 496bp 705bp,width=0.49\linewidth , 
keepaspectratio=true]{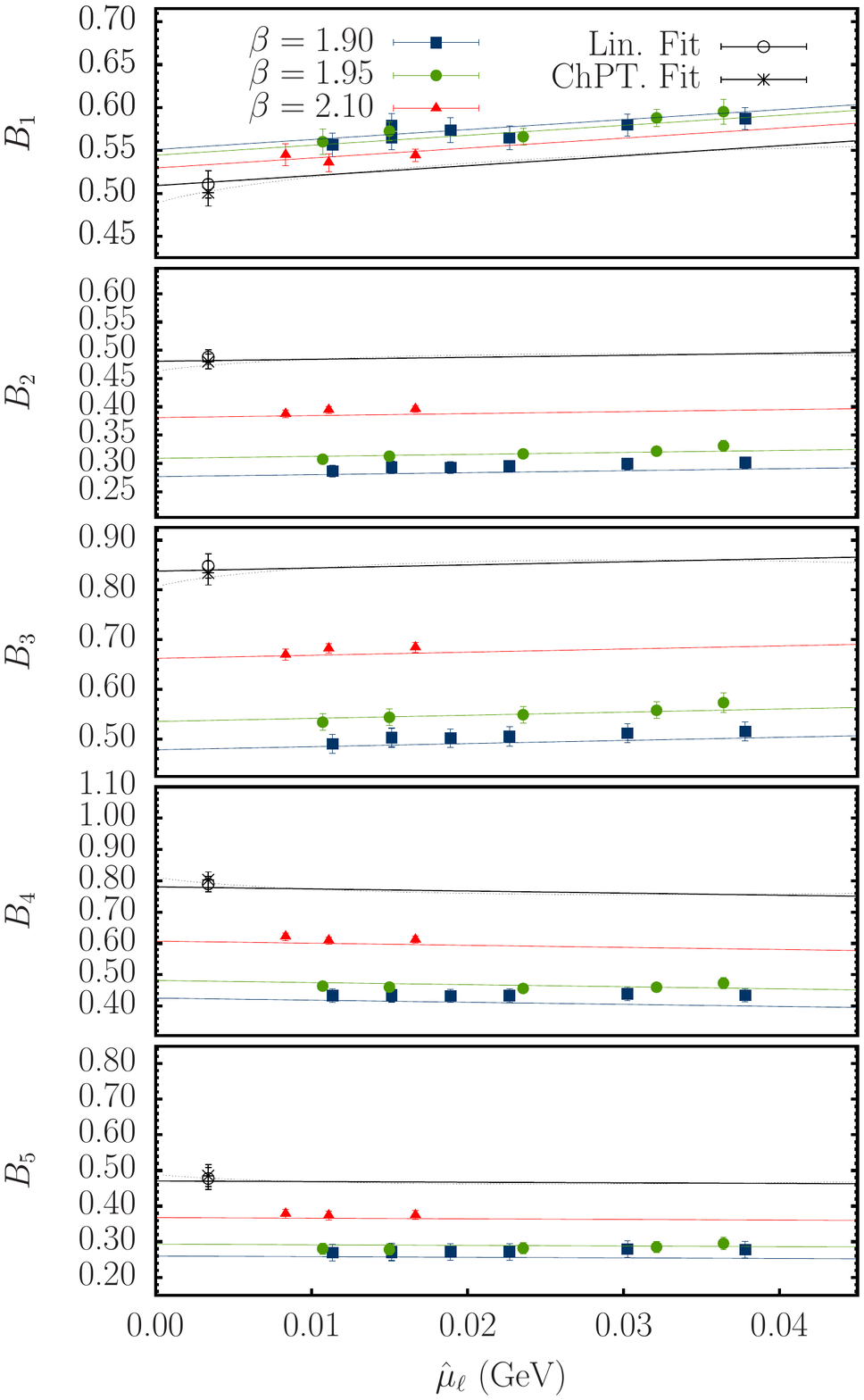} &
\hspace*{-0.1cm}\vspace*{0.1cm}\includegraphics[bb=171bp 497bp 496bp 705bp,width=0.49\linewidth , 
keepaspectratio=true]{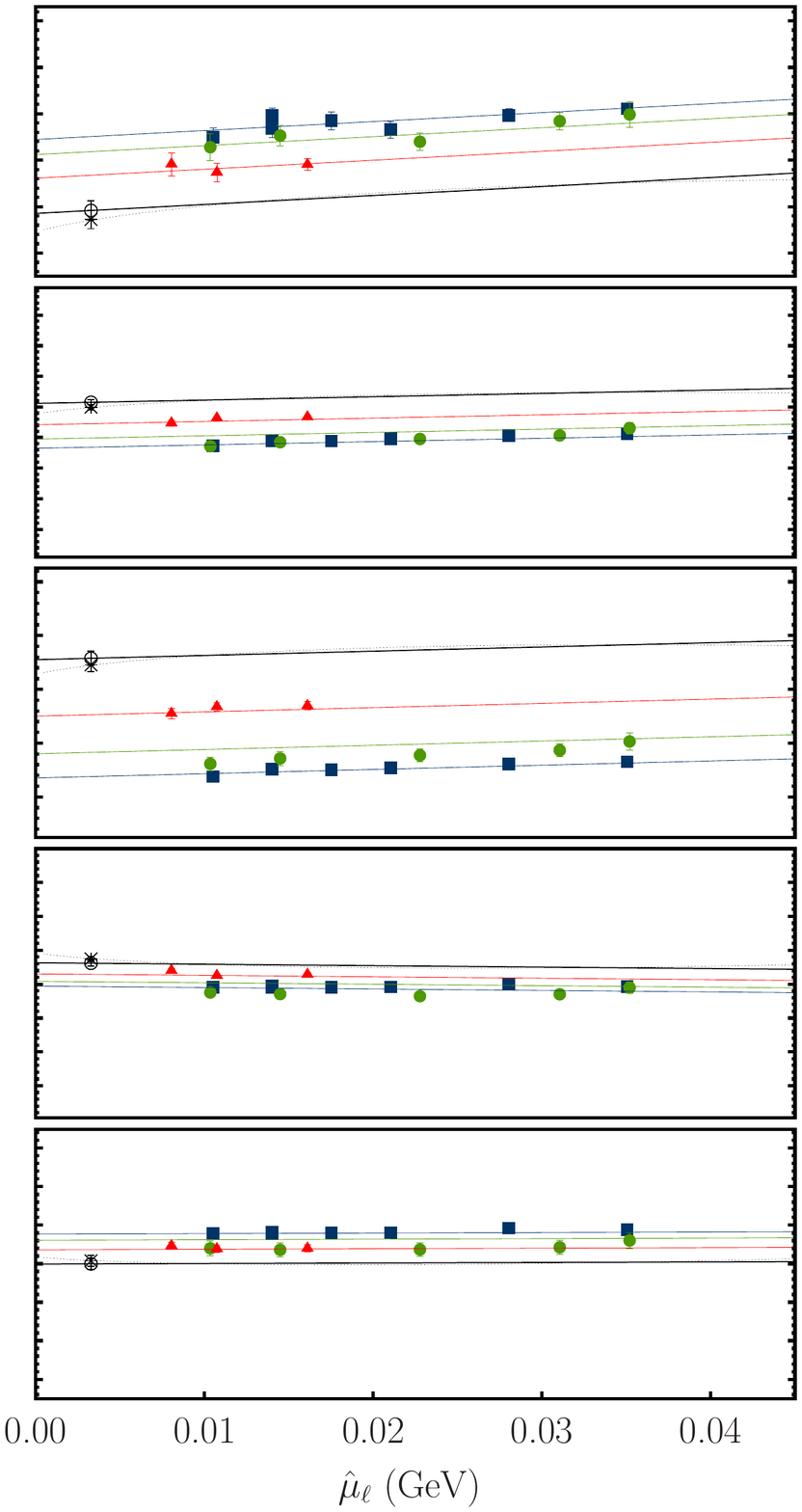} \tabularnewline
\end{tabular}
\vspace*{8cm}
\caption{\label{fig:K-continuumlimit} Combined chiral and continuum
extrapolation for the five $B_{i}$ of the $\overline{K}^0-K^0$ case. 
Bag-parameters are renormalized in the $\overline{\rm{MS}}$
scheme of~\cite{mu:4ferm-nlo} at the scale of 3~GeV. Left and right panels 
correspond to M1-type and M2-type four-fermion RCs, respectively, following the 
nomenclature of Ref.~\cite{Constantinou:2010gr}.
In each panel open circles and stars represent the value at the physical 
point corresponding to the linear and NLO ChPT fit, respectively.
 }
\end{figure}

\
\begin{figure}[!h]
\vspace*{1cm}
\begin{tabular}{cc}
\hspace*{0.1cm}\vspace*{0.1cm}\includegraphics[bb=171bp 497bp 496bp 705bp,width=0.49\linewidth , 
keepaspectratio=true]{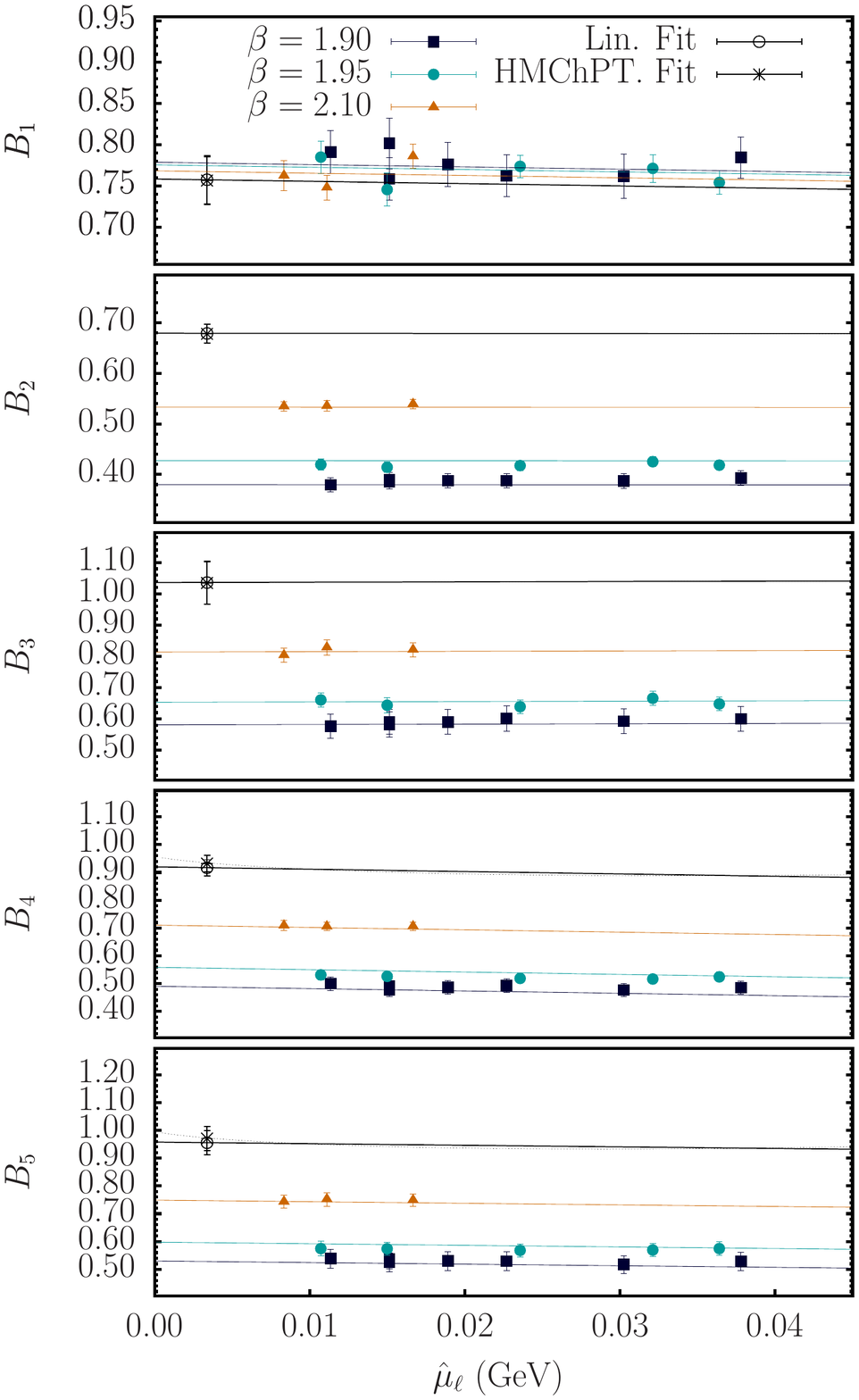} &
\hspace*{-0.1cm}\vspace*{0.1cm}\includegraphics[bb=171bp 497bp 496bp 705bp,width=0.49\linewidth , 
keepaspectratio=true]{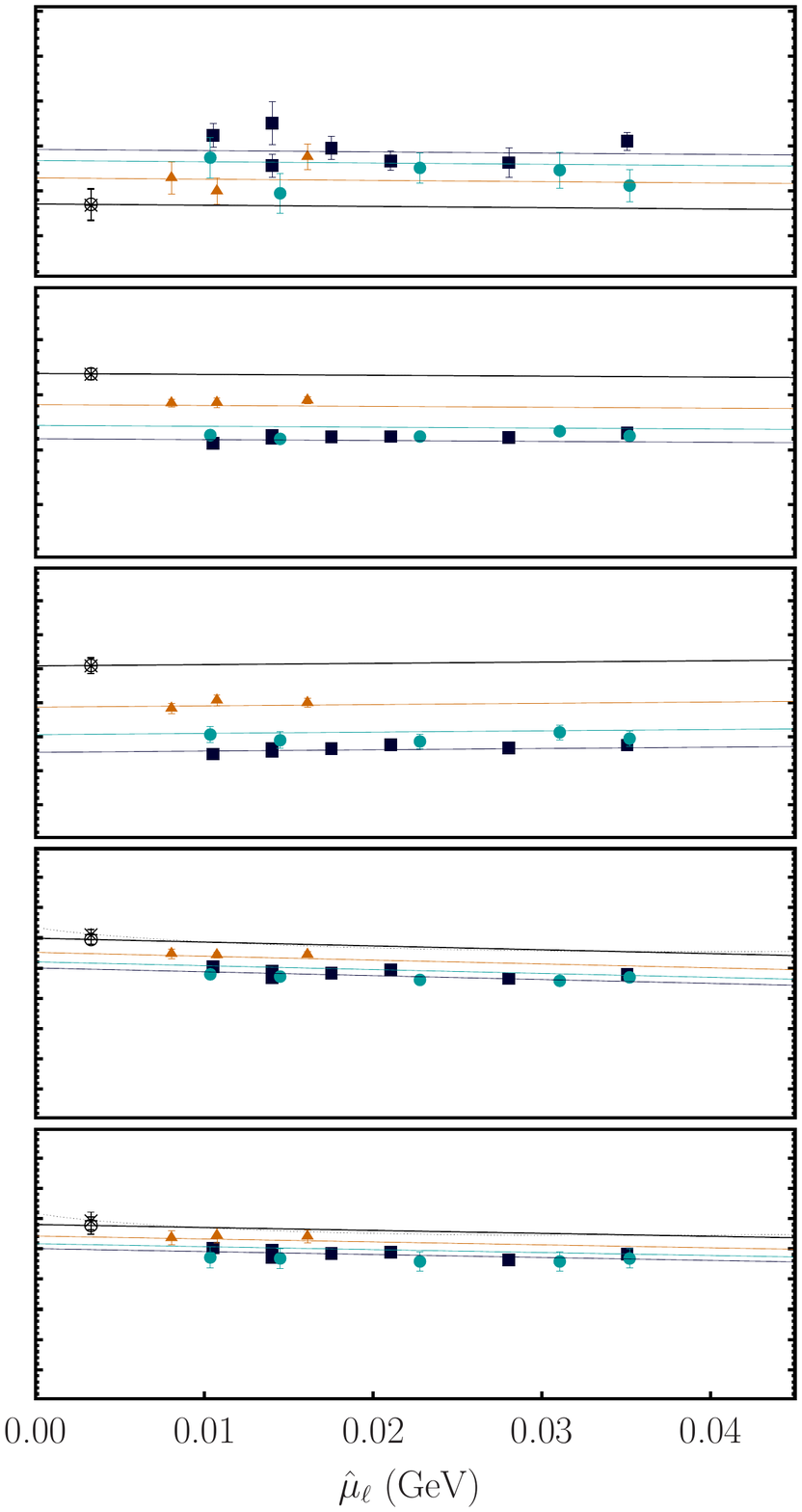} \tabularnewline
\end{tabular}
\vspace*{8cm}
\caption{\label{fig:D-continuumlimit} Combined chiral and continuum
extrapolation for the five $\overline{D}^0-D^0$ bag-parameters, $B_{i}$, renormalized in the $\overline{\rm{MS}}$
scheme of~\cite{mu:4ferm-nlo} at the scale of 3~GeV. Left and right panels correspond to M1-type 
and M2-type four-fermion RCs, respectively, following the nomenclature of Ref.~\cite{Constantinou:2010gr}.
In each panel open circles and stars represent the value at the physical point 
corresponding to the linear and NLO HMChPT fit, 
respectively. For $B_1, B_2$ and $B_3$ the polynomial (linear) and the HMChPT 
fits are practically indistinguishable. 
 }
\end{figure}

In Figs.~\ref{fig:K-continuumlimit} and~\ref{fig:D-continuumlimit} we display the combined chiral and 
continuum fit of the $B_{i}$ ($i=1, \ldots, 5$) neutral $K$ and $D$ meson bag-parameters renormalized 
in the $\overline{\rm{MS}}$ scheme of Ref.~\cite{mu:4ferm-nlo} at the scale of 3~GeV. We report in left and 
right panels, respectively, data corresponding to the two ways, namely M1 and M2, of determining the RCs 
proposed in Ref.~\cite{Constantinou:2010gr}, differing in the manner O$(a^2)$ lattice artefacts 
are treated. More details will be given in the Appendices~\ref{sec:RCs-comput-setup} and~\ref{app:RCs4f}. 
For each $B_i$ the results using either the polynomial or the chiral fit ans\"atze defined in 
Eqs.~(\ref{eq:linfit}), (\ref{eq:ChPT}) and~(\ref{eq:HMChPT}) are compatible among themselves within less than one standard deviation. 
We also notice that the use of M1 and M2-type RCs, though leading to rather  
different discretisation effects, provide compatible continuum limit determinations 
for the bag-parameters within 1-2 standard deviations.         
Moreover, by comparing results from two lattice volumes, $24^3 \times 48$ and $32^3 \times 64$, at $\beta=1.90$  
at one value of the sea quark mass ($a\mu_{sea}=0.0040$)  we notice no systematic finite volume 
effect on the values of the bag parameters.
The results for $B_i$ agree within at most one standard deviation in the worst case, while for the majority of the cases they are 
practically indistinguishable.

\section{Final results and error budget}
\label{sec:results}

In this section we present the final results for the bag-parameters 
and we discuss the error budget of statistical and systematic uncertainties. 

In our analysis we combine results obtained by using several possible ways to account for systematic effects related to 
the RCs determination, chiral extrapolation and discretisation uncertainties. We have analysed a number of 32 estimates for $B_1$ and 
64 estimates for $B_i$ with $i > 1$, see below for details.   

In particular, we have examined in detail the impact on the final values of the bag-parameters of  various possible 
sources of systematic error related to the computation of the RCs. 
We would like to mention that a large part of the uncertainties in the RI-MOM 
calculation of the RCs affects the cutoff systematics in the error budget. 

As described in Appendix~\ref{app:RCs4f}, we have computed the $5 \times 5$ four-fermion RCs in the RI$'$-MOM scheme 
using two different methods to deal with cutoff effects, which following Ref.~\cite{Constantinou:2010gr} we label M1 and M2. 
The M1 method consists in removing O$((a\tilde{p})^2)$ effects in the matrix elements used to extract the RCs by performing 
a fit in an appreciably large fixed  window of the $(a\tilde{p})^2$ momentum variable. In the M2  method the 
RCs are determined as weighted averages of RCs estimators over a $\tilde p^2$-interval (fixed in physical units) and common to all 
the gauge configuration ensembles. 
To control possible systematic effects due to the choice of the momentum interval two sets of momentum intervals have 
been compared leading to fully compatible results. 

Note that in the mixed action setup of~\cite{Frezzotti:2004wz} the off-diagonal wrong chirality mixing elements of 
the $5\times 5$ renormalization matrix are only O$(a^2)$ cutoff effects. If the latter are ignored, the lattice RCs 
matrix shows the same mixing pattern as in the formal continuum theory.
To check to what extent discretisation systematics can affect the final values of the bag-parameters, we compared 
the numbers obtained by simply ignoring the off-diagonal RC matrix elements with what one gets by including O$(a^2)$ 
mixing effects.  

The analysis of systematic uncertainties due to the use of polynomial and (HM)ChPT fit ans\"atze, see Eqs~(\ref{eq:linfit}-
\ref{eq:HMChPT}), is performed with reference to the so-called ``golden" bag-combinations~\cite{Bae:2013tca}  

\begin{equation}
\begin{array}{lll}
G_{23}=\dfrac{B_{2}}{B_{3}}, &  & G_{45}=\dfrac{B_{4}}{B_{5}}\\
\\
G_{24}=B_{2}B_{4}, &  & G_{21}=\dfrac{B_{2}}{B_{1}}\, .
\end{array}
\label{eq:golden-ratio}
\end{equation}
Since these quantities are constructed in a way that chiral logarithmic terms cancel up to 
NLO~\footnote{Strictly speaking in the case of $D$ this is not so for the combination $G_{24}$ as the 
NLO logarithmic terms do not cancel out completely. However, since the $G_{24}$ data show anyway a good 
linear behaviour vs.\ the light quark mass, we tried a linear fit ansatz even in this case.}, they are 
expected to follow an almost linear behaviour as a function of $\hat{\mu}_{\ell}$. Using the 
parametrization~(\ref{eq:golden-ratio}) we obtain additional estimates for $B_{2, \ldots, 5}$ 
without having to fit chiral logarithmic behaviours.   

Finally, in order to estimate systematic uncertainties due to cutoff effects for $B_{2, \ldots, 5}$ we have also  
carried out the scaling analysis of quantities which are found, empirically,  
to be affected by reduced discretisation errors. Therefore, if the M1-type RCs are employed, in the $K$ case we consider
\begin{equation}
B_{1}\times B_{2},\,\,\,\, B_{2}/B_{3},\,\,\, B_{3}/B_{4},\,\,\,\, B_{4}/B_{5}\, ,
\label{eq:GNUR1}
\end{equation}
while in the $D$ case we take 
\begin{equation}
B_{2}/B_{1},\,\,\,\, B_{2}/B_{3},\,\,\, B_{3}/B_{4},\,\,\,\, B_{4}/B_{5}\, .
\label{eq:GNUR1-D}
\end{equation}
If the M2-type RCs are employed, in the $K$ case they are
\begin{equation}
B_{1}\times B_{2},\,\,\,\, B_{1}\times B_{3},\,\,\, B_{1}\times B_{4},\,\,\,\, B_{1}/B_{5}\, ,
\label{eq:GNUR2}
\end{equation}
while in the $D$ case we take
\begin{equation}
B_{1}\times B_{2},\,\,\,\, B_{1}\times B_{3},\,\,\, B_{1}\times B_{4},\,\,\,\, B_{1} \times B_{5}\, .
\label{eq:GNUR2-D}
\end{equation} 
Naturally in this kind of analysis all the BSM bag-parameters 
will eventually turn out to be expressed in terms of $B_1$ which, 
however, among all the others is the quantity that is less affected by discretization effects.

\begin{figure}[!t]
\hspace*{3.5cm} \includegraphics[bb=191bp 497bp 496bp 705bp,width=0.59\linewidth , 
keepaspectratio=true]{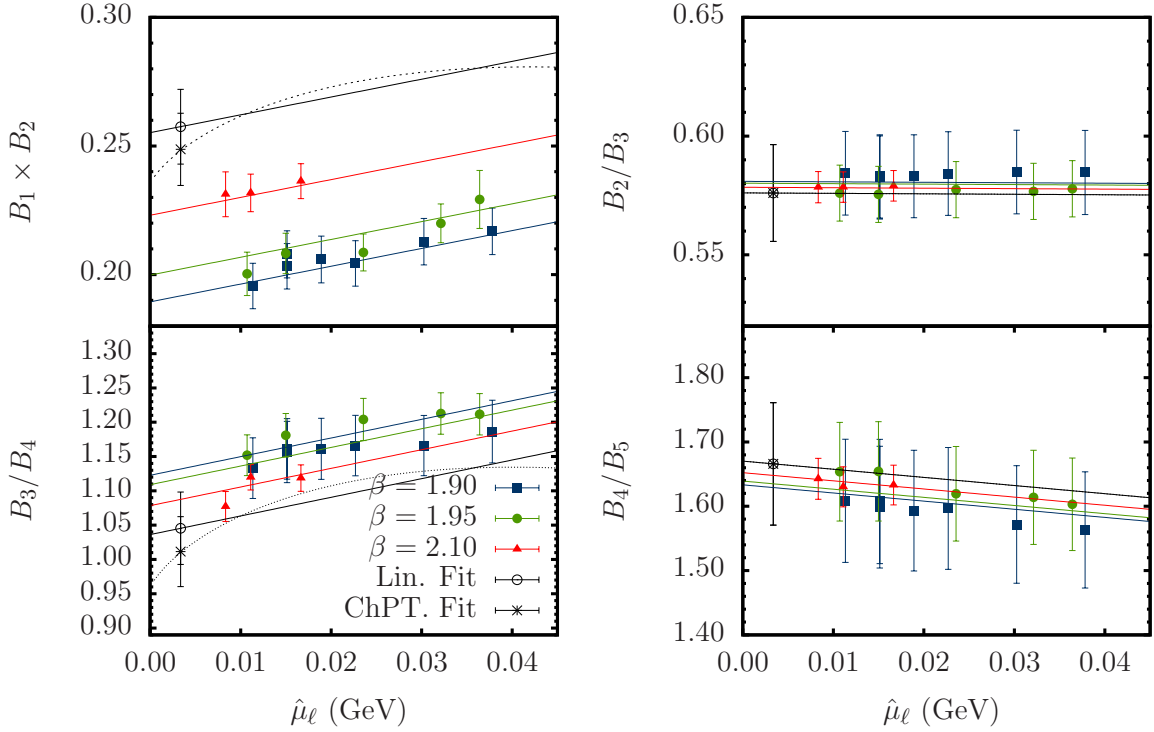} 
\vspace*{2cm}
\caption{\label{fig:K-continuumlimit-NUR1} Combined chiral and continuum
extrapolation in the $\overline{K}^0-K^0$ case of the combinations defined in Eq.~(\ref{eq:GNUR1}). We use M1-type RCs in 
the $\overline{\rm{MS}}$
scheme of~\cite{mu:4ferm-nlo} at 3~GeV. For the combinations shown in the right panels the polynomial (linear) 
and the NLO ChPT fit ansatz coincide.}
\end{figure}

To summarise we have carried out\footnote{The total number of different analyses for the SM bag-parameter $B_1$ 
we have considered is given by the product $32=2\times 4\times 2\times2$. These numbers refer to   
the  two fit ans\"atze for the chiral extrapolation, the four ways of 
combining the M1 and M2 kinds of RCs estimates 
needed for the renormalization of the four- and two-fermion operators, the two choices of the $p^2$-interval, 
and finally a factor of 2 for 
including the off-diagonal scale independent O$(a^2)$ matrix elements $\Delta_{ij}$ in the construction 
of the renormalized operators or setting them equal to zero. As for the number of BSM bag-parameter analysis, 
owing the alternative ways of parametrizing chiral (Eq.~\ref{eq:golden-ratio}) and lattice spacing fit anz\"atze 
(Eqs.~(\ref{eq:GNUR1}), (\ref{eq:GNUR2}) and~Eqs.~(\ref{eq:GNUR1-D}), (\ref{eq:GNUR2-D})), the above number must 
multiplied by 2, thus giving in total 64 kinds of analysis.} 32 kinds of analysis for $B_1$ and 64 for the BSM 
bag-parameters, $B_{2, \ldots, 5}$. 

Statistical errors have been evaluated using the jackknife method. We have verified that for all the gauge 
configuration ensembles 16 jackknife bins are enough to have autocorrelations well under control. Fit cross 
correlations are taken into account by generating 1000 bootstrap samples for each gauge configuration ensemble. 
The RCs computation has been performed on a different set of  $N_f=4$ gauge configuration ensembles. 
The  error on each RC has been propagated  assuming RCs to be gaussian distributed 
with the central values and the standard deviations reported in Tables~\ref{tab:RCs-bilinear} and \ref{tab:RCs-4f-MS}.

Our total statistical uncertainty includes the statistical errors on the bare matrix elements, the statistical 
uncertainty of the RCs and the propagated error coming from the combined continuum and chiral fit extrapolation.

For each bag-parameter the central value is determined by the average over 
the corresponding set of results. Note that, since all our analyses are characterised by comparable fit quality, 
we combine the results from different analyses assuming the same weight for all of them. Therefore for the final 
central values as well as for the statistical and systematic uncertainties we make use of the formulae 
(as already done in Ref.~\cite{Carrasco:2014cwa}):
\begin{eqnarray}
\overline{x}  &=& \dfrac{1}{N}\displaystyle\sum_{i=1}^{N}x_{i} , \\ \label{eq:xbar}
\sigma^2    &=&\dfrac{1}{N}\displaystyle\sum_{i=1}^{N}\sigma_{i}^2+\dfrac{1}{N}\sum_{i=1}^{N}(x_{i}-\overline{x})^2 , \label{eq:sigma}
\end{eqnarray}
where $x_i$ and $\sigma_i$ are the central value and the variance of the $i$-th analysis and $N$ is the total number 
of analyses, i.e.\ $N=32$ for $B_1$ and $N=64$ for $B_{2, \ldots, 5}$. From the first term of the r.h.s.\ of 
Eq.~(\ref{eq:sigma}) we read off the statistical error while the second, which represents the spread among the 
results of different analyses, provides an  estimate for the total systematic uncertainty. 

In Tables~\ref{tab:Bi-K-all} and~\ref{tab:Bi-D-all}  we have collected our final results for $B_i$ $(i=1, \ldots, 5)$ 
evaluated in the ${\overline{\rm{MS}}}$ and RI$'$ schemes. The final uncertainty is given by summing in 
quadrature the statistical and the systematic errors following Eq.~(\ref{eq:sigma}) . 

In Fig.~\ref{fig:BK-sist} and~Fig.~\ref{fig:BD-sist} we illustrate the distribution of the results 
for each $B_i$ in the $K$ and $D$ case, respectively. 
Had we chosen the median of the results to represent the central value, we would have obtained numbers fully compatible 
(within better than one -statistical- standard deviation) with the results collected in Tables~\ref{tab:Bi-K-all} and~\ref{tab:Bi-D-all}. 
It has also been  checked that the width of the interval which selects the 68\% of the area around the average 
(or the median) is in all cases very close to the value 
provided by Eq.~(\ref{eq:sigma}). We consider this a nice test of the validity and usefulness of 
our way of estimating the total error\footnote{The fit quality for the vast majority of the 32 analyses 
for $B_1$ and 
the 64 analyses for $B_{i=2, \ldots, 5}$ is good while only for a small number of cases -- in particular, 6 (7) out of 64 analyses for $B_2$ and 10 (11) out 
64 analyses for $B_4$ for the neutral Kaon (D) case -- we noticed poor fit quality. 
Nevertheless, we decided to attribute to the results of all analyses 
the same weight since this choice led to somewhat more conservative estimates of the systematic uncertainties. 
Infact if we had opted for a $\chi^2$-weighted analysis 
strategy  the shifts of the central values for all $B_i$ would have been minimal and the finally estimated systematic errors 
smaller (by a few percent for $B_{i=1, 2, 4}$ up to almost 30\% for $B_{i=3, 5}$) than the ones presented in 
Tables \ref{tab:error-budget-K} and \ref{tab:error-budget-D}.}.

\begin{figure}[!h]
\hspace*{-1cm}\includegraphics[scale=0.75]{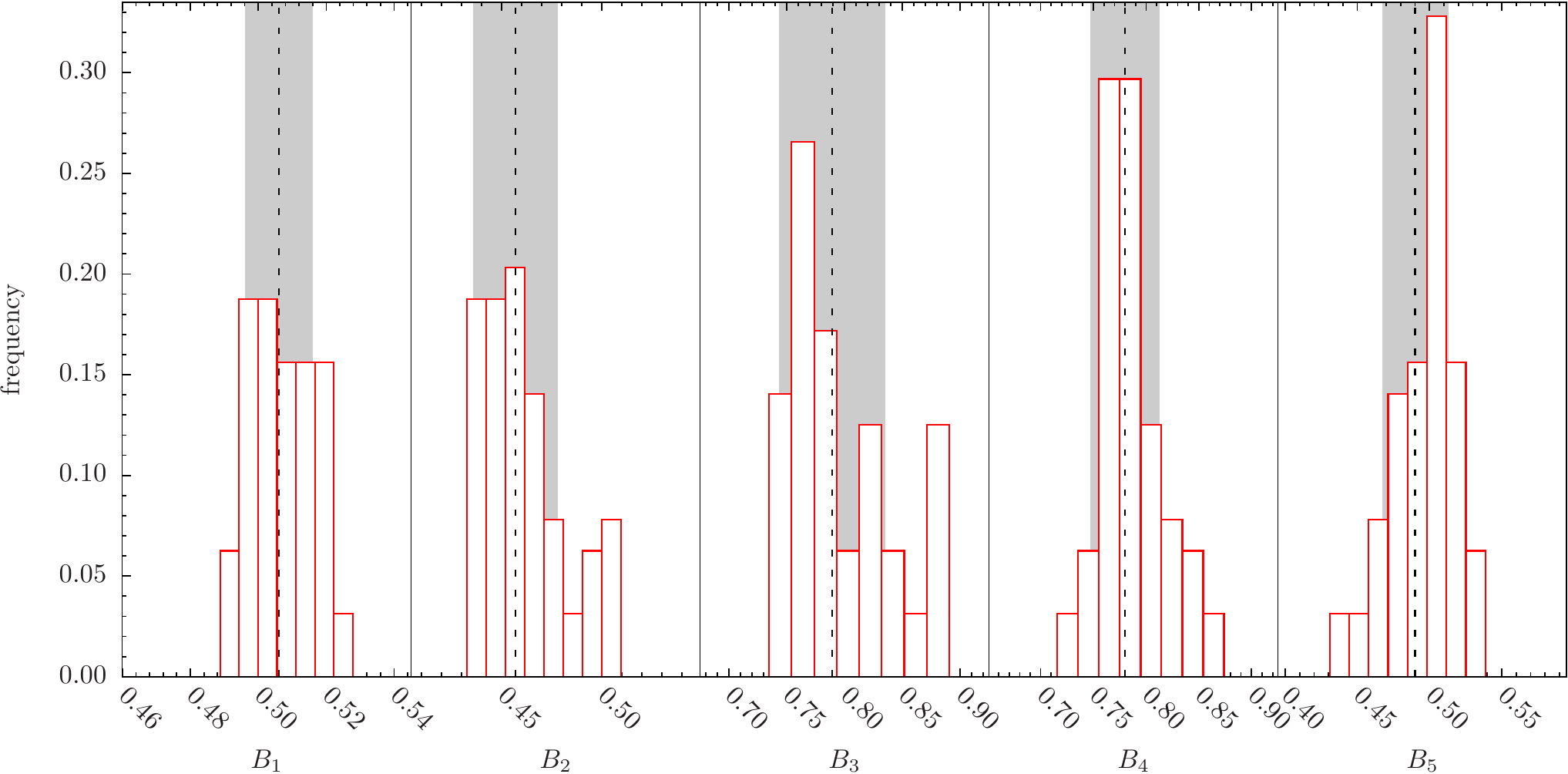} 
\caption{\label{fig:BK-sist} Distribution of $B_{1, \ldots, 5}$ results for the neutral $K$-mixing renormalized in 
the $\overline{\rm{MS}}$ scheme of Ref.~\cite{mu:4ferm-nlo} at the scale of 3 GeV. 
The solid vertical line marks the central value (average) while the gray band indicates 
the systematic error determined from Eq.~(\ref{eq:sigma}). }
\end{figure}

\begin{figure}[!h]
\hspace*{-1cm}\includegraphics[scale=0.75]{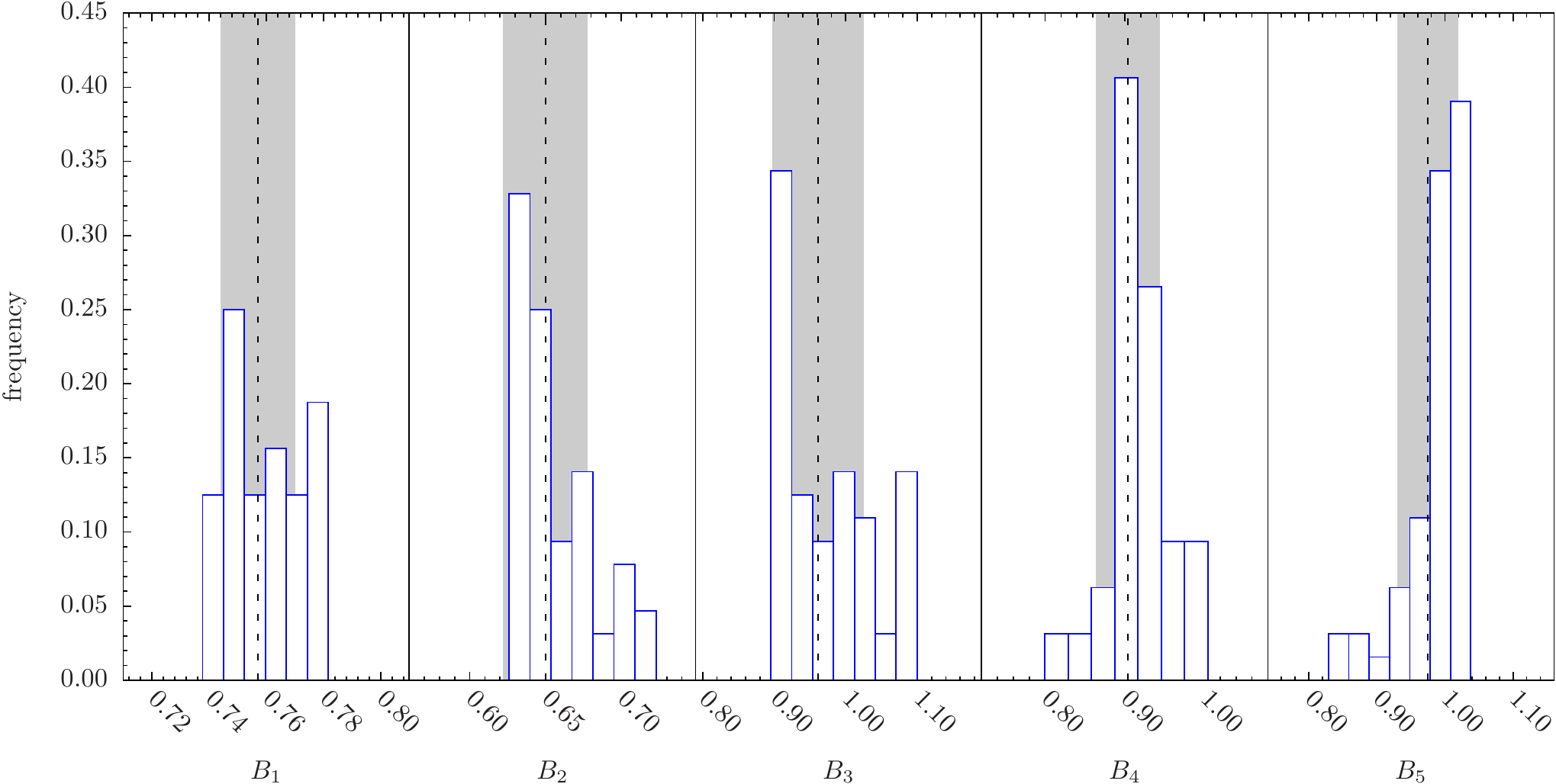} 
\caption{\label{fig:BD-sist} Same as in Fig.~\ref{fig:BK-sist} but for the neutral $D$-mixing. }
\end{figure}

In Tables~\ref{tab:error-budget-K} and~\ref{tab:error-budget-D} we report the detailed error budget of our 
determination of the $K$ and $D$ bag parameters. The numbers represent the percentage of the main sources 
of uncertainty in our calculation. The total percentage error is reported in the last rows of Tables~\ref{tab:error-budget-K} 
and~\ref{tab:error-budget-D}. 

Under the label ``stat+fit+RCs" we lump together the error coming from the statistical uncertainties of 
correlators, the interpolation/extrapolation of the simulated quark masses to the physical values, the 
extrapolation to the continuum limit, as well as the statistical uncertainties of the RCs.

Under the label ``syst. chiral" we give our estimates of the chiral fit uncertainty. This has been 
determined using the different ways we have used to perform the chiral extrapolation, namely comparing 
results coming from the use of Eqs.~(\ref{eq:linfit}), (\ref{eq:ChPT}) and~(\ref{eq:HMChPT}) as well as Eqs.~(\ref{eq:golden-ratio}).

In the row labeled ``syst. discr." we give our final estimate of the systematic uncertainties 
related to the choice of the fit ansatz of discretisation artifacts. The uncertainty is taken as the 
spread of the bag parameter results obtained with the use of Eqs.~(\ref{eq:GNUR1}) to~(\ref{eq:GNUR2-D}) 
and of the different ways (M1 or M2) of computing the relevant RCs.

Finally, in the last row of each table we quote our estimate for the systematic uncertainty due to the perturbative matching of the 
RI$'$ and $\overline{\rm{MS}}$ schemes. We recall that anomalous dimensions of the four-fermion operators are known up to NLO.
Our associated systematic error has been estimated by considering   
the difference between the values obtained at NLO and LO at the scale of 3~GeV for each one of the  bag-parameters 
and multiplying it with the value ($\sim 0.25$) that $\alpha_s^{\overline{\rm{MS}}}(3~ \rm{GeV})$  takes at the same scale.   

Finite volume effects, as mentioned in the previous section, are practically negligible at the level of our precision.

In Figs~\ref{Fig:error-budget-K} and~\ref{Fig:error-budget-D} we 
graphically show the error budget associated to the lattice computation {\it i.e.} without including the 
systematic error due to the perturbative conversion from RI$'$ to $\overline{\rm{MS}}$ scheme, 
see Tables~\ref{tab:error-budget-K} and~\ref{tab:error-budget-D}.

\vspace*{0.6cm}

{\renewcommand{\arraystretch}{1.3}
\begin{table}
\begin{centering}
\begin{tabular}{|l|l|l|l|l|l|}
\hline 
\multicolumn{6}{|c|}{$\overline{K}^0-K^0$}\tabularnewline
\hline 
\hline
\hline 
source of uncertainty (\%) & $B_{1}$ & $B_{2}$ & $B_{3}$ & $B_{4}$ & $B_{5}$\tabularnewline
\hline 
\hline 
stat+fit+RCs & 2.5 & 2.4 & 2.7 & 2.7 & 5.4\tabularnewline
\hline 
syst. chiral & 0.8 & 1.3 & 1.1 & 1.8 & 2.6\tabularnewline
\hline 
syst. discr. & 2.0 & 4.7 & 5.8  & 3.8 & 4.1\tabularnewline
\hline 
RI$'$-$\overline{\rm{MS}}$ matching & 0.5 & 2.5 & 1.8 & 3.9 & 2.3 \tabularnewline
\hline
Total        & 3.4 & 6.0 & 6.7 & 6.3 & 7.6\tabularnewline 
\hline 
\end{tabular}\caption{\label{tab:error-budget-K} 
Full error budget of the $B_{1, \ldots, 5}$ estimates for the neutral $K$-mixing.} 
\par\end{centering} 
\end{table}
 
\renewcommand{\arraystretch}{1.3}
\begin{table}
\begin{centering}
\begin{tabular}{|l|l|l|l|l|l|}
\hline 
\multicolumn{6}{|c|}{$\overline{D}^0-D^0$}\tabularnewline
\hline 
\hline
\hline 
source of uncertainty (\%) & $B_{1}$ & $B_{2}$ & $B_{3}$ & $B_{4}$ & $B_{5}$\tabularnewline
\hline 
\hline 
stat+fit+RCs & 2.9 & 2.9 & 4.4 & 3.5 & 5.1\tabularnewline
\hline 
syst. chiral & 0.2 & 0.4 & 0.6 & 2.3 & 2.6\tabularnewline
\hline 
syst. discr. & 2.1 & 4.3 & 6.7 & 3.8 & 3.3\tabularnewline
\hline 
RI$'$-$\overline{\rm{MS}}$ matching & 0.5 & 2.5 & 1.7 & 3.9 & 1.1 \tabularnewline
\hline
Total        & 3.6 & 5.8 & 8.2 & 6.9 & 6.7\tabularnewline
\hline 
\end{tabular}\caption{\label{tab:error-budget-D}   Full error budget of the $B_{1, \ldots, 5}$ estimates for the neutral $D$-mixing.} 
\par\end{centering} 
\end{table}
\vspace*{1cm}

 \begin{figure}[!h]
\vspace*{0cm}
\begin{tabular}{cccccc}
\hspace*{-0.8cm}\includegraphics[scale=0.28]{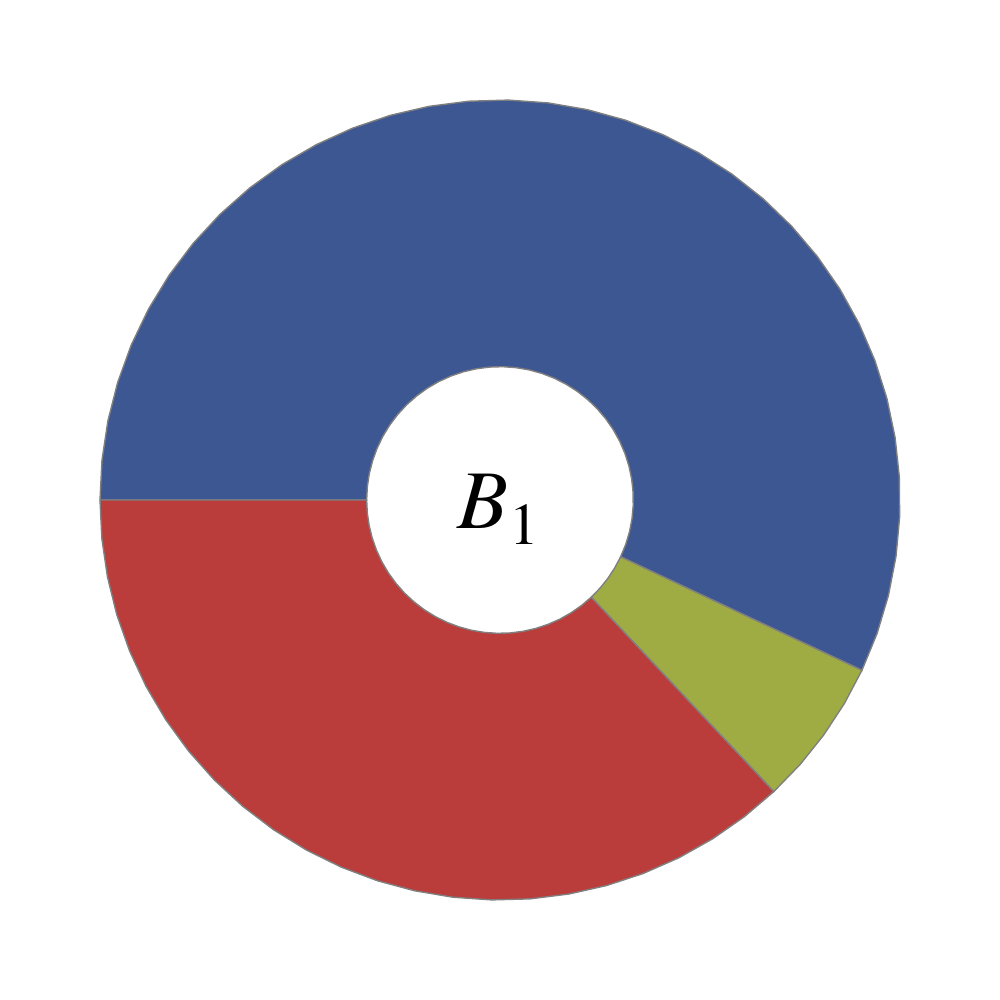}\hspace*{-0.82cm} 
& \includegraphics[scale=0.28]{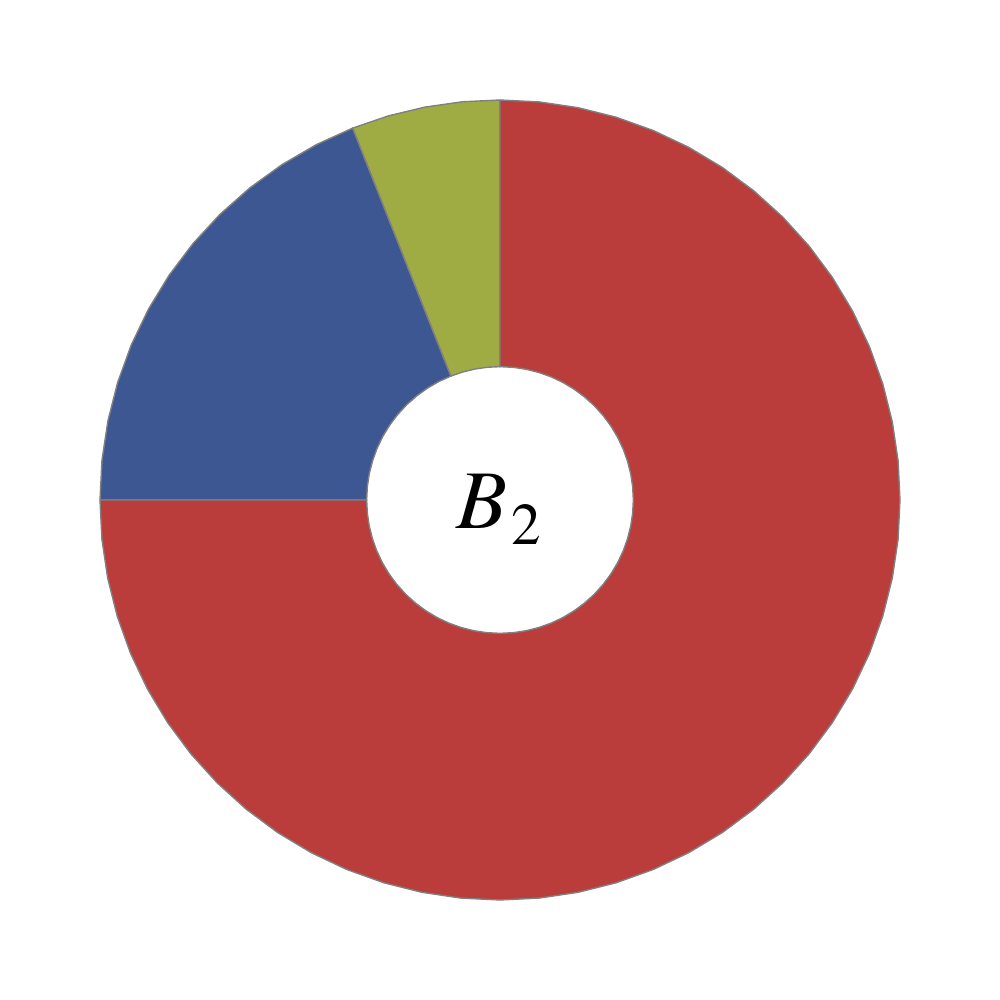}\hspace*{-0.82cm} 
& \includegraphics[scale=0.28]{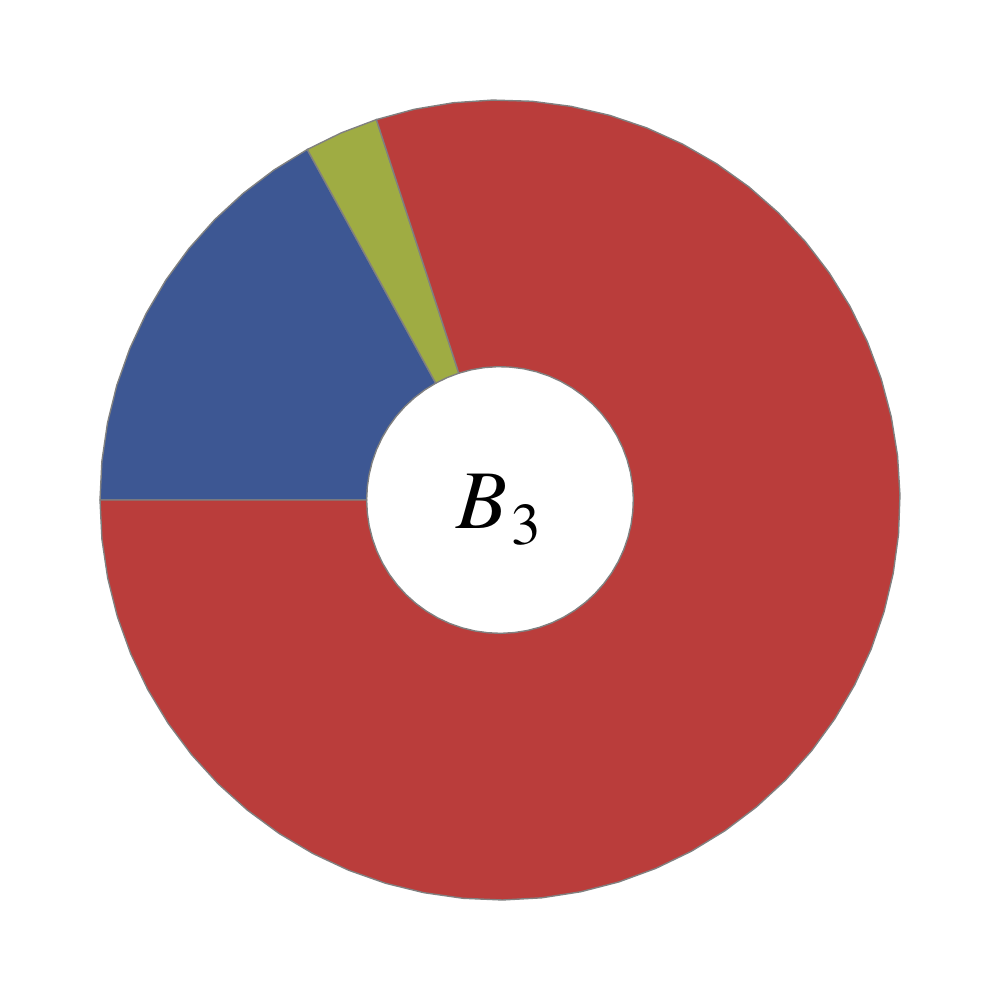}\hspace*{-0.82cm} 
& \includegraphics[scale=0.28]{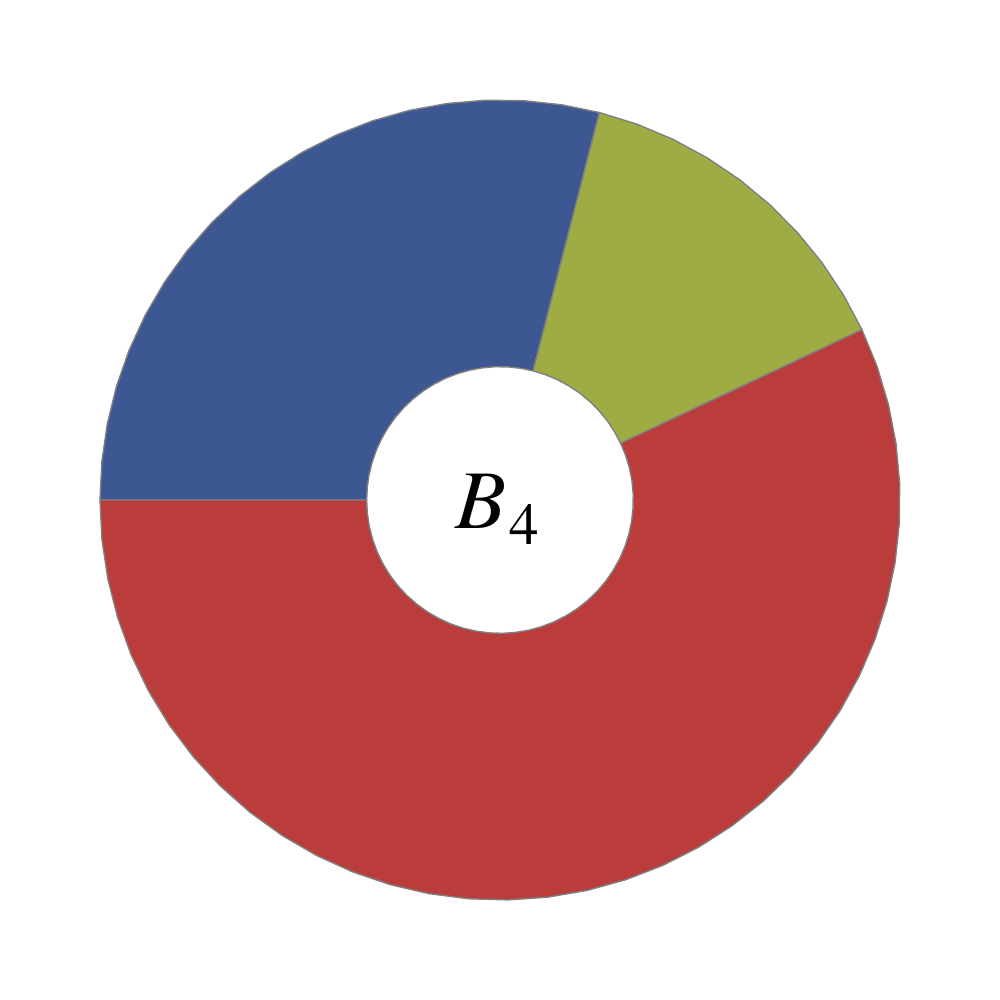}\hspace*{-0.82cm} 
& \includegraphics[scale=0.28]{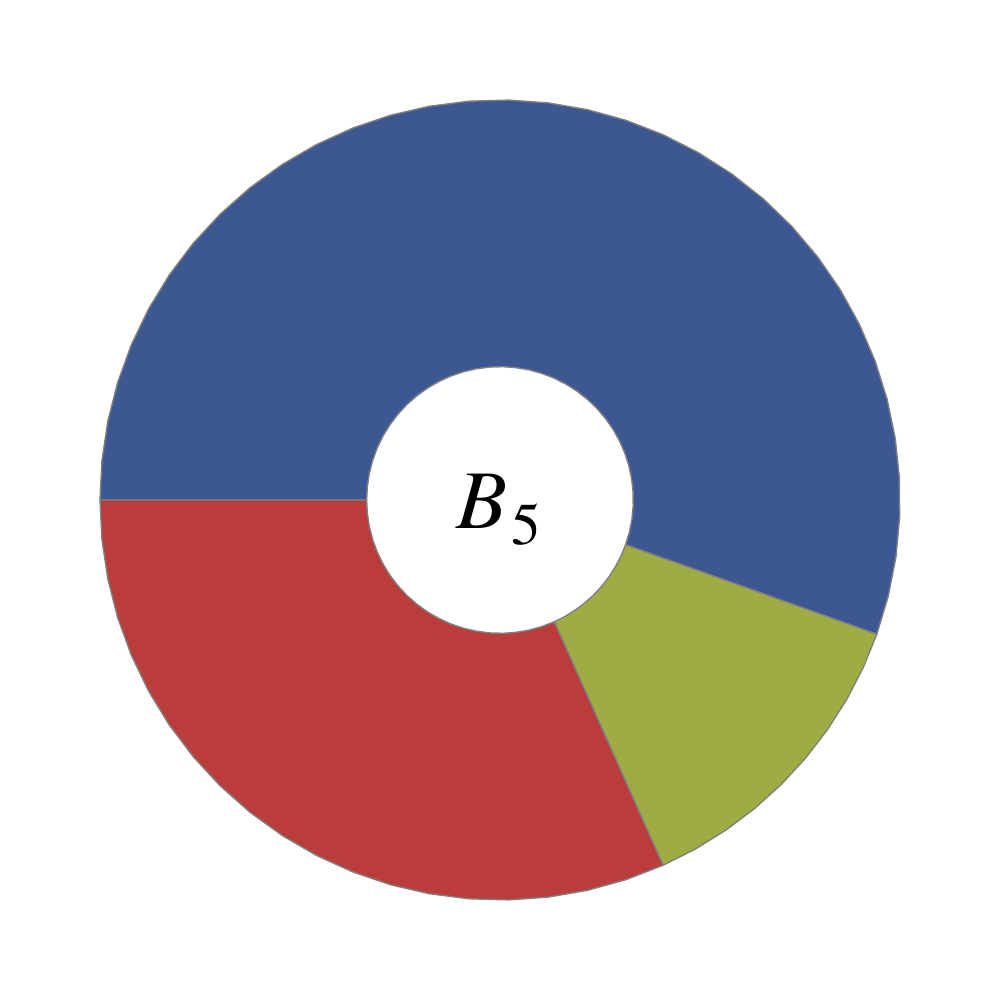}\hspace*{-0.55cm} 
&\includegraphics[bb=0bp -95bp 288bp 170bp,scale=0.2]{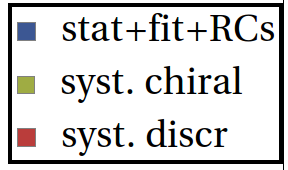}\tabularnewline
\end{tabular}
\caption{\label{Fig:error-budget-K} Graphical representation of the 
error budget owing to the lattice computation ({\it i.e.} without including the estimate for the systematic uncertainty due to 
the perturbative matching between the RI$'$ and $\overline{\rm{MS}}$ schemes) 
for the $K$ bag-parameters, as reported in Table~\ref{tab:error-budget-K}.}
\end{figure}

\begin{figure}[!h]
\begin{tabular}{cccccc}
\hspace*{-0.8cm}\includegraphics[scale=0.28]{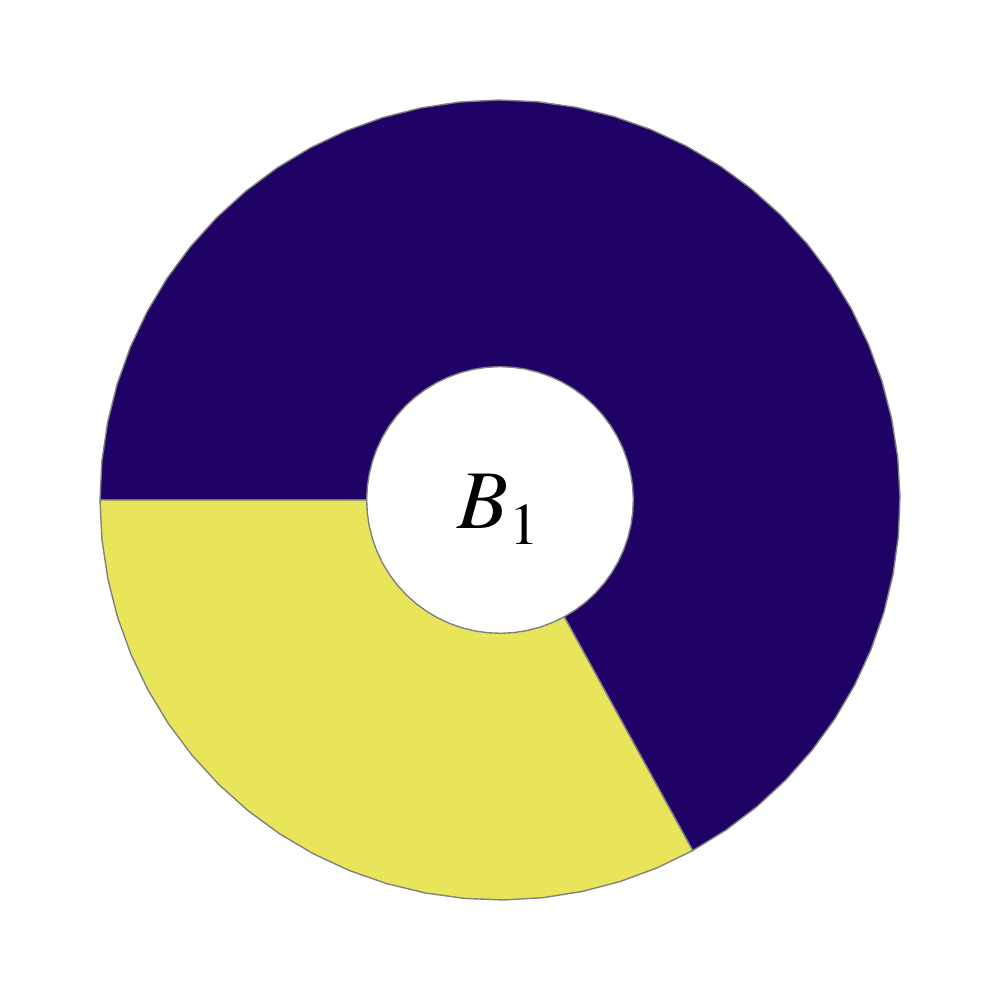}\hspace*{-0.82cm} 
& \includegraphics[scale=0.28]{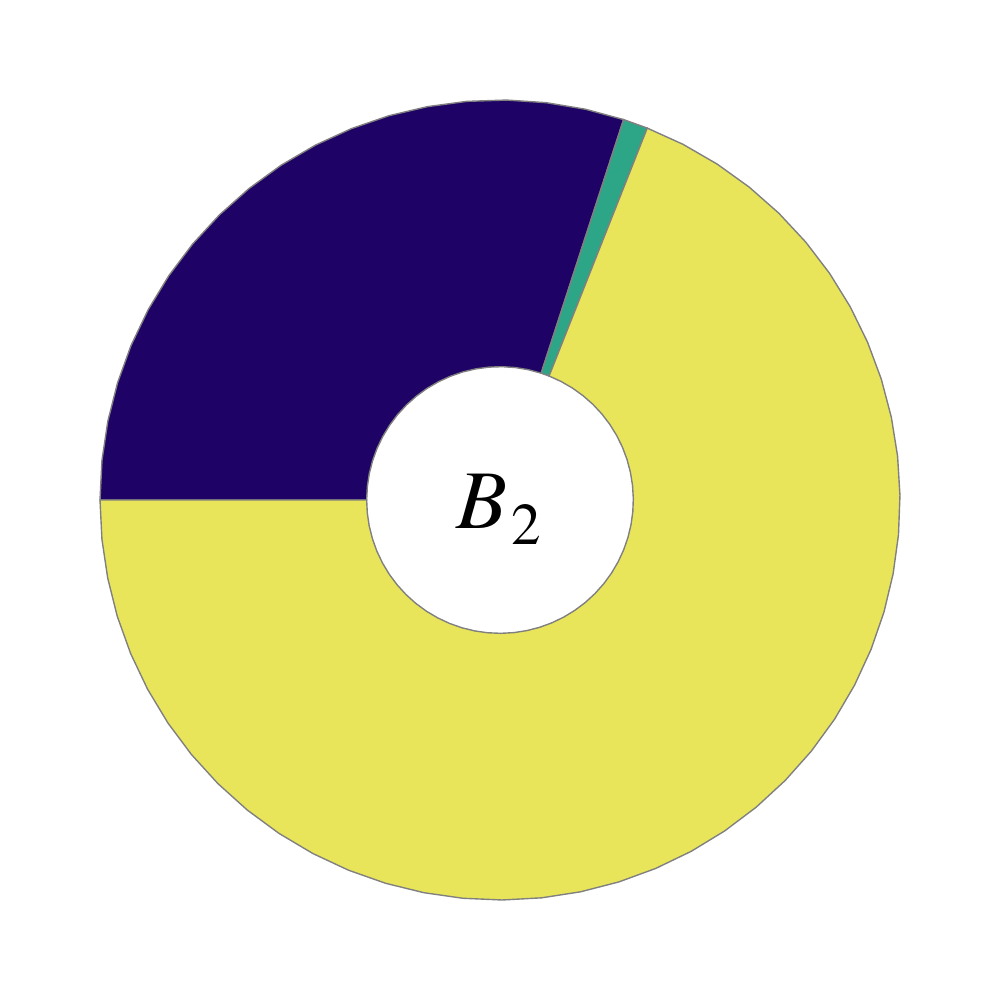}\hspace*{-0.82cm} 
& \includegraphics[scale=0.28]{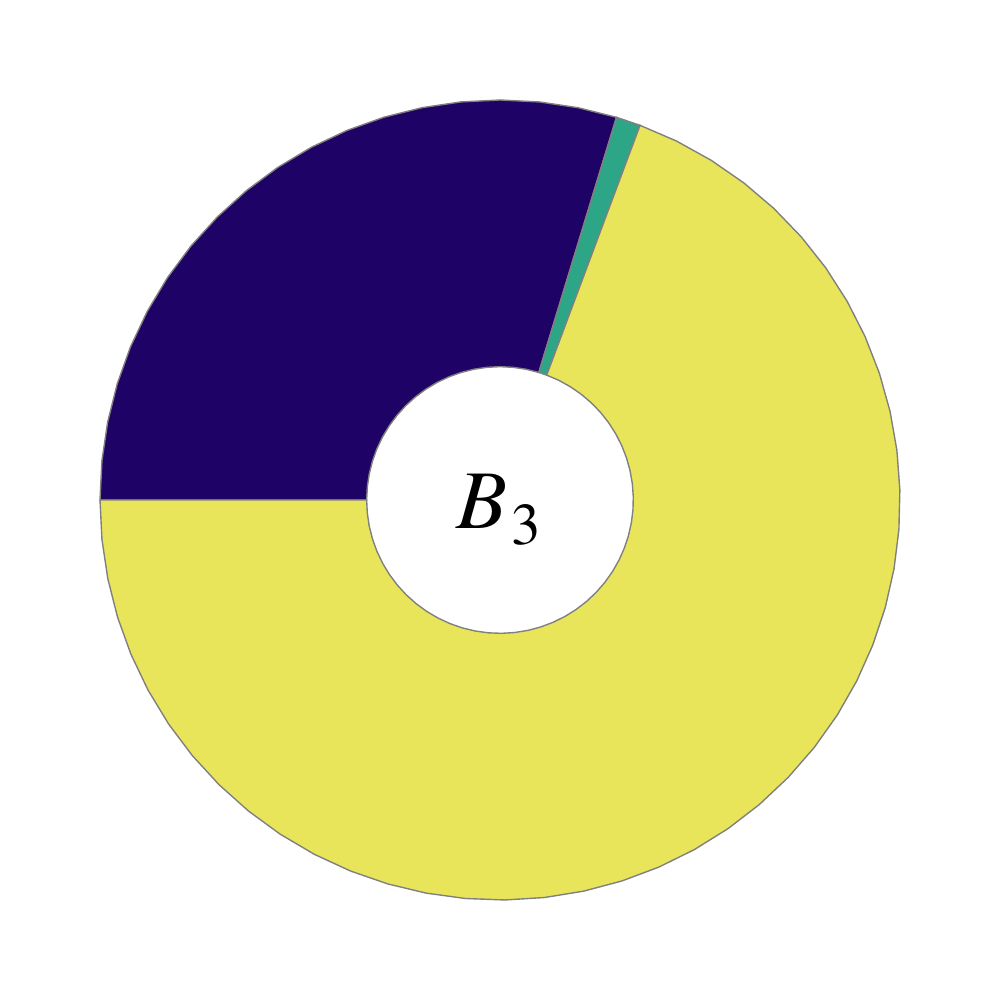}\hspace*{-0.82cm} 
& \includegraphics[scale=0.28]{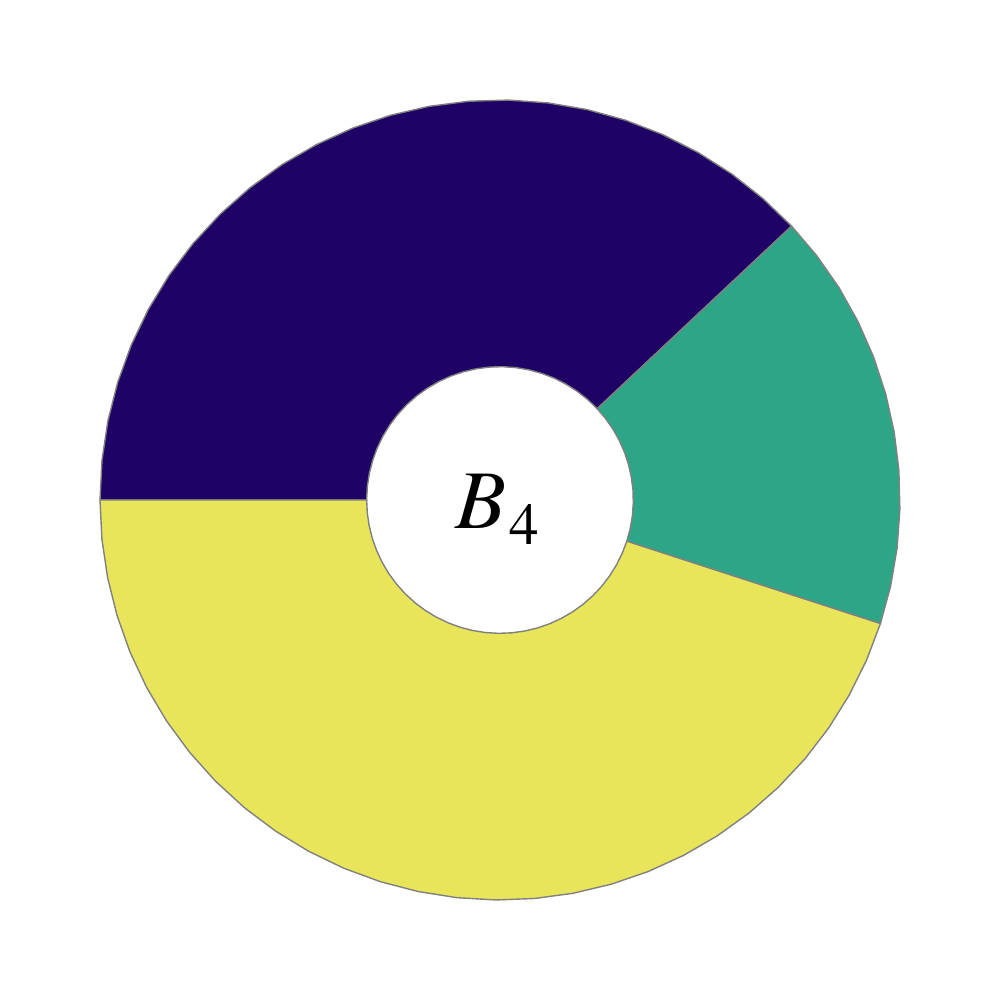}\hspace*{-0.82cm} 
& \includegraphics[scale=0.28]{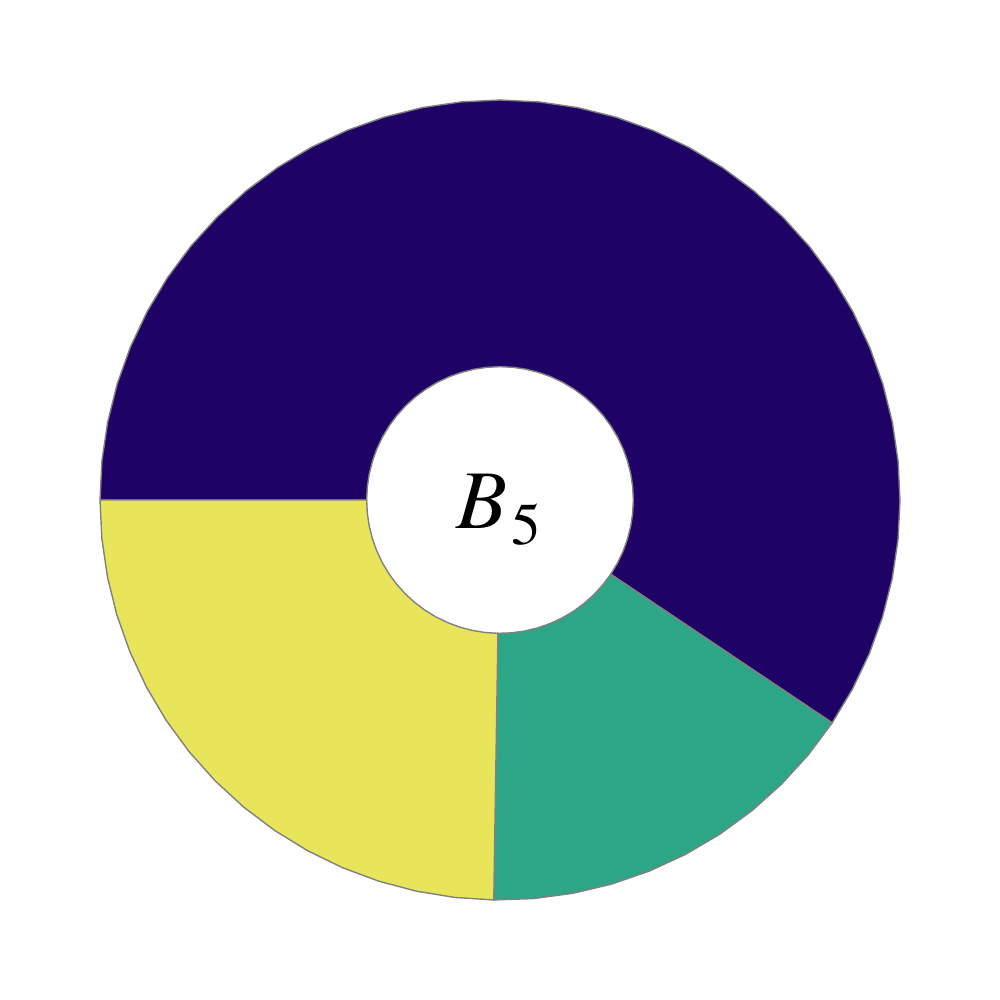}\hspace*{-0.55cm} 
&\includegraphics[bb=0bp -95bp 288bp 170bp,scale=0.2]{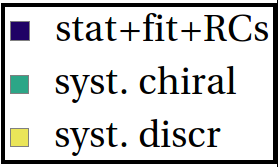}\tabularnewline
\end{tabular}
\caption{\label{Fig:error-budget-D} Same as in Fig~\ref{Fig:error-budget-K} for the $D$ bag-parameters. }
\end{figure}

\vspace*{2.cm}
\noindent {\bf Acknowledgements}\\
We warmly thank our colleagues of the ETM Collaboration for fruitful discussions.
\noindent We acknowledge the CPU time provided by the PRACE Research Infrastructure under the projects
PRA027 ``QCD Simulations for Flavor Physics in the Standard Model and Beyond'' and PRA067 ``First Lattice 
QCD study of B-physics with four flavors of dynamical quarks" on the BG/P and BG/Q systems at the J\"ulich 
SuperComputing Center (Germany) and at CINECA (Italy), and by the agreement between INFN and CINECA 
under the specific initiative INFN-LQCD123 on the Fermi BG/Q system at CINECA (Italy).
V. L., S. S. and C. T. thank MIUR (Italy) for partial support under Contract No. PRIN 2010-2011.

\begin{appendices}
\section{Computational setup for the RCs}
\label{sec:RCs-comput-setup}
\numberwithin{equation}{section}
\setcounter{equation}{0}

As we use a mass-independent renormalization scheme, the calculation of the RCs of two- and four-fermion 
operators and in particular of operators with non-vanishing anomalous dimension must be performed in the  massless quark limit. 

For this purpose we have  produced dedicated sets of $N_f=4$ Wilson twisted-mass degenerate dynamical quark gauge 
configurations with the 
same gluon action as the one used in the non-degenerate case and for a 
number of moderately light 
sea masses. For each ensemble with given sea quark mass parameters we have also computed the RCs estimators 
at several values of the 
valence parameters. Naturally the RCs computed with either $N_f=2+1+1$ or $N_f=4$ ensembles would yield 
identical numbers in the chiral limit.

The $N_f=4$ ensembles are generated at values of the twist angle somewhat different from $\pi/2$ (maximal twist), 
i.e.\ at $m_0 \neq m_{cr}$. The reason is that for small values of $m_0-m_{cr}$ large autocorrelation times have been noticed for 
simulations performed at two out of the three values of the inverse gauge coupling ($\beta=1.90$ and 1.95) we use. Although an off 
maximal-twist setup does not lead to automatic O$(a)$-improvement, one can prove~\cite{FrezzoRoss1} that for any hadronic observable 
the average over results obtained at opposite values of the PCAC quark mass is actually O$(a)$-improved. Naturally, the need of 
performing the average leads to doubling the CPU time-cost of the calculation, which however remains quite affordable as we are 
dealing with simulations at non-zero standard Wilson and twisted mass.

In the Appendix~A of Ref.~\cite{Carrasco:2014cwa} we have presented in detail the $N_f=4$ operator renormalization 
procedure for the cases of quark field and quark bilinears. 
Nevertheless, for the reader's convenience and to fix our notations, we briefly summarise here the main parts of our 
$N_f=4$ computational setup. 
 
We employ the Iwasaki action for the gluons while the $N_f=4$ fermionic action in the so-called twisted basis reads
\be
   S_{tm}^{\rm sea} = a^4 \sum_{x, f}  \bar\chi_f^{\rm sea} \Big{[}\gamma \cdot \tilde\nabla + W_{cr}  +
                                  (m_{0, f}^{\rm sea} - m_{cr}) + ir_f^{\rm sea} \mu_f^{\rm sea} \gamma_5 \Big{]}\chi_f^{\rm sea} ~ , \\
    \label{eq:SEA-LTB}
\ee
where $f = u, d, s, c$, $\gamma \cdot \tilde\nabla = \gamma_\mu (\nabla_\mu + \nabla_\mu^*)/2$ and 
$W_{cr} = -(a/2) \nabla_\mu^*\nabla_\mu + m_{cr}$. We have also set  
\bea
    r_d^{\rm sea} &=& -r_u^{\rm sea},  ~~~ r_c^{\rm sea} = -r_s^{\rm sea} \nn \\ 
    \mu_u^{\rm sea} &=& \mu_d^{\rm sea} = \mu_s^{\rm sea} = \mu_c^{\rm sea} \equiv \mu^{\rm sea}\,  .
    \label{eq:mu-sea}
\eea
Note that the form of the action~(\ref{eq:mu-sea}) guarantees the positivity of the fermion determinant. The valence fermion action takes the form  
\be
    S^{\rm val} = a^4 \sum_{x, f} \bar\chi_f^{\rm val} \Big{[} \gamma \cdot \tilde\nabla -
                          \frac{a}{2} \nabla_\mu^*\nabla_\mu + m_{0, f}^{\rm val} +
                          ir_f^{\rm val} \mu_f^{\rm val} \gamma_5 \Big{]} \chi_f^{\rm val} \, .
    \label{eq:VAL-LTB}
\ee

In our notations the sea and valence sectors the various $r_f^{\rm val,sea}$-Wilson parameters take values equal 
to $\pm 1$ and the twisted masses $a\mu_f^{\rm val,sea}$ are non-negative quantities.  

In Table~\ref{tab:Nf4_simul} we report the simulation details and the quark mass parameters relevant for the $N_f=4$ 
gauge ensembles defined above. For each value of the sea (twisted) 
quark mass, $a\mu^{\rm{sea}}$, we have generated two gauge ensembles which are denoted by the letter 
``m" or ``p" added to their label, 
and correspond to (nearly) opposite values of the PCAC quark mass, $am_{\rm{PCAC}}^{\rm{sea}}$. 

Moreover, for each of the sea gauge ensembles quark propagators have been computed for a number 
of valence quark twisted masses, $a\mu^{\rm{val}}$. 
The measured value of the valence PCAC quark mass, $am_{\rm{PCAC}}^{\rm{val}}$, for each 
sea ensemble of the ``m" or ``p" type is 
given in the last column of Table~\ref{tab:Nf4_simul}.      

Based on Ref.~\cite{Frezzotti:2004wz} the definition of the renormalized quark mass parameters in our partially quenched setup is 
\bea
    {\cal{M}}^{\rm sea, val} & = & Z_P^{-1} {\cal{M}}_0^{\rm sea, val} = Z_P^{-1} \sqrt{ (Z_A\, m_{PCAC}^{\rm sea, val})^2 +
                                                   (\mu^{\rm sea, val})^2 } ~ ,  \nonumber \\
    {\mbox{tg}} ~ \theta_f^{\rm sea, val} & = & \frac{Z_A\, m_{PCAC}^{\rm sea, val}}{r_{f}^{\rm sea, val} \mu^{\rm sea, val}} ~ ,
    \label{eq:SEAM+VALMpar}
\eea
where $Z_A$ is the RC of the (flavor non-singlet) axial current and $m_{PCAC}^{\rm sea, val}$ denotes the PCAC quark mass 
computed from correlators in the sea and valence sector, respectively. The angles $\theta_f^{\rm sea}$ and $\theta_f^{\rm val}$ 
are determined from the formulae ${\cal{M}}^{\rm sea/val} \cos(\theta_f^{\rm sea/val}) = r_f^{\rm sea/val} \mu^{\rm sea/val}$ and 
${\cal{M}}^{\rm sea/val} \sin(\theta_f^{\rm sea/val}) = Z_A m_{PCAC}^{\rm sea/val}$, respectively. If convenient, quark 
mass parameters in the valence sector may be chosen to be different from their sea counterparts. 

\begin{table}[!h]
\begin{center}
\scalebox{0.85}{
\begin{tabular}{|l|c|c|c|c|c|c|}
\hline 
& {\scriptsize $a\mu^{\textrm{sea}}$} & {\scriptsize $am_{\textrm{PCAC}}^{\textrm{sea}}$} & {\scriptsize $am_{0}^{\textrm{sea}}$} 
& {\scriptsize $\theta^{\textrm{sea}}$} & {\scriptsize $a\mu^{\textrm{val}}$} & {\scriptsize $am_{\textrm{PCAC}}^{\textrm{val}}$}
\tabularnewline
\hline 
\hline 
\multicolumn{7}{|c}{{\scriptsize $\beta=1.90$ ($L=24$, $T=48$)}}\tabularnewline
\hline 
{\scriptsize A4m} & {\scriptsize 0.0080} & {\scriptsize -0.0390(01)} & {\scriptsize 0.0285(01)} & {\scriptsize -1.286(01)} & 
{\scriptsize \{0.0060, 0.0080, 0.0120, } & {\scriptsize -0.0142(02)}\tabularnewline
{\scriptsize A4p} &  & {\scriptsize 0.0398(01)} & {\scriptsize 0.0290(01)} & {\scriptsize +1.291(01)} & {\scriptsize 0.0170, 0.0210
,0.0260\}} & {\scriptsize +0.0147(02)}\tabularnewline
\hline 
{\scriptsize A3m} & {\scriptsize 0.0080} & {\scriptsize -0.0358(02)} & {\scriptsize 0.0263(01)} & {\scriptsize -1.262(02)} & 
{\scriptsize \{0.0060, 0.0080, 0.0120, } & {\scriptsize -0.0152(02)}\tabularnewline
{\scriptsize A3p} &  & {\scriptsize 0.0356(02)} & {\scriptsize 0.0262(01)} & {\scriptsize +1.260(02)} & {\scriptsize 0.0170, 0.0210
,0.0260\}} & {\scriptsize +0.0147(03)}\tabularnewline
\hline 
{\scriptsize A2m} & {\scriptsize 0.0080} & {\scriptsize -0.0318(01)} & {\scriptsize 0.0237(01)} & {\scriptsize -1.226(02)} & 
{\scriptsize \{0.0060, 0.0080, 0.0120, } & {\scriptsize -0.0155(02)}\tabularnewline
{\scriptsize A2p} &  & {\scriptsize +0.0310(02)} & {\scriptsize 0.0231(01)} & {\scriptsize +1.218(02)} & {\scriptsize 0.0170, 0.021
0,0.0260\}} & {\scriptsize +0.0154(02)}\tabularnewline
\hline 
{\scriptsize A1m} & {\scriptsize 0.0080} & {\scriptsize -0.0273(02)} & {\scriptsize 0.0207(01)} & {\scriptsize -1.174(03)} & 
{\scriptsize \{0.0060, 0.0080, 0.0120, } & {\scriptsize -0.0163(02)}\tabularnewline
{\scriptsize A1p} &  & {\scriptsize +0.0275(04)} & {\scriptsize 0.0209(01)} & {\scriptsize +1.177(05)} & {\scriptsize 0.0170, 0.021
0,0.0260\}} & {\scriptsize +0.0159(02)}\tabularnewline
\hline 
\multicolumn{7}{|c}{{\scriptsize $\beta=1.95$ ($L=24$, $T=48$)}}\tabularnewline
\hline 
{\scriptsize B1m} & {\scriptsize 0.0085} & {\scriptsize -0.0413(02)} & {\scriptsize 0.0329(01)} & {\scriptsize -1.309(01)} & 
{\scriptsize \{0.0085, 0.0150, 0.0203,} & {\scriptsize -0.0216(02)}\tabularnewline
{\scriptsize B1p} &  & {\scriptsize +0.0425(02)} & {\scriptsize 0.0338(01)} & {\scriptsize +1.317(01)} & {\scriptsize{} 0.0252, 0.02
98\}} & {\scriptsize +0.0195(02)}\tabularnewline
\hline 
{\scriptsize B7m} & {\scriptsize 0.0085} & {\scriptsize -0.0353(01)} & {\scriptsize 0.0285(01)} & {\scriptsize -1.268(01)} & 
{\scriptsize \{0.0085, 0.0150, 0.0203,} & {\scriptsize -0.0180(02)}\tabularnewline
{\scriptsize B7p} &  & {\scriptsize +0.0361(01)} & {\scriptsize 0.0285(01)} & {\scriptsize +1.268(01)} & {\scriptsize{} 0.0252, 0.02
98\}} & {\scriptsize +0.0181(01)}\tabularnewline
\hline 
{\scriptsize B8m} & {\scriptsize 0.0020} & {\scriptsize -0.0363(01)} & {\scriptsize 0.0280(01)} & {\scriptsize -1.499(01)} & 
{\scriptsize \{0.0085, 0.0150, 0.0203,} & {\scriptsize -0.0194(01)}\tabularnewline
{\scriptsize B8p} &  & {\scriptsize +0.0363(01)} & {\scriptsize 0.0274(01)} & {\scriptsize +1.498(01)} & {\scriptsize{} 0.0252, 0.02
98\}} & {\scriptsize +0.0183(02)}\tabularnewline
\hline 
{\scriptsize B3m} & {\scriptsize 0.0180} & {\scriptsize -0.0160(02)} & {\scriptsize 0.0218(01)} & {\scriptsize -0.601(06)} & 
{\scriptsize \{0.0060,0.0085,0.0120,0.0150,} & {\scriptsize -0.0160(02)}\tabularnewline
{\scriptsize B3p} &  & {\scriptsize +0.0163(02)} & {\scriptsize 0.0219(01)} & {\scriptsize +0.610(06)} & {\scriptsize 0.0180,0.0203,
0.0252,0.0298\}} & {\scriptsize +0.0162(02)}\tabularnewline
\hline 
{\scriptsize B2m} & {\scriptsize 0.0085} & {\scriptsize -0.0209(02)} & {\scriptsize 0.0182(01)} & {\scriptsize -1.085(03)} & 
{\scriptsize \{0.0085, 0.0150, 0.0203,} & {\scriptsize -0.0213(02)}\tabularnewline
{\scriptsize B2p} &  & {\scriptsize +0.0191(02)} & {\scriptsize 0.0170(02)} & {\scriptsize +1.046(06)} & {\scriptsize{} 0.0252, 0.02
98\}} & {\scriptsize +0.0191(02)}\tabularnewline
\hline 
{\scriptsize B4m} & {\scriptsize 0.0085} & {\scriptsize -0.0146(02)} & {\scriptsize 0.0141(01)} & {\scriptsize -0.923(04)} & 
{\scriptsize \{0.0060,0.0085,0.0120,0.0150,} & {\scriptsize -0.0146(02)}\tabularnewline
{\scriptsize B4p} &  & {\scriptsize +0.0151(02)} & {\scriptsize 0.0144(01)} & {\scriptsize +0.940(07)} & {\scriptsize 0.0180,0.0203,
0.0252,0.0298\}} & {\scriptsize +0.0151(02)}\tabularnewline
\hline 
\multicolumn{7}{|c}{{\scriptsize $\beta=2.10$ ($L=32$, $T=64$)}}\tabularnewline
\hline 
{\scriptsize C5m} & {\scriptsize 0.0078} & {\scriptsize -0.00821(11)} & {\scriptsize 0.0102(01)} & {\scriptsize -0.700(07)} & 
{\scriptsize \{0.0048,0.0078,0.0119,} & {\scriptsize -0.0082(01)}\tabularnewline
{\scriptsize C5p} &  & {\scriptsize +0.00823(08)} & {\scriptsize 0.0102(01)} & {\scriptsize +0.701(05)} & {\scriptsize 0.0190,0.0242
,0.0293\}} & {\scriptsize +0.0082(01)}\tabularnewline
\hline 
{\scriptsize C4m} & {\scriptsize 0.0064} & {\scriptsize -0.00682(13)} & {\scriptsize 0.0084(01)} & {\scriptsize -0.706(09)} & 
{\scriptsize \{0.0039,0.0078,0.0119,} & {\scriptsize -0.0068(01)}\tabularnewline
{\scriptsize C4p} &  & {\scriptsize +0.00685(12)} & {\scriptsize 0.0084(01)} & {\scriptsize +0.708(09)} & {\scriptsize 0.0190,0.0242
,0.0293\}} & {\scriptsize +0.0069(01)}\tabularnewline
\hline 
{\scriptsize C3m} & {\scriptsize 0.0046} & {\scriptsize -0.00585(08)} & {\scriptsize 0.0066(01)} & {\scriptsize -0.794(07)} & 
{\scriptsize \{0.0025,0.0046,0.0090,0.0152,} & {\scriptsize -0.0059(01)}\tabularnewline
{\scriptsize C3p} &  & {\scriptsize +0.00559(14)} & {\scriptsize 0.0064(01)} & {\scriptsize +0.771(13)} & {\scriptsize 0.0201,0.0249
,0.0297\}} & {\scriptsize +0.0056(01)}\tabularnewline
\hline 
{\scriptsize C2m} & {\scriptsize 0.0030} & {\scriptsize -0.00403(14)} & {\scriptsize 0.0044(01)} & {\scriptsize -0.821(17)} & 
{\scriptsize \{0.0013,0.0030,0.0080,0.0143,} & {\scriptsize -0.0040(01)}\tabularnewline
{\scriptsize C2p} &  & {\scriptsize +0.00421(13)} & {\scriptsize 0.0045(01)} & {\scriptsize +0.843(15)} & {\scriptsize 0.0195,0.0247
,0.0298\}} & {\scriptsize +0.0042(01)}\tabularnewline
\hline 
\end{tabular}
}
\caption{Simulation details and quark mass parameters of the $N_f=4$ gauge ensembles employed in the RCs computation.}
\label{tab:Nf4_simul}
\end{center}
\end{table}

Our RC-estimators are evaluated at the values $p_{\mu}=\left(2\pi/L_{\mu}\right)n_{\mu}$ of the momenta, where 
\bea
    n_{\mu} & = & \left(\left[0,2\right],\left[0,2\right],\left[0,2\right],\left[0,3\right]\right) \nonumber \\
                  &    & \left(\left[2,3\right],\left[2,3\right],\left[2,3\right],\left[4,7\right]\right) , ~~~ 
                  \mbox{for }\beta=1.95,  \nonumber \\[2mm] 
    n_{\mu} & = & \left(\left[0,2\right],\left[0,2\right],\left[0,2\right],\left[0,3\right]\right) \nonumber \\
                  &    & \left(\left[2,5\right],\left[2,5\right],\left[2,5\right],\left[4,9\right]\right) , ~~~ 
                  \mbox{for }\beta=1.90 ~ \mbox{~ and ~} 2.10
\label{eq:set_momenta}                  
\eea
with $L_{\mu}$ the lattice size along the direction $\mu$ (with $L_4 \equiv T$ and $L_{1,2,3} \equiv L$). 
Quark fields obey anti-periodic time boundary conditions implemented by a constant shift,  
$\Delta p_4 = \pi / L_4$, in the time component of the four-momentum. 
Notice also that our final analysis of the RCs estimators has been carried out 
using ``democratic" four-momentum values that satisfy the condition 
\be
    \Delta_4(p) \equiv \frac{\sum_{\mu} \tilde p_{\mu}^4}{(\sum_{\mu} \tilde p_{\mu}^2)^2} < 0.29, 
\label{eq:democr-moment}
\ee
with
\be
     \tilde p_{\mu} \equiv \frac{1}{a} \sin (ap_{\mu}) ~ .
\label{eq:latmoment}
\ee
In order to come up with smoother discretisation errors we have subtracted from the Green's functions entering the RI-MOM 
computation the perturbative cutoff effects up to order O$(a^2 g^2)$ 
(see Refs~\cite{Constantinou:2009tr, Constantinou:2010zs}).

As it has been stated above, following the general proof given in Ref.~\cite{FrezzoRoss1}, which 
in the Appendix A.2 of Ref.~\cite{Carrasco:2014cwa} 
has been exemplified for the case of quark bilinear RCs 
estimators, the average over RCs 
estimators computed on ensembles produced with opposite values of the sea and valence PCAC quark mass enables to remove all the odd 
integer power cutoff effects and hence the ${\cal O}(a)$ discretisation errors~\footnote{The proof for the case of the RCs of the four-fermion operators 
required in our mixed action setup (see Section~\ref{subsec:bag-parameters}) is closely analogous to the one for the case of 
quark bilinear RCs.}. Based on the definition of the angle $\theta$ in 
terms of the PCAC quark mass given in Eq.~(\ref{eq:SEAM+VALMpar}), 
we generally refer to this procedure as $\theta$-average O$(a)$-improvement.

The evaluation of the RCs for the quark bilinear operators, namely $Z_A$, $Z_V$, $Z_P$ and $Z_S$ as well as the RC of the quark wave 
function, $Z_q$, has been done in Ref.~\cite{Carrasco:2014cwa}. Our final numbers at each value of $\beta$ have been labeled 
as M1 or M2 RCs. 
As explained in detail in~\cite{Carrasco:2014cwa}, they correspond to different ways in which the cutoff effects are treated. 
For convenience
 all the results are again reported in the Table~\ref{tab:RCs-bilinear} of Appendix~\ref{app:RCs4f} of the present work. 
Note that as for $Z_V$, in the present analysis we have made use of the much more precise Ward Takahashi Identity 
determination~\cite{Constantinou:2010gr}.

\section{RI-MOM computation of RCs of the four-fermion operators}
\label{app:RCs4f}
\numberwithin{equation}{section}
\setcounter{equation}{0}

The RI$'$-MOM renormalization procedure we used for the four-fermion operators has been explained in the Appendices A 
and B of Ref.~\cite{Bertone:2012cu}. From the conceptual and operational point of view a great part of the computational 
details are very similar between the $N_f=2$ case of Ref.~\cite{Bertone:2012cu} and the present $N_f=4$ case, except for the fact 
that in the latter one has to compute RCs estimators in the ``m" and ``p" ensembles separately before taking their $\theta$-average. 
In this section we will fix our notations making extensive use of the description of Ref.~\cite{Bertone:2012cu}. We will 
however recall some essential points of the computation in order to make easier for the reader to follow the presentation 
of the analysis and our results. 

In computing RCs it is convenient to work in a basis where the operators $O^{MA}_{i[\pm]}$ with $i=2, \ldots, 5$ 
defined in Eq.~(\ref{OMAPM_v2}), 
are Fierz transformed as suggested in Ref.~\cite{Donini:1999sf}. 
Using here a generic labeling (that can be obviously adapted to the case of interest $(1, 2, 3, 4)\rightarrow (h, \ell, h', \ell')$) 
the operator basis now reads\footnote{The quark fields $q_1, q_2, q_3$ and $q_4$ are valence fields with the lattice action specified in Eq.~(\ref{eq:OS}) - {\it i.e.}
they are written in the physical basis of maximally twisted LQCD.}
\begin{eqnarray}
&&Q^{MA}_{1[\pm]}= 2\big{\{}\big{(}[\bar q_1\gamma_\mu q_2][\bar q_3\gamma_\mu q_4]+
[\bar q_1\gamma_\mu \gamma_5 q_2][\bar q_3\gamma_\mu \gamma_5 q_4]\big{)}\pm \big{(}2\leftrightarrow 4\big{)}\big{\}}\nonumber \\
&&Q^{MA}_{2[\pm]}=2\big{\{}\big{(}[\bar q_1\gamma_\mu q_2][\bar q_3\gamma_\mu q_4]-
[\bar q_1\gamma_\mu \gamma_5 q_2][\bar q_3\gamma_\mu \gamma_5 q_4]\big{)}\pm \big{(}2\leftrightarrow 4\big{)}\big{\}}\nonumber \\
&&Q^{MA}_{3[\pm]}= 2\big{\{}\big{(}[\bar q_1 q_2][\bar q_3 q_4]-[\bar q_1\gamma_5 q_2][\bar q_3\gamma_5 q_4]\big{)}\pm \big{(}2\leftrightarrow 4\big{)}\big{\}}\nonumber \\
&&Q^{MA}_{4[\pm]}= 2\big{\{}\big{(}[\bar q_1 q_2][\bar q_3 q_4]+[\bar q_1\gamma_5 q_2][\bar q_3\gamma_5 q_4]\big{)}\pm \big{(}2\leftrightarrow 4\big{)}\big{\}}\nonumber \\
&&Q^{MA}_{5[\pm]}=2\big{\{}\big{(}[\bar q_1\sigma_{\mu\nu}q_2][\bar q_3\sigma_{\mu\nu}q_4]\big{)}\pm \big{(}2\leftrightarrow 4\big{)}\big{\}}\,\,\, (\rm{for}\, \mu > \nu) ,
\label{QMAPM}
\end{eqnarray}
where color indices are meant to be contracted within each square parenthesis, 
``MA" stands for ``Mixed Action" and $\sigma_{\mu\nu} = [\gamma_\mu,\gamma_\nu]/2$. 
The transformation matrix between the two operator bases (Eqs~(\ref{OMAPM_v2}) and (\ref{QMAPM})) is given by
\begin{equation}
O^{MA}_{i[\pm]}=\Lambda_{ij}^{[\pm]}Q^{MA}_{j[\pm]}\, ,\qquad\quad\Lambda^{[\pm]} = 
\left(\begin{array}{ccccc}
1 & 0 & 0 & 0 & 0 \\
0 & 0 & 0 & 1 & 0 \\
0 & 0 & 0 & \mp 1/2 & \pm 1/2 \\
0 & 0 & 1 & 0 & 0 \\
0 & \mp 1/2 & 0 & 0 & 0 \\
\end{array} \right)
\label{REL}
\end{equation}

To simplify the notation, in the rest of the Appendix we drop the superscript ``MA"
and subscript ``$\pm$" and denote the operators~(\ref{QMAPM}) simply with the symbol $Q_{i}$. Then in a self-evident matrix 
notation the renormalization pattern of the bare operators $\mathbf{Q}^{(b)}$ takes the form
\be
\mathbf{Q}^{\rm ren}\, =\, \mathbf{Z_Q}\, \left[\, \mathbf{I}\, +\, \pmb{\Delta}\, \right]\, \mathbf{Q}^{(b)}
\label{Qrenpatt}
\ee
where $\mathbf{Z_Q}$ is a scale-dependent renormalization matrix which has the same block-diagonal form as the formal 
continuum one. The wrong chirality mixings are parametrized by $\pmb{\Delta}$ which is a sparse off-diagonal and UV-finite matrix  with the structure  
\be
\pmb{\Delta}\, =\, \left[\, \begin{array}{ccccc}
0&\Delta_{12}&\Delta_{13}&\Delta_{14}&\Delta_{15}\\
\Delta_{21}&0&0&\Delta_{24}&\Delta_{25}\\
\Delta_{31}&0&0&\Delta_{34}&\Delta_{35}\\
\Delta_{41}&\Delta_{42}&\Delta_{43}&0&0\\
\Delta_{51}&\Delta_{52}&\Delta_{53}&0&0\\
\end{array} \right]  \; .
\ee
The renormalization pattern (\ref{Qrenpatt}) can be proved (following Appendix A of Ref.~\cite{Bertone:2012cu}) 
for the renormalization of $\mathbf{Q}$ defined out of maximal twist, for 
both positive and negative $m_{{\rm PCAC}}$ masses - we refer to the ensembles ``p" and ``m" discussed in the previous Appendix. At this level of course lattice 
artifacts are still O$(a)$. By the $\theta$-averaging procedure discussed above, however, we obtain O$(a)$ improved RC-estimators for which Eq.~(\ref{Qrenpatt}) 
holds with only O$(a^2)$ lattice artifacts. These RC-estimators are used for renormalizing the bare matrix element we computed at maximal twist.   
 
For completeness we summarise the main technical points of the RC calculations.
We start by computing in the Landau gauge the four-point Green's function of the 
operators, ${Q_i}$ between external quark states with the momenta given in Eqs~(\ref{eq:set_momenta})-(\ref{eq:latmoment}). 

The RI$'$-MOM renormalization condition is imposed by requiring the projected amputated functions be equal to their 
tree-level value. This last step is conveniently and compactly implemented 
with the construction of the so-called dynamic matrix defined by the equation ${\mathbf D} = \mathbf P \mathbf \Lambda$, where 
$\mathbf P$ and $\mathbf \Lambda$ are the matrices of the spin projectors and the amputated Green's functions, respectively;
see also Appendix B.1 of Ref.~\cite{Bertone:2012cu}.    

The further steps of the analysis are the following.
\renewcommand\labelitemi{$\textendash$}
\begin{itemize}[noitemsep, leftmargin=*]
\itemsep0.3em 
\item  We subtract from the dynamic matrix and the quark form factor (relevant for $Z_q$) 
the perturbative O$(a^2 g^2_{\rm{boost}})$ corrections computed in the massless 
theory~\cite{Constantinou:2009tr, Constantinou:2010zs}. 
The boosted coupling is defined as $g^2_{\rm{boost}} = 6/(\beta \langle P \rangle)$ 
where $\langle P \rangle$ is the average plaquette value.

\item We subtract from the projected amputated four-fermion correlators 
the contribution of the Goldstone boson pole (GB-pole) at each value  
of the  momenta defined in Eqs~(\ref{eq:set_momenta})-(\ref{eq:latmoment}). 
This step is performed by carrying out the chiral limit extrapolation in the valence sector.  
The GB-pole contribution is realised by the presence of terms that go like\footnote{For each combination 
of two valence quark masses in each gauge ensemble of Table~\ref{tab:Nf4_simul} we have computed and used the corresponding 
value of the pseudoscalar mass, $m_{ps}$.} 
 $(1/m_{ps}^2)^n$, being each one of them suppressed by corresponding powers of the inverse square momentum, $(1/\tilde{p}^2)^n$. 
 For each sea quark ensemble and for each value of the momentum we fit each one of the dynamic matrix elements 
 in terms of the pseudoscalar mass 
 in the valence sector, using the ansatz 
\begin{equation}
D_{ij}(a^2 \tilde p^2; m_{ps}^2) = {\cal A}(a^2 \tilde p^2) + {\cal B}(a^2 \tilde p^2) m_{ps}^2 +
{\cal C}(a^2 \tilde p^2) / m_{ps}^2 + {\cal D}(a^2 \tilde p^2) / (m_{ps}^2)^2
\label{eq:GBP}\end{equation}

In Fig.~\ref{fig:GBP-b190-p2small} we show the GB-pole subtraction of the block-diagonal elements  
of the dynamic matrix corresponding to the continuum-like matrix elements 
in the case of the coarsest lattice ($\beta=1.90$) at relatively small momenta where systematic effects are expected to be larger. 
We observe, as it is also expected, that the GB-pole contribution is getting suppressed as the value of the momentum increases. 
This feature can be noticed comparing Figs~\ref{fig:GBP-b190-p2small} and~\ref{fig:GBP-b190-p2large} both referring to $\beta=1.90$ 
at two different values of the momentum.
Moreover typical GB-pole subtraction fits in the finest lattice are given in the panels of Fig.~\ref{fig:GBP-b210}. 

Important information concerning the double GB-pole subtraction can be revealed from plots as for example those presented in the panels  
of Fig.~\ref{fig:DGP} which refer to the ensembles A1p and A1m of the coarsest lattice spacing ($\beta=1.90$). 
We form the product of $(a^2 \tilde p^2)^2$ with 
the fit parameter, ${\cal D}(a^2 \tilde p^2)$, of the double GB-pole term ({\it cf.} Eq.~(\ref{eq:GBP})),  
then  plotting it against  $(a^2 \tilde p^2)$. Two observations are in order. 
First, we notice that the product ${\cal D}(a^2 \tilde p^2) \times (a^2 \tilde p^2)^2 $ 
takes almost constant value for large enough values of momentum 
that lie in the momentum intervals we have used in order to extract our RCs estimates. 
This finding also serves as a confirmation of the good fit quality in analysing the double GB-pole term. 
Second, we find that  the double GB-pole contribution is negligible for the case 
of $D_{11}$, in nice agreement with theoretical expectations, 
see Ref.~\cite{Constantinou:2010qv}, whereas it is 
different from zero for several of the $D_{ij}$ with $i,j > 1$. 
So  in our final analysis we have adopted single GB-pole subtraction ({\it i.e.} set ${\cal D}=0$ 
in the fit ansatz of Eq.~(\ref{eq:GBP})) for the former and 
single and double GB-pole subtraction ({\it cf.} Eq.~(\ref{eq:GBP})) for the latter cases.

\begin{figure}[!h]
\begin{center}
\vspace*{0.5cm}
\begin{tabular}{cc}
\hspace*{0.1cm}\vspace*{0.2cm}\includegraphics[bb=171bp 497bp 496bp 705bp,width=0.29\linewidth , 
keepaspectratio=true]{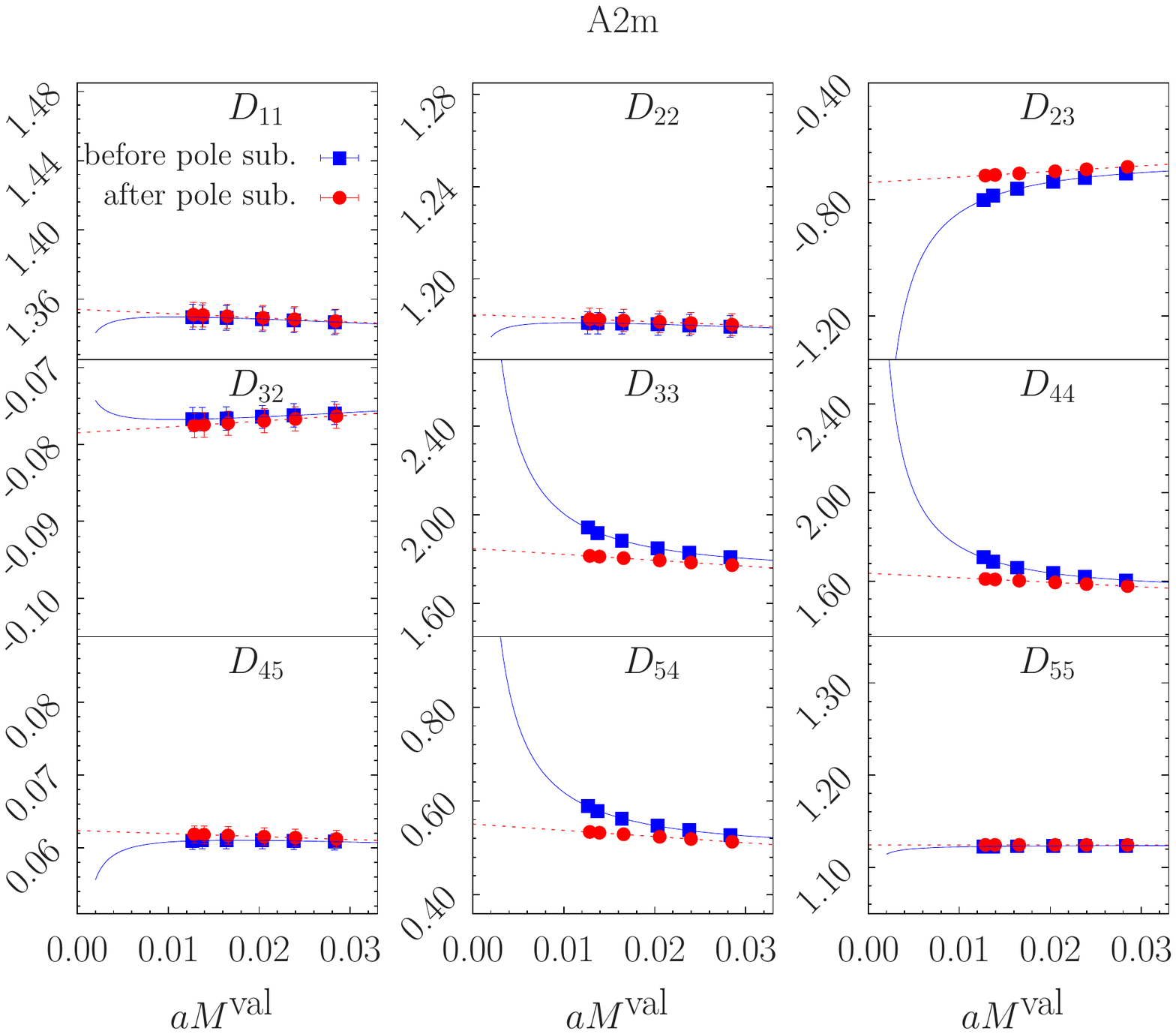} & 
\hspace*{3.cm}\vspace*{0.2cm}\includegraphics[bb=171bp 497bp 496bp 705bp,width=0.29\linewidth ,
keepaspectratio=true]{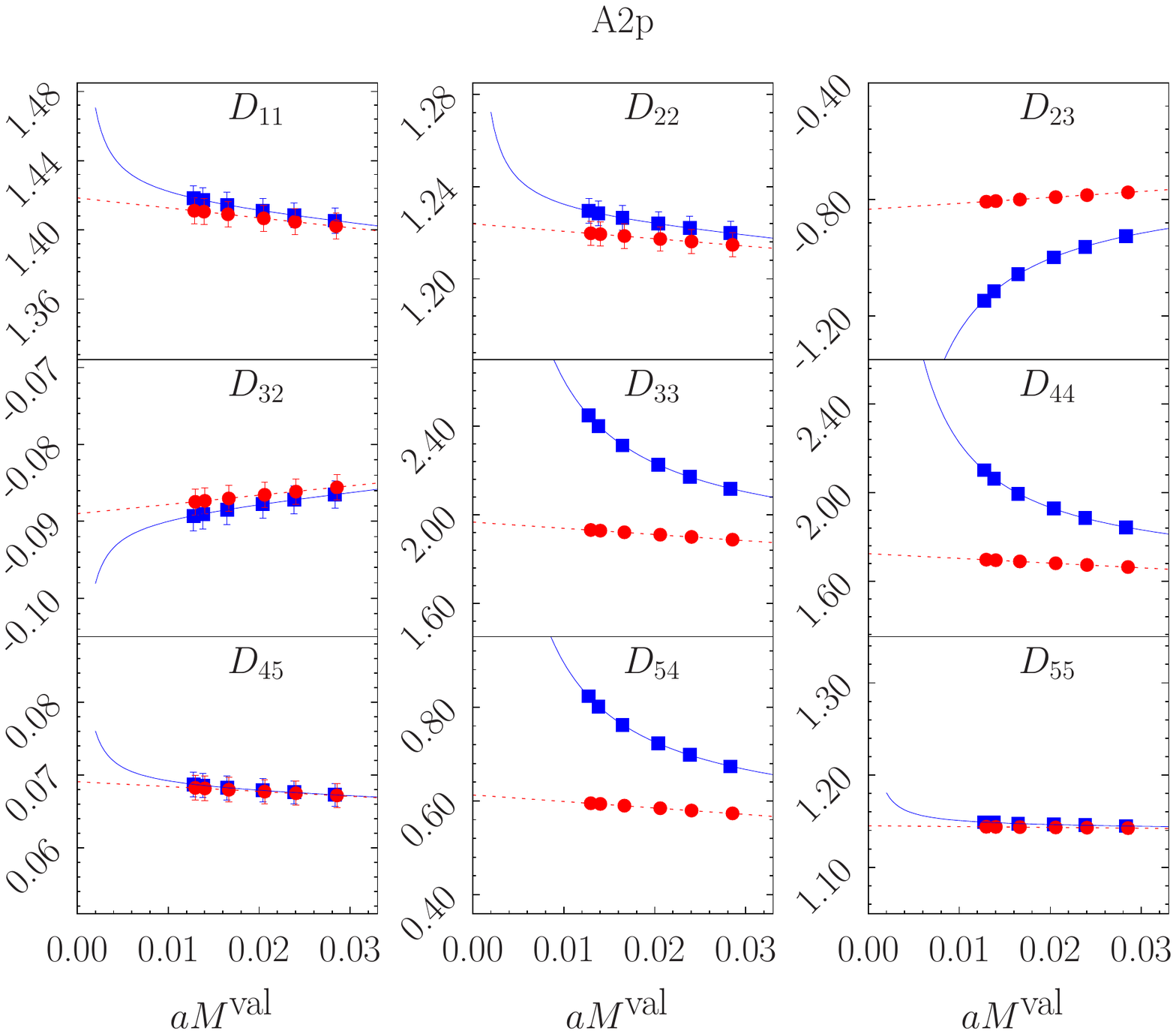} \tabularnewline

\end{tabular}
\end{center}
\vspace*{1.5cm}
\caption{\label{fig:GBP-b190-p2small}
Fitting procedure for the GB-pole subtraction  on the block diagonal elements of the dynamic matrix. 
We show two examples from the coarsest lattice, in particular for the ensemble A2m (left) and A2p (right), at a relatively small value of momentum, namely 
$(a\tilde{p})^2 \simeq 1.57$.}
\end{figure}

\begin{figure}[!h]
\begin{center}
\vspace*{0.5cm}
\begin{tabular}{cc}
\hspace*{0.1cm}\vspace*{0.2cm}\includegraphics[bb=171bp 497bp 496bp 705bp,width=0.29\linewidth , 
keepaspectratio=true]{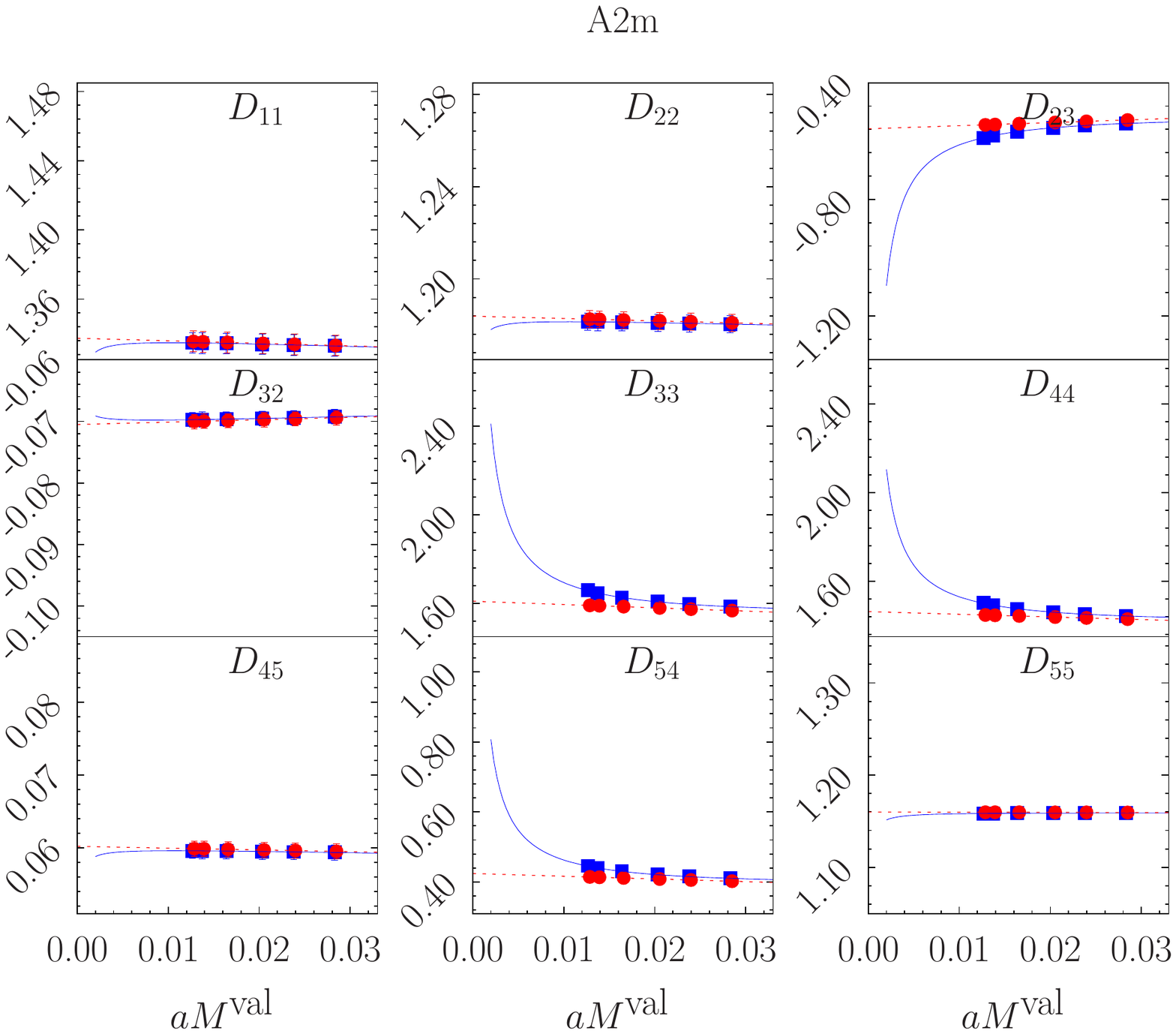} & 
\hspace*{3.cm}\vspace*{0.2cm}\includegraphics[bb=171bp 497bp 496bp 705bp,width=0.29\linewidth ,
keepaspectratio=true]{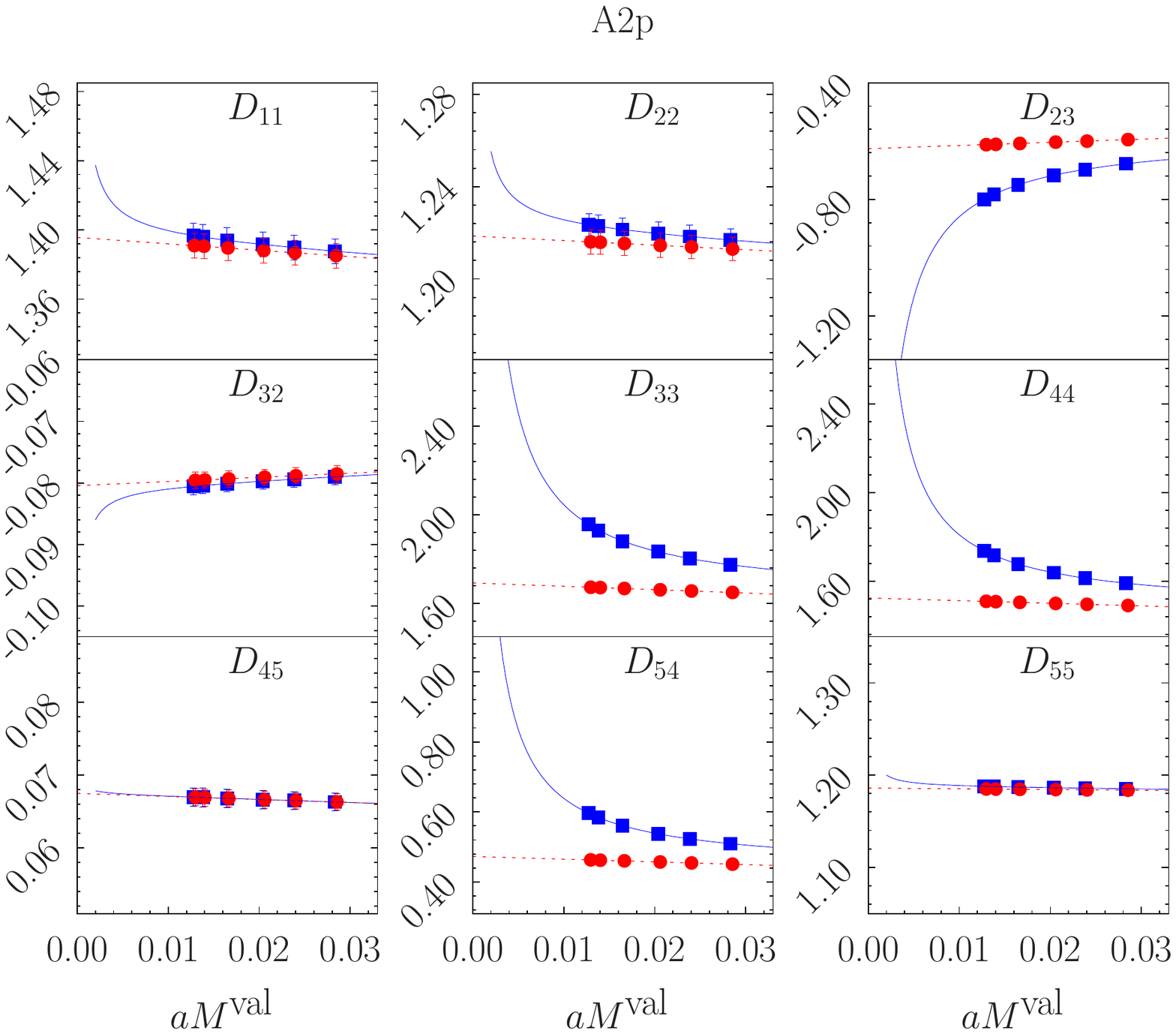} \tabularnewline

\end{tabular}
\end{center}
\vspace*{1.5cm}
\caption{ \label{fig:GBP-b190-p2large}
 Same as in Fig.~\ref{fig:GBP-b190-p2small} but at a larger value of momentum, namely $(a\tilde{p})^2 \simeq 2.19$.  }
\end{figure}  

\begin{figure}[!h]
\begin{center}
\vspace*{0.5cm}
\begin{tabular}{cc}
\hspace*{0.1cm}\vspace*{0.2cm}\includegraphics[bb=171bp 497bp 496bp 705bp,width=0.29\linewidth , 
keepaspectratio=true]{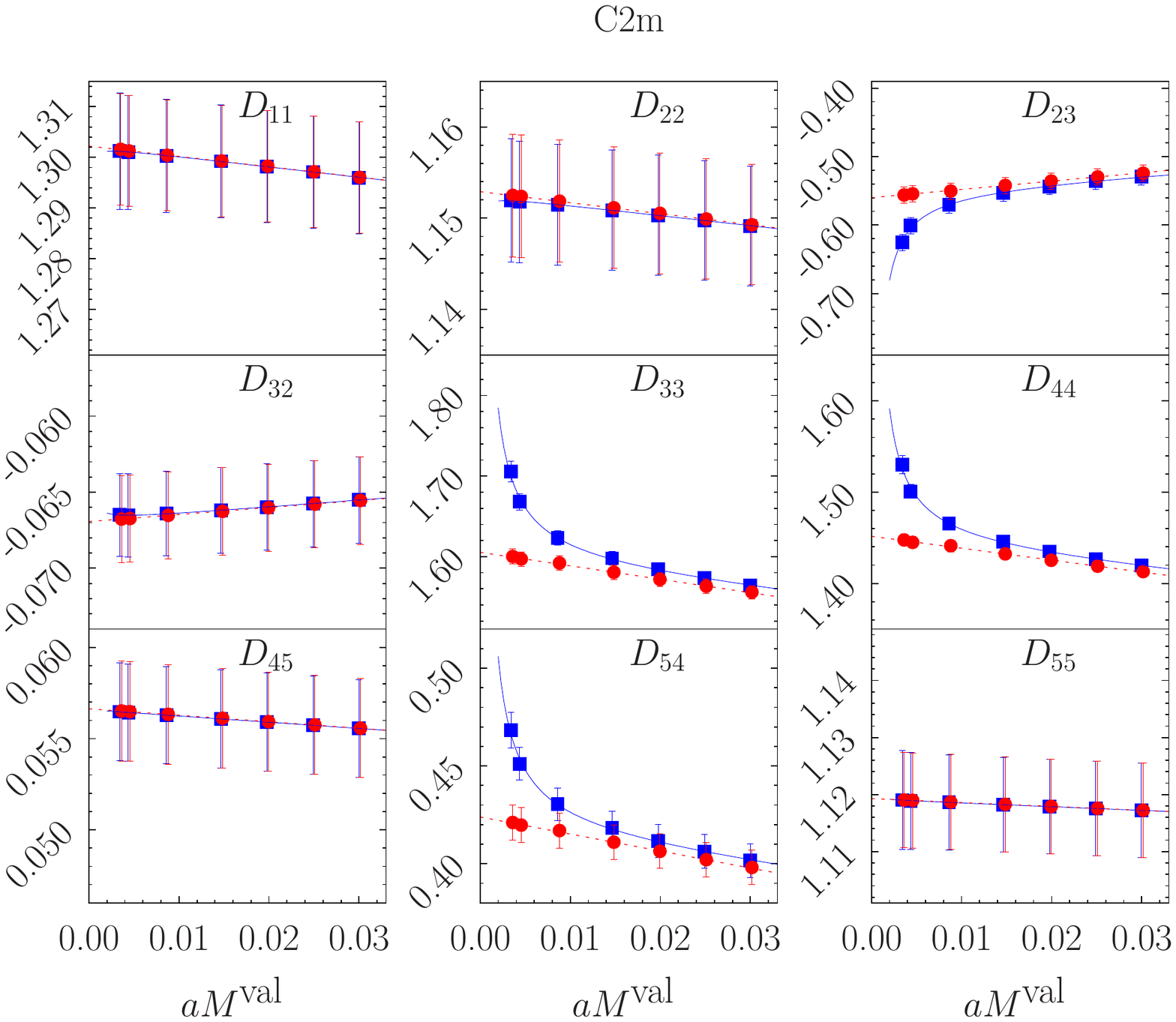} & 
\hspace*{3.cm}\vspace*{0.2cm}\includegraphics[bb=171bp 497bp 496bp 705bp,width=0.29\linewidth ,
keepaspectratio=true]{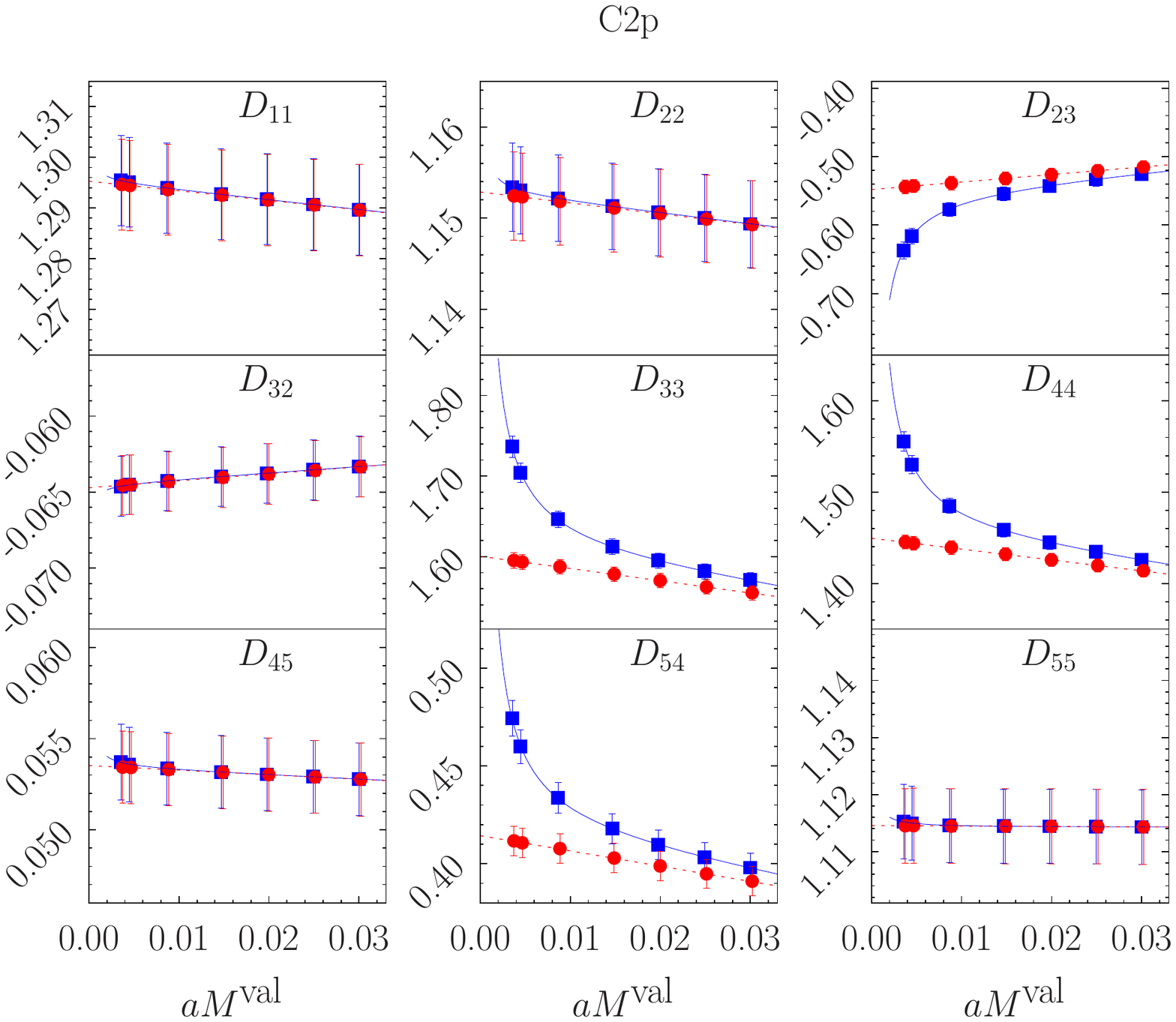} \tabularnewline

\end{tabular}
\end{center}
\vspace*{1.5cm}
\caption{\label{fig:GBP-b210}  
Same as in Fig.\ref{fig:GBP-b190-p2small} at $\beta=2.10$, for the ensemble  C2m (left) and C2p (right) and $(ap)^2 \simeq 1.57$.}
\end{figure}

\begin{figure}[!h]
\hspace*{-0.5cm}\includegraphics[scale=0.70]{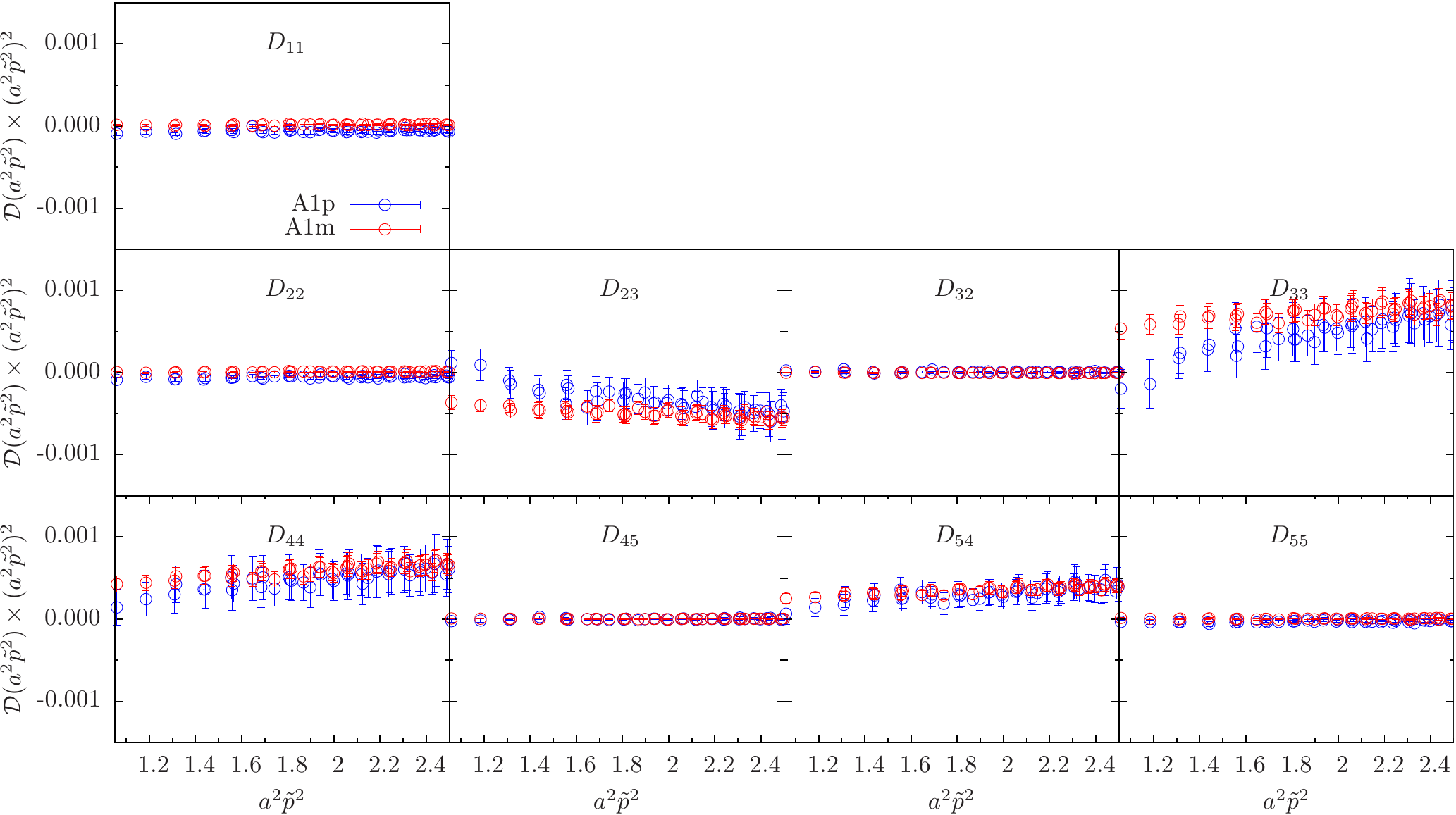} 
\caption{\label{fig:DGP} For each of block-diagonal matrix element we form the product of $(a^2 \tilde p^2)^2$  with the fit parameter 
${\cal D}(a^2 \tilde p^2)$ that is associated to the term of the double GB-pole ({\it cf.} Eq.~(\ref{eq:GBP})) and we plot it 
against  $(a^2 \tilde p^2)$ for the ensembles A1p and A1m of $\beta=1.90$. 
  }
\end{figure}

\item After applying the  $\theta$-average of the RC-estimators at each $\beta$ and at each value of momentum defined by 
Eqs~(\ref{eq:democr-moment}) and (\ref{eq:latmoment}), which is required in order to achieve O$(a)$-improvement,  
we carry out the sea chiral limit for each element of the 
 $5 \times 5$ renormalization matrix. 
A simple polynomial (linear) fit ansatz  fits the data smoothly. In Fig.~\ref{fig:Z4F-seachiral}
we illustrate an example of the chiral extrapolation in the sea 
using almost the same  value of momentum at the three $\beta$'s.

\begin{figure}[!h]
\includegraphics[scale=0.7]{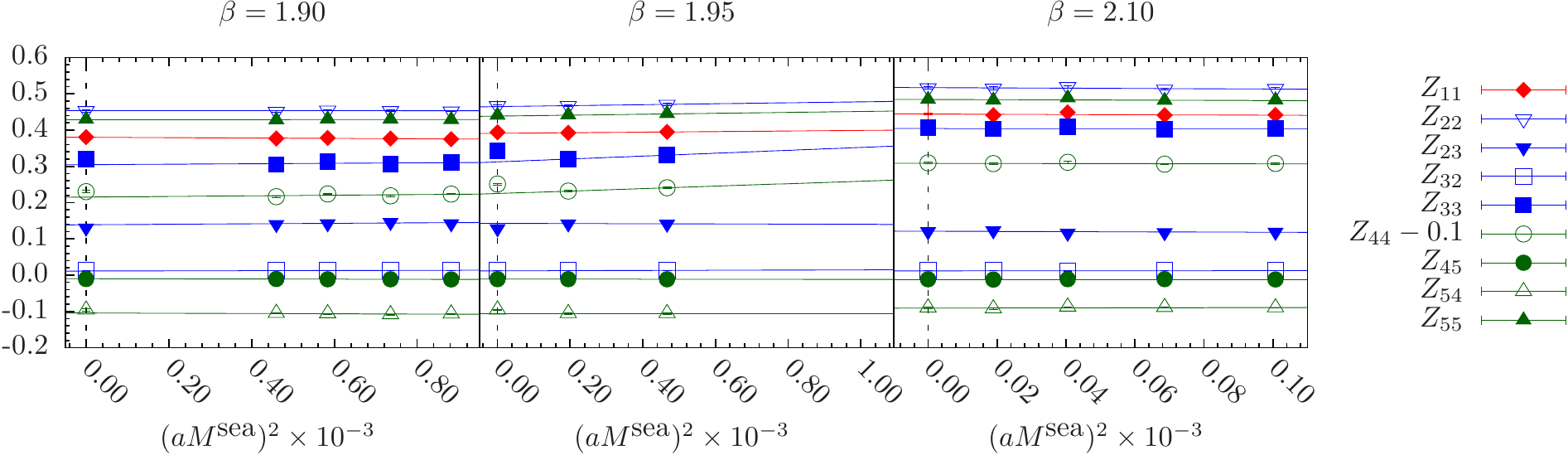} 
\caption{\label{fig:Z4F-seachiral} Example of the sea chiral extrapolation of the RC estimators at three values of $\beta$ 
computed at $(a\tilde{p})^2 \simeq 1.57$.
 }
\end{figure}

\item Having performed the chiral limit extrapolations  our estimates for the 
RCs take the form
$Z_{ij}^{RI'}(\tilde {p}^2; (a\tilde p)^2) \equiv Z_{ij}^{RI'}(\tilde {p}^2; (a\tilde p)^2, aM^{\rm{sea, val}}=0)$, 
where the first argument refers to 
the scale. Moreover we obtain the scale independent off-diagonal elements, namely 
$\Delta_{ij}^{RI'}((a\tilde p)^2) \equiv \Delta_{ij}^{RI'}((a\tilde p)^2, aM^{\rm{sea, val}}=0)$. 
The estimators $Z_{ij}^{RI'}(\tilde {p}^2; (a\tilde p)^2)$ can be evolved 
to a common scale $p_0$ using the running formula for the operators $Q_i$ known up to 
NLO~\cite{rm1:4ferm-nlo, mu:4ferm-nlo} obtaining thus  estimates of the form $Z_{ij}^{RI'}(p_0^2; (a\tilde p)^2)$.  
To  be able to carry out a controlled study of the systematic discretisation errors on the renormalized  bag-parameters, 
we apply the two methods proposed in Ref.~\cite{Constantinou:2010gr}. We recall that each of these two methods, 
called M1 and M2, prescribes a  different 
treatment of the cutoff effects. Method M1 consists in fitting $Z^{RI'}_{ij}(p_{0}^{2};\, (a \tilde{p})^2)$ 
to the linear ansatz
\be
\label{MethodM1}
Z^{RI'}_{ij}(p_{0}^{2};\, (a \tilde{p})^2)\, =\, Z^{RI'}_{ij}(p_{0}^{2})\, +\, \lambda_{i j}\times ( a \tilde{p})^2
\ee
in some large momentum region. For better controlling the systematics  we have made two choices, namely 
$(a\tilde{p})^2 \in [1.5, 2.2]$ and 
$(a\tilde{p})^2 \in [1.8, 2.2]$. As expected, thanks to subtraction of the perturbative O$(a^2 g^2)$ effects, 
the slopes $\lambda_{i j}$  show very smooth dependence on $\beta$. In fact we parametrize the slopes in Eq.~(\ref{MethodM1}) as 
$\lambda_{ij}=\lambda_{ij}^{(0)}+\lambda_{ij}^{(1)}g^2$ and we perform a simultaneous linear extrapolation to 
$(a\tilde{p})^2=0$ at all values 
of $\beta$. Then we may convert the extrapolated results $Z_{ij}^{RI'}(p_0)$ to any scale and scheme as for example 
the $\overline{\rm MS}$ scheme using NLO running. 
In Fig.~\ref{fig:Z4F-p2fit} we show the best linear fits of $Z_{ij}$ (in the $\overline{\rm{MS}}$ scheme  at the scale of 3 GeV) 
using Eq.~(\ref{MethodM1}).    
M2 method works in a quite different way from M1. It consists in getting the average RC value in a narrow window of momenta which is fixed in physical units 
for all values of $\beta$. We have carried out the M2-type analysis for two choices of momentum interval, namely $\tilde p^2 \in [10:13]$ GeV$^2$ and 
$\tilde p^2 \in [11:14]$ GeV$^2$. These rather high values of momentum offer the possibility 
to gain more confidence in the absence of hadronic state contaminations and in the validity of the RC-evolution using the 
NLO anomalous dimension at the price, however, of taking no special care to end up with reduced  O$(a^2)$ cutoff effects.          
\begin{figure}[!h]
\includegraphics[scale=0.75]{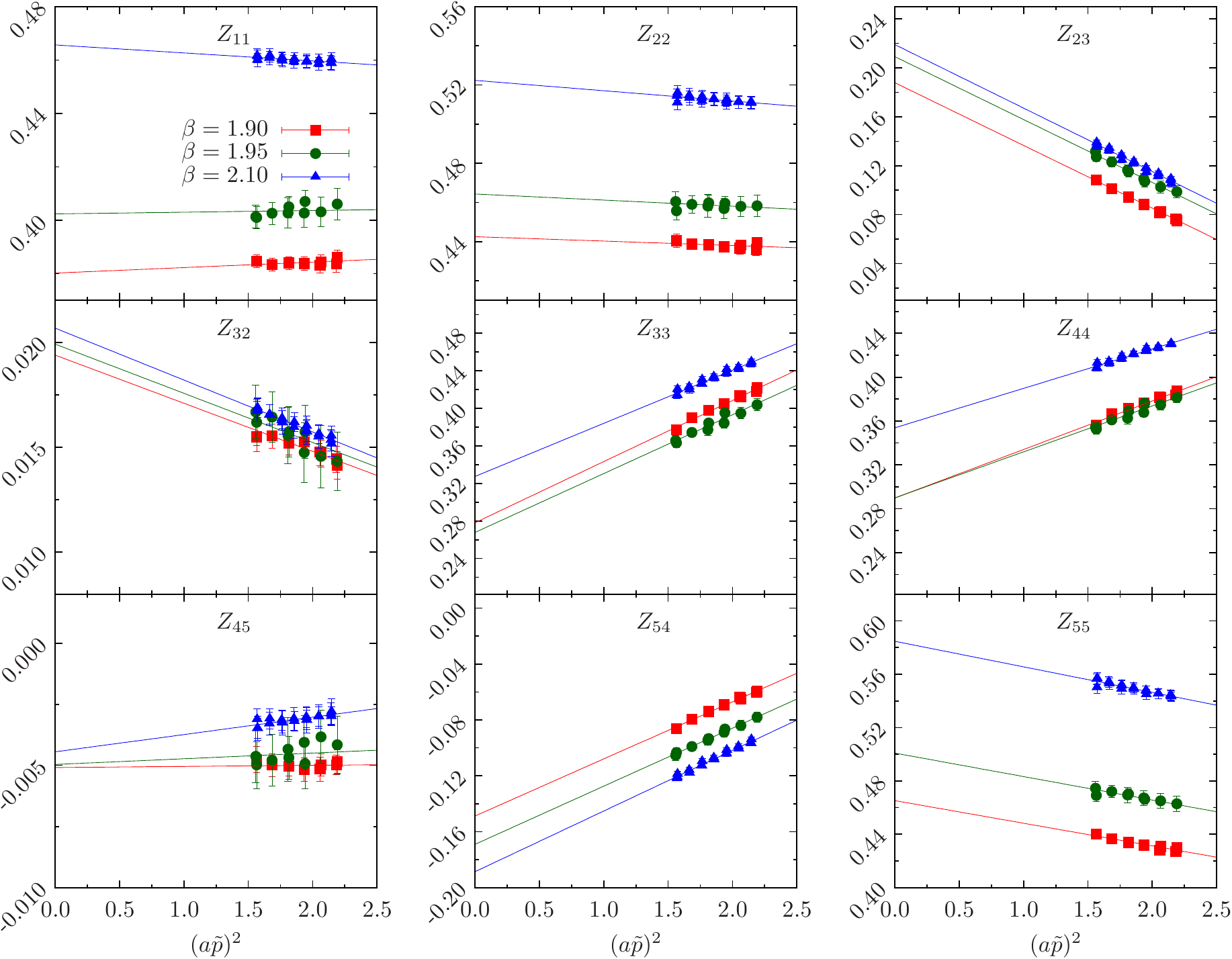} 
\caption{\label{fig:Z4F-p2fit}  Best linear fits ({\it cf.} Eq.~\ref{MethodM1}) in the momentum interval 
$(a\tilde{p})^2 \in [1.8:2.2]$ for the block-diagonal RCs, $Z_{ij}$, 
at each value of $\beta$, renormalized in the $\overline{\rm{MS}}$ scheme at 
the scale of 3 GeV. 
 }
\end{figure}

In order to get an immediate view of the impact of the RCs cutoff effects on the continuum limit estimates for the bag-parameters,
we have performed a scaling test for all $B_i$ ($i=1, \ldots, 5$) which, for the case of the neutral $K$-mixing,
is illustrated in Figs~\ref{fig:BK-scaling}.
By using M1 or M2-type RCs we get the bag-parameter estimates at some fixed reference value of the light
quark mass and then we extrapolate them
linearly in $a^2$ to the continuum limit. From the relevant panels of Fig.~\ref{fig:BK-scaling} 
it can be noticed
that for both RCs-types~\footnote{Notice that we have considered four cases corresponding to all possible combinations of
taking M1- and M2-type RCs for the four- and two-fermion operators.} the $a^2$-scaling behaviour is good
and the extrapolated continuum limit results are compatible within 1 or 2 standard deviations, depending on the case.  \\
The situation presented in Figs \ref{fig:BK-scaling} 
has to be considered as purely indicative but representative of the 
fact that at some arbritrary -reference- value of the light quark mass, 
using all four possible combinations of the RCs, the results for $B_{i=1, \ldots, 5}$ 
converge to continuum limit  values that are compatible within each other. 
We find this result very reassuring since M1- and M2-type RCs 
are computed in such a way that the corresponding cutoff effects are much different, though of O$(a^2)$. 
In this sense, by using two types of RCs we gain confidence that  
systematic effects due to the RCs RI-MOM computation and discertisation effects are under control.  
We also recall that for our final results we do not rely on plots like the ones presented in Fig.~19, but  
we perform combined chiral and continuum limit fits as those descibed  
by  the fit ans\"atze of Eqs~(17)-(19) of Section 2.  \\
Finally, in Fig.~\ref{fig:Delta-p2} we depict the behaviour of the scale independent 
off-diagonal matrix elements $\Delta_{ij}$ for $\beta=2.10$, while in Fig.~\ref{fig:Delta-scale} we plot the final estimates for $\Delta_{ij}$ against $a^2$.   
We note that in all cases  $\Delta_{ij}$ get small values around zero. 
For reader's convenience we collect in the last row  of Fig.~\ref{fig:Delta-scale} the $\Delta_{ij}$ for which 
we observe relatively larger values.

As we have anticipated in Section~\ref{sec:results} in order to take into account 
possible residual cutoff effects, in our final set of analyses we have included 
results for the bag-parameters computed both with $\Delta_{ij}=0$ and 
$\Delta_{ij} \neq 0$.

\begin{figure}[!h]
\includegraphics[scale=0.75]{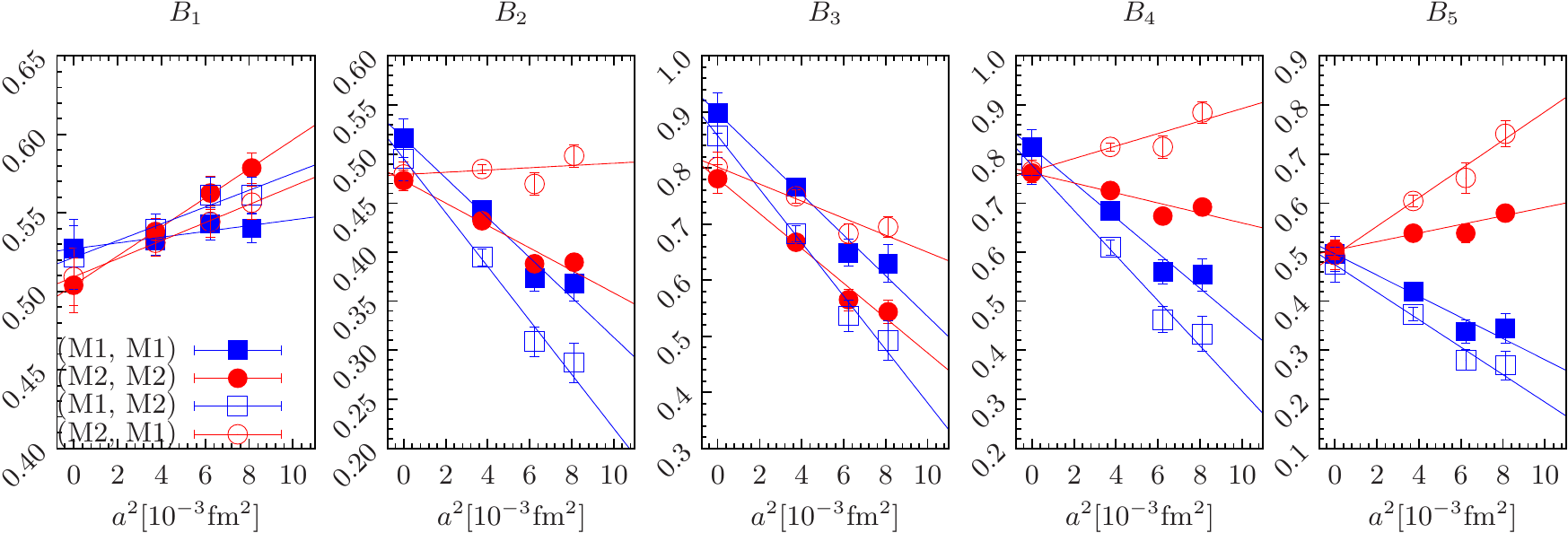} 
\caption{\label{fig:BK-scaling} $B_{1, \ldots, 5}$ estimates computed at the physical strange quark value and at a fixed reference quark 
mass $\mu_{\ell}^{ref} = 12.0~ \mbox{MeV}$ plotted against $a^2$ for the three values of the lattice spacing.  
We compare the scaling behaviour of the bag-parameter estimates computed 
with the four possible combinations of M1- and M2-type for the four- and two-fermion RCs. 
The best linear fit in $a^2$ and the corresponding CL value for each RCs combination is also shown. }
\end{figure}

\begin{figure}[!h]
\includegraphics[scale=0.63]{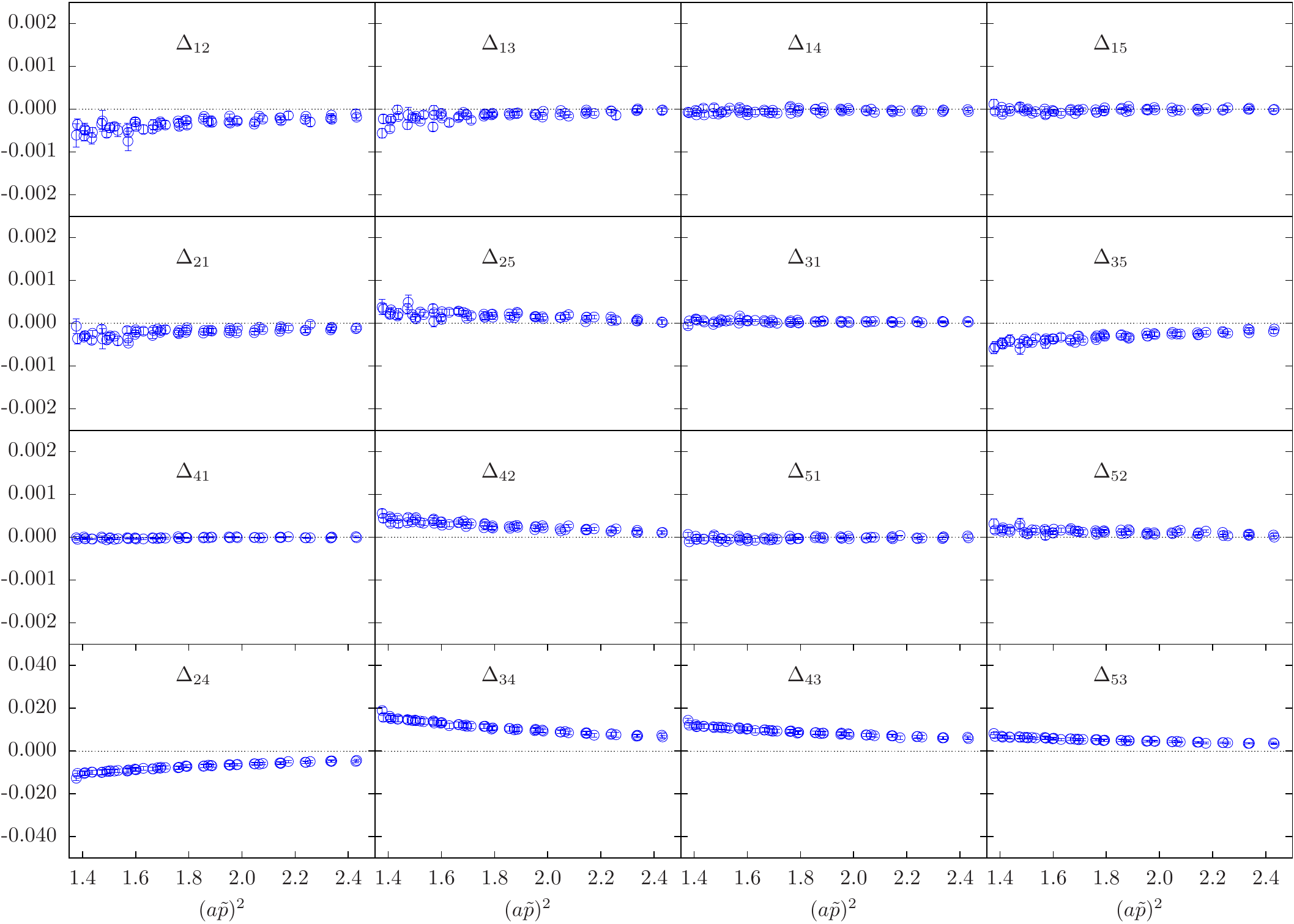} 
\caption{\label{fig:Delta-p2} The behaviour of the off-diagonal mixing coefficients $\Delta_{ij}$ as a function of  $(a\tilde p)^2$ for $\beta=2.10$.  
}
\end{figure}

\begin{figure}[!h]
\includegraphics[scale=0.63]{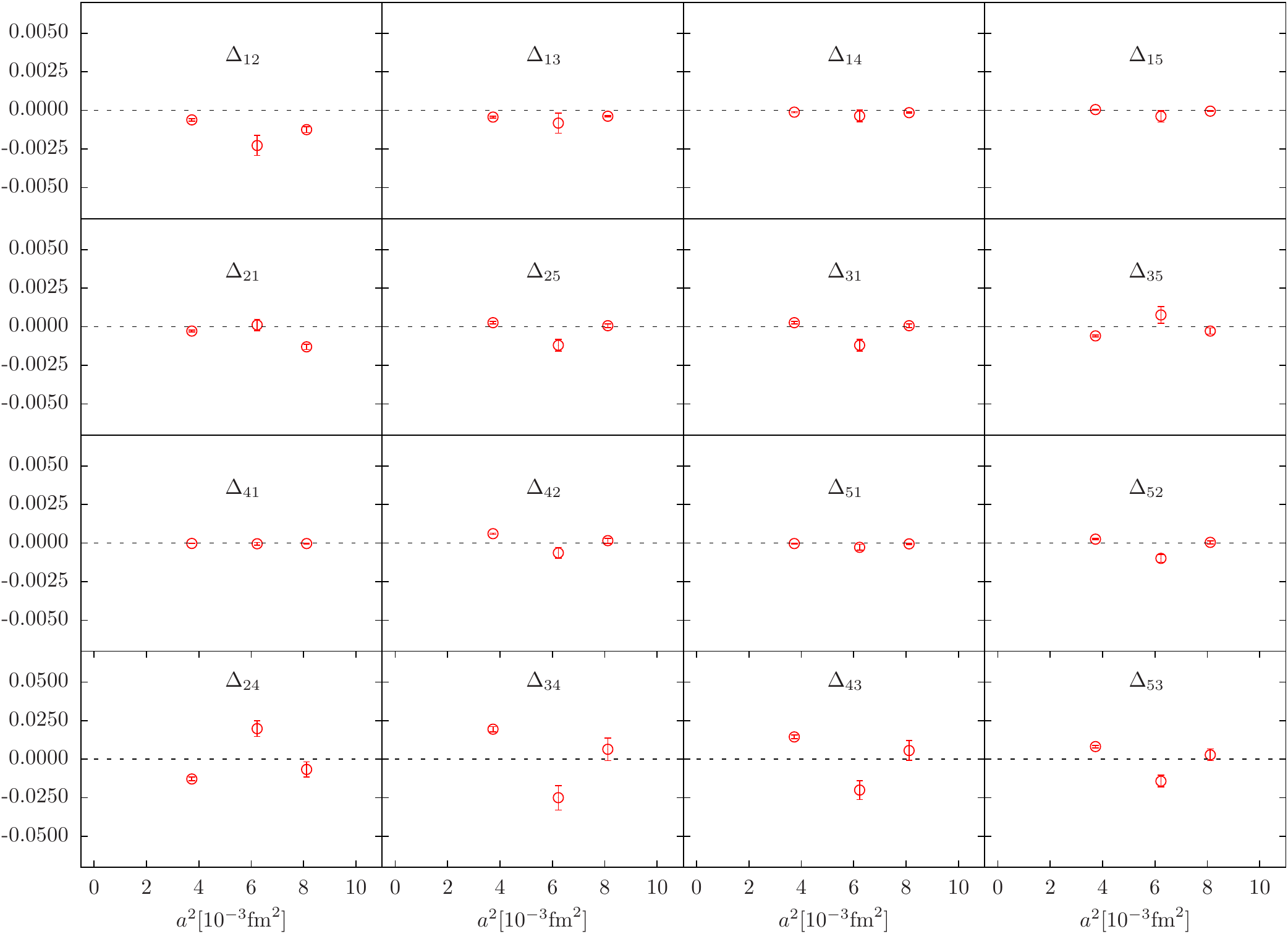} 
\caption{\label{fig:Delta-scale}  $\Delta_{ij}$ computed with the M2-method against $a^2$.  
 }
\end{figure}

\end{itemize}

We collect for convenience the RCs for the bilinear quark operators, calculated in Ref.~\cite{Carrasco:2014cwa}, 
in Table~\ref{tab:RCs-bilinear},
while in Tables~\ref{tab:RCs-4f-RI} and \ref{tab:RCs-4f-MS} 
we summarise the RCs values for the four-fermion operators.

\begin{table}[!h]
\begin{center}
\renewcommand{\arraystretch}{1.20}
\begin{tabular}{||c|c|c|c|c|c|c||}  \hline
$\beta$ & Method & $Z_V$ & $Z_A$ & $Z_P$ & $Z_S$ & $Z_q$ \\ \hline \hline
                  & M1  & 0.587(04) & 0.731(08) & 0.587(08) &  0.830(14) & 0.705(05) \\
1.90              & M2  & 0.608(03) & 0.703(02) & 0.637(06) &  0.974(04) & 0.720(02) \\
                  & WTI & \hspace*{0.2cm}0.5920(04) & - & - & - & - \\
 \hline
                  & M1  & 0.603(03) & 0.737(05) & 0.566(05) &  0.812(09) & 0.719(04) \\
1.95              & M2  & 0.614(02) & 0.714(02) & 0.606(03) &  0.913(03) & 0.727(01) \\
                  & WTI & \hspace*{0.2cm}0.6095(03) & - & - & - & -\\
\hline
                  & M1  & 0.655(03) & 0.762(04) & 0.572(02) &  0.777(06) & 0.759(04)\\
2.10              & M2  & 0.657(02) & 0.752(02) & 0.605(02) &  0.832(04) & 0.760(02) \\
                  & WTI & \hspace*{0.2cm}0.6531(02) & - & - & - & - \\
 \hline
\end{tabular}
\end{center}
\caption{ Bilinear RCs published  in Ref.~\cite{Carrasco:2014cwa}. 
The scale independent $Z_V$, $Z_A$ and the scale dependent $Z_P$, $Z_S$ and $Z_q$  
are obtained with the methods M1 and M2. 
The scale dependent RCs are expressed in the $\overline{\rm MS}$ scheme at the scale of 3 GeV.  
$Z_V$ is also obtained performing a very accurate computation employing the Ward-Takahashi identity (WTI), 
for details see Section 2.3 of Ref.~\cite{Constantinou:2010gr}). 
}
\label{tab:RCs-bilinear}
 \end{table}

\begin{table}[!t]
\begin{tabular}{|c|c|c|c|c|c|c|}
\hline
RI$'$ (3 GeV) & \multicolumn{2}{c|}{$\beta=1.90$} & \multicolumn{2}{c|}{$\beta=1.95$} & \multicolumn{2}{c|}{$\beta=2.10$} \\ \hline \hline
$Z_{ij}$ &     M1    &   M2      &   M1      &    M2     &   M1      &    M2       \\ \hline 
$Z_{11}$ &0.373(07) & 0.383(03) &0.395(05) & 0.398(05) &0.454(04) & 0.455(03) \tabularnewline
$Z_{22}$ &0.450(07) & 0.450(04) &0.474(06) & 0.469(05) &0.536(07) & 0.528(04) \tabularnewline
$Z_{23}$ &0.200(07) & 0.073(03) &0.222(05) & 0.122(05) &0.236(03) & 0.175(02) \tabularnewline
$Z_{32}$ &0.015(02) & 0.009(01) &0.015(01) & 0.010(01) &0.015(00) & 0.013(00) \tabularnewline
$Z_{33}$ &0.247(11) & 0.382(04) &0.237(07) & 0.337(05) &0.285(06) & 0.335(02) \tabularnewline
$Z_{44}$ &0.277(08) & 0.368(03) &0.277(06) & 0.344(04) &0.333(05) & 0.362(02) \tabularnewline
$Z_{45}$ &-0.012(01) & -0.010(00) &-0.012(01) & -0.010(01) &-0.012(01) & -0.011(00) \tabularnewline
$Z_{54}$ &-0.146(05) & -0.054(02) &-0.166(04) & -0.090(03) &-0.187(03) & -0.141(02) \tabularnewline
$Z_{55}$ &0.435(07) & 0.403(03) &0.471(05) & 0.439(05) &0.552(07) & 0.533(04) \tabularnewline
\hline
\end{tabular}

\caption{ Typical values for the four-fermion operator RCs at  three values of $\beta$. For M1 method linear extrapolation to 
$(a\tilde{p})^2$ has been performed using data in the interval $(a\tilde{p})^2 \in [1.8, 2.2]$, 
while for method M2 we have used data from the 
narrow momentum window determined by $\tilde p^2 \in [11:14]$ GeV$^2$. RCs are expressed in the 
in the RI$'$ scheme at the scale of 3 GeV.  
}
\label{tab:RCs-4f-RI}
 \end{table}

\begin{table}[!h]
\begin{tabular}{|c|c|c|c|c|c|c|}
\hline
${\overline{\rm{MS}}}$ (3 GeV) & \multicolumn{2}{c|}{$\beta=1.90$} & \multicolumn{2}{c|}{$\beta=1.95$} & \multicolumn{2}{c|}{$\beta=2.10$} \\ \hline \hline
$Z_{ij}$ &     M1    &   M2      &   M1      &    M2     &   M1      &    M2       \\ \hline 
$Z_{11}$ &0.379(07) & 0.389(03) &0.402(05) & 0.404(06) &0.462(04) & 0.462(03) \tabularnewline
$Z_{22}$ &0.440(07) & 0.440(03) &0.463(06) & 0.458(05) &0.524(07) & 0.516(04) \tabularnewline
$Z_{23}$ &0.182(08) & 0.050(03) &0.204(05) & 0.101(05) &0.216(03) & 0.153(02) \tabularnewline
$Z_{32}$ &0.020(02) & 0.013(01) &0.020(02) & 0.014(01) &0.021(01) & 0.017(00) \tabularnewline
$Z_{33}$ &0.293(13) & 0.453(04) &0.281(08) & 0.399(06) &0.339(07) & 0.398(03) \tabularnewline
$Z_{44}$ &0.304(09) & 0.405(03) &0.303(07) & 0.378(05) &0.364(06) & 0.397(03) \tabularnewline
$Z_{45}$ &-0.006(01) & -0.004(00) &-0.006(01) & -0.004(01) &-0.005(01) & -0.003(00) \tabularnewline
$Z_{54}$ &-0.143(06) & -0.042(02) &-0.163(04) & -0.081(03) &-0.183(03) & -0.134(02) \tabularnewline
$Z_{55}$ &0.460(08) & 0.426(04) &0.497(06) & 0.464(06) &0.584(08) & 0.564(05) \tabularnewline
\hline
\end{tabular}
\caption{ Same as in Table~\ref{tab:RCs-4f-RI}, but in the $\overline{\rm{MS}}$ scheme of \cite{mu:4ferm-nlo} 
at the scale of 3 GeV. 
}
\label{tab:RCs-4f-MS}
 \end{table}

\end{appendices}

\clearpage
\bibliographystyle{mybibstyle}
\bibliography{lattice}

\end{document}